\documentclass[12pt,preprint]{aastex}

\slugcomment{Astrophysical Journal}

\shorttitle{Infrared spectroscopy of ULIRGs}
\shortauthors{Imanishi et al.}

\begin{document}


\title{A Spitzer IRS Low-Resolution Spectroscopic Search for Buried AGNs
in Nearby Ultraluminous Infrared Galaxies 
-- A Constraint on Geometry between Energy Sources and Dust --}  


\author{Masatoshi Imanishi\altaffilmark{1}}
\affil{National Astronomical Observatory, 2-21-1, Osawa, Mitaka, Tokyo
181-8588, Japan}

\author{C. C. Dudley}
\affil{Naval Research Laboratory, Remote Sensing Division, 
Code 7211, Building 2, Room 240B, 4555 Overlook Ave SW, 
Washington DC 20375-5351, U.S.A.}

\author{Roberto Maiolino}
\affil{INAF - Osservatorio Astronomico di Roma,
via di Frascati 33, I-00040 Monte Porzio Catone, Roma, Italy}

\author{Philip R. Maloney}
\affil{formerly Center for Astrophysics and Space Astronomy, 
University of Colorado, Boulder, CO 80309-0389, U.S.A.}

\author{Takao Nakagawa}
\affil{Institute of Space and Astronautical Science, Japan Aerospace
Exploration Agency, 3-1-1 Yoshinodai, Sagamihara, Kanagawa 229-8510,
Japan} 

\and 

\author{Guido Risaliti\altaffilmark{2}}
\affil{INAF - Osservatorio Astrofisico di Arcetri, L.go E. Fermi 5,
I-50125 Firenze, Italy} 

\altaffiltext{1}{Department of Astronomy, School of Science, Graduate
University for Advanced Studies, Mitaka, Tokyo 181-8588}

\altaffiltext{2}{Harvard-Smithsonian Center for Astrophysics, 60
Garden Street, Cambridge, MA 02138, U.S.A.}

\begin{abstract}
We present the results of {\it Spitzer} IRS low-resolution infrared 5--35
$\mu$m spectroscopy of nearby ultraluminous infrared galaxies (ULIRGs)
at $z <$ 0.15.  
We focus on the search for the signatures of buried active galactic
nuclei (AGNs) in the complete sample of ULIRGs classified optically
as non-Seyferts (LINERs or HII-regions). 
In addition to polycyclic aromatic hydrocarbon (PAH) emission features 
at 6.2 $\mu$m, 7.7 $\mu$m, and 11.3 $\mu$m, the conventional
tool of starburst-AGN separation, we use the optical depths of the 9.7
$\mu$m and 18 $\mu$m silicate dust absorption features to infer the
geometry of energy sources and dust at the nuclei of these ULIRGs,
namely, whether the energy sources are spatially well mixed with dust
(a normal starburst) or are more centrally concentrated than the dust
(a buried AGN).  
Infrared spectra of at least 30\%, and possibly 50\%, of the observed
optical non-Seyfert ULIRGs are naturally explained by emission
consisting of (1) energetically insignificant, modestly obscured
(A$_{\rm V}$ $<$ 20--30 mag) 
PAH-emitting normal starbursts, and (2) energetically dominant,
highly dust-obscured, centrally concentrated energy sources with no
PAH emission. We interpret the latter component as a buried AGN.   
The fraction of ULIRGs showing some buried AGN signatures is higher in
LINER ULIRGs than in HII-region ULIRGs.
Most of the luminous buried AGN candidates are found in ULIRGs with cool
far-infrared colors.  
Where the absorption-corrected intrinsic AGN luminosities are derivable
with little uncertainty, they are found to be of the order of
10$^{12}$L$_{\odot}$, accounting for the bulk of the ULIRGs' luminosities.  
The 5--35 $\mu$m spectroscopic starburst/AGN classifications are 
generally consistent with our previous classifications based on 3--4 $\mu$m 
spectra for the same sample.
\end{abstract}


\section{Introduction}

Galaxies which radiate very large luminosities as infrared dust
emission (L$_{\rm IR}$ $\gtrsim$ 10$^{12}$L$_{\odot}$) are called
ultraluminous infrared galaxies (ULIRGs; Sanders \& Mirabel 1996).
The huge infrared luminosities mean that: (1) very powerful energy
sources, starbursts and/or active galactic nuclei (AGNs), are present
hidden behind dust; (2) energetic radiation from the energy sources is
absorbed by the surrounding dust; and (3) the heated dust grains
re-emit this energy as infrared thermal radiation. In the local
universe ($z <$ 0.3), the ULIRG population dominates the bright end of
the luminosity function \citep{soi87}. The importance of the ULIRG
population i.e. the comoving infrared energy density, increases rapidly
with increasing redshift \citep{lef05}. Identifying the dust obscured
energy sources of nearby ULIRGs is thus very important, not only to
unveil the nature of nearby ULIRGs, but also to clarify the obscured
AGN-starburst connections in the high redshift ($z >$ 0.5) ULIRG
population, where detailed investigations are difficult with
existing facilities.

In nearby ULIRGs, it is now widely accepted that the dominant energy
sources are compact ($<$500 pc), highly dust-obscured nuclear cores,
rather than extended ($>$kpc), weakly-obscured starburst activity in
the host galaxies \citep{soi00,fis00}. The most important issue is to
determine whether the obscured compact cores are powered by very
compact starbursts and/or AGNs, from which strong emission is produced
in very compact regions around the central accreting supermassive
blackholes.

If a luminous AGN is present hidden behind dust in a {\it torus}
geometry, as inferred for the classical Seyfert population
(e.g., Antonucci 1993), a large amount of energetic AGN radiation can
escape along the direction perpendicular to the torus, and 
form the so-called narrow line regions (NLRs; Antonucci 1993). 
NLRs are the sources of strong forbidden emission lines seen in the
optical--infrared, and their relative strengths are different from
those of a starburst because the spectrum at the ionizing high energy
UV range ($\lambda$ $<$ 912 $\rm \AA$) is harder in an AGN than in a
starburst. An AGN obscured by a torus-shaped dust
distribution, with a {\it well-developed NLR}, is thus distinguishable
from a starburst galaxy relatively easily, through optical spectroscopy
(classified as Seyfert 2; Veilleux \& Osterbrock 1987) or
high-resolution infrared spectroscopy \citep{gen98,arm07}.

Since the gas and dust in an AGN have angular momentum with respect to
the central supermassibe blackhole, an axisymmetric spatial distribution
is more natural than spherical geometry. In this case, the column densities
can be high in certain directions (the torus directions), but low in
others (along the torus axis). The classical Seyfert-type AGN is
naturally understood as the population where most of the torus axis
direction is transparent to the AGN's ionizing UV--soft-X-ray
radiation. However, compared to the classical Seyfert AGNs, ULIRGs are
known to contain a substantially larger amount of gas and dust in
their nuclei \citep{sam96}. For a fixed angular momentum, the
increased amount of gas and dust results in higher column densities in
all directions. If high density gas/dust blocks the bulk of the AGN's
ionizing radiation at very small radii ($<<$10 pc) even in the torus
axis direction, then no significant NLR will be present.  It is very
likely that any luminous AGNs in many ULIRGs exist in this {\it buried}
condition, with no obvious Seyfert signatures in the optical spectra
(hereafter ``optical non-Seyfert''). 
Studying only AGNs with well-developed NLRs provides a very incomplete
picture of the AGN population, and will miss many {\it buried} AGNs,
whose understanding may be essential to clarify the true nature of
ULIRGs.  

An often-used indicator to separate a buried AGN from a normal
starburst is the emission from polycyclic aromatic hydrocarbons
(PAHs), seen in infrared spectra at $\lambda_{\rm rest}$ = 3--25
$\mu$m in the rest-frame \citep{gen98,imd00}. In a normal starburst,
UV emitting HII-regions and molecular gas and dust are spatially well
mixed.  PAHs are excited by far-UV photons from stars, and strong PAH
emission is produced in photo-dissociation regions (PDRs), the
interfaces between HII-regions and molecular gas \citep{sel81}, unless
the metallicity is very low \citep{hou04b}.  In a normal starburst,
the equivalent width of the PAH emission is insensitive to level of
dust extinction, as both the PAH emission and the nearby continuum are
similarly attenuated.  Thus, a normal starburst should always show
large equivalent width PAH emission, regardless of dust extinction.
Near an AGN, the PAHs are destroyed by strong X-ray radiation from the
AGN \citep{voi92,sie04}.  If the PAHs are sufficiently shielded by
obscuration from the X-ray emission of the central AGN, the PAHs can
survive.  However, the PAH-exciting far UV emission from the AGN will
also be attenuated by the obscuration, so in a pure AGN without
starburst activity, the PAH emission will be very weak. Instead, a
PAH-free continuum from hot, submicron-sized dust grains heated by the
AGN is observed.  In a starburst/AGN composite galaxy, PAH emission from
the starburst regions is seen, but its equivalent width will be smaller
than in a pure starburst, because of the dilution by PAH-free
continuum produced by the AGN.  Thus, this PAH-based method can be
used to disentangle a buried AGN from a normal starburst at an
obscured ULIRG's core.  Since the PAH emission features are spectrally
very broad, low-resolution infrared spectroscopy is adequate.  In
addition, ULIRGs are by definition bright in the infrared, so that
this infrared PAH technique can be applied to a large number of ULIRGs
in a reasonable amount of telescope time, and thus can be a very
effective tool to systematically identify the hidden energy sources
of ULIRGs' cores.

Using low-resolution infrared spectroscopy, we also have another
powerful tool to distinguish buried AGNs from normal starbursts in the
cores of ULIRGs, by investigating the {\it geometry} between energy
sources and the gas and dust. In a normal starburst, the energy
sources (stars) and gas/dust are spatially well mixed (Figure 1a)
\citep{pux91,mcl93,for01}, while in a buried AGN, the energy source
(the central accreting supermassive blackhole) is very compact and is
more centrally concentrated than the surrounding gas and dust (Figure 1b)
\citep{soi00,im03,sie04}. The difference in geometry is reflected in
two different ways in the observed low-resolution infrared spectra.
The first is that while the absolute optical depths of dust absorption
features in the 3--10 $\mu$m range cannot exceed a certain threshold 
in a normal starburst with mixed dust/source geometry (Figure 1a), 
they can be arbitrarily large in a buried AGN (Figure 1b)
\citep{im03,idm06,lev07}. 
The second is that a strong dust temperature gradient is found
in a buried AGN (Figure 2), but not in a normal starburst.
This dust temperature gradient can be investigated by comparing 
optical depths of dust absorption features at different infrared 
wavelengths \citep{dud97,ima00,idm06}.

In this paper we present low-resolution 5--35 $\mu$m spectra of nearby
ULIRGs at $z <$ 0.15, to search for the signatures of optically
elusive \citep{mai03} buried AGNs in the cores of ULIRGs and to
determine the energetic contribution of buried AGNs.  Throughout this
paper, $H_{0}$ $=$ 75 km s$^{-1}$ Mpc$^{-1}$, $\Omega_{\rm M}$ = 0.3,
and $\Omega_{\rm \Lambda}$ = 0.7 are adopted.  The physical scale of
1$''$ is 0.34 kpc in the nearest source at $z =$ 0.018, 1.72 kpc at $z
=$ 0.1, and 2.44 kpc in the furtherest source at $z =$ 0.15.

\section{Targets}

Our main targets are optically non-Seyfert ULIRGs at $z <$ 0.15 in the
{\it IRAS} 1 Jy sample \citep{kim98}.  In this 1 Jy sample, there are
69 ULIRGs at $z <$ 0.15.  Based on the optical classifications by
Veilleux et al. (1999a; their Table 2), 28 ULIRGs are classified
optically as LINERs and 20 ULIRGs are classified optically as HII
regions.  These 48 (=28+20) ULIRGs are selected, comprising $\sim$70\%
of the 1 Jy sample at $z <$ 0.15.  
The remaining $\sim$30\% ULIRGs are classified
optically as Seyferts (17 sources) and optically unclassified (4
sources) \citep{vei99a}.  Table 1 summarizes the basic information and
{\it IRAS}-based infrared properties of the selected LINER and
HII-region ULIRGs.  Low-resolution infrared 5--35 $\mu$m spectra for
this complete sample are presented in this paper.  For most sources,
ancillary ground-based infrared 3--4 $\mu$m spectra are available
\citep{idm06}. 

\section{Observations and Data Analysis}

Observations of all the 48 ULIRGs were performed using the Infrared
Spectrograph (IRS) \citep{hou04a} onboard Spitzer Space Telescope
\citep{wer04}.  All four modules, Short-Low 2 (SL2; 5.2--7.7 $\mu$m)
and 1 (SL1; 7.4--14.5 $\mu$m), Long-Low 2 (LL2; 14.0--21.3 $\mu$m) and
1 (LL1; 19.5--38.0 $\mu$m) were used, to obtain full 5--35 $\mu$m
low-resolution (R $\sim$ 100) spectra.  13 out of the 48 sources were
in the GTO target list (PI. J. Houck, pid = 105), and their archival
spectra were analyzed.  The remaining 35 sources were observed through
our GO-1 program (PI = M. Imanishi, pid = 2306) and another GO-1
program (PI = S. Veilleux, pid = 3187).  The GO-1 program by Dr.
Veilleux covered only SL2 and SL1.  For these ULIRGs, LL2 and LL1
spectra were taken in our GO-1 program.

For our program, each observation consists of 60 sec $\times$ 2 cycles
for SL and 30 sec $\times$ 4 cycles for LL.  The observation was
performed both at A and B beam positions, and so total exposure time
was 240 sec for both SL and LL.  For the programs by Drs. Houck and
Veilleux, the total exposure times are similar or smaller for bright
sources.  Table 2 tabulates the observing log of the 48 ULIRGs.
The slit width is 3$\farcs$6 or 2 pixels for SL2 (1$\farcs$8
pixel$^{-1}$) and is 3$\farcs$7 or $\sim$2 pixels for SL1 (1$\farcs$8
pixel$^{-1}$).  For LL2 and LL1, the slit widths are 10$\farcs$5 and
10$\farcs$7, respectively, which correspond to $\sim$2 pixels for both
LL2 (5$\farcs$1 pixel$^{-1}$) and LL1 (5$\farcs$1 pixel$^{-1}$).  

The latest pipeline-processed data products at the time of our data
analysis (S11--14, pbcd files) were used.  Frames taken at the A
position were subtracted from those taken at the B position, to remove
background emission, mostly the zodiacal light.  Then, spectra were
extracted in a standard manner using the task ``apall'' in IRAF
\footnote{ IRAF is distributed by the National Optical Astronomy
Observatories, which are operated by the Association of Universities
for Research in Astronomy, Inc. (AURA), under cooperative agreement
with the National Science Foundation.}.
Apertures with 4--6 and 4--5 pixels were employed for SL and LL data, 
respectively, depending on the spatial extent of 
individual sources.  Then, spectra extracted for the A and B positions
were summed.  Wavelength calibration was made based on the files of
the {\it Spitzer} pipeline processed data, named ``b0\_wavsamp.tbl''
and ``b2\_wavsamp.tbl'' for SL and LL, respectively.  It is believed
to be accurate within 0.1 $\mu$m.  A small level of possible error in
wavelength calibration will not affect our main discussions.  Since
emission from all ULIRGs is dominated by spatially compact sources at
the observed wavelength, flux calibration was performed using the {\it
  Spitzer} pipeline processed files ``b0\_fluxcon.tbl'' (SL) and
``b2\_fluxcon.tbl'' (LL).  For SL1 spectra, data at $\lambda_{\rm
  obs}$ $>$ 14.5 $\mu$m in the observed frame are invalid (Infrared
Spectrograph Data Handbook Version 1.0) and so were discarded.  For
LL1 spectra, we used only data at $\lambda_{\rm obs}$ $<$ 35 $\mu$m,
because the scatter is large at $\lambda_{\rm obs}$ $>$ 35 $\mu$m.  No
defringing was attempted.

For flux-calibration, we adopted the values of the pipeline processed
data.  We made no attempt to re-calibrate our spectra using {\it IRAS}
measurements at 12 $\mu$m and 25 $\mu$m, primarily because (1) only
upper limits are provided for {\it IRAS} 12 $\mu$m photometry in a
sizable fraction ($>$60\%; 31/48) of the observed ULIRGs and for 25
$\mu$m data in some fraction ($\sim$20\%; 10/48) of them (Table 1),
and (2) the validity of re-calibrating IRS fixed-aperture slit
spectra, using the large aperture photometry of {\it IRAS}, is not
obvious for ULIRGs which possibly show a non-negligible fraction of
spatially extended dim emission outside the IRS slit.  Hence, the
absolute flux calibration is dependent on the accuracy of the pipeline
processed data, which is taken to be $<$20\% for SL and LL (Infrared
Spectrograph Data Handbook).  This level of possible flux uncertainty
will not affect our main discussions significantly.  In fact, in the
majority of ULIRGs with {\it IRAS} detection at 25 $\mu$m, the {\it
  Spitzer} IRS 25 $\mu$m flux agrees within 30\% to the {\it IRAS} 25
$\mu$m data, and the {\it Spitzer} IRS flux is generally smaller.  In
a small fraction of ULIRGs, {\it Spitzer} IRS 25 $\mu$m flux is $\sim$50\%
of {\it IRAS} 25 $\mu$m data, possibly because extended dim emission
is important.  For ULIRGs with {\it IRAS} non-detection at 25 $\mu$m,
the measured {\it Spizter} IRS flux at 25 $\mu$m is always smaller than
the upper limit of the {\it IRAS} data.

For a fraction of the ULIRGs, slight flux discrepancies between SL1
and LL2 were discernible, ranging from 10\% to 40\%.  When the
discrepancy is seen, the SL1 flux (3$\farcs$7 wide slit) is usually
smaller than the LL2 flux (10$\farcs$5), suggesting the presence of
extended emission in these ULIRGs.  In these cases we adjusted the
smaller SL1 (and SL2) flux to match the larger LL2 flux.

The main part of the 9.7 $\mu$m silicate absorption feature extends
$\lambda_{\rm rest}$ = 8--13 $\mu$m in the rest wavelength (e.g.,
Chiar \& Tielens 2006). For ULIRGs at $z >$ 0.11, the feature is not
fully covered with SL1 spectra at $\lambda_{\rm obs}$ = 7.4--14.5
$\mu$m. Even for ULIRGs at $z <$ 0.11, the continuum data at
$\lambda_{\rm rest}$ = 13--14 $\mu$m, which are required to estimate
the strength of the 9.7 $\mu$m silicate absorption, fall beyond
$\lambda_{\rm obs} >$ 14.5 $\mu$m, outside the wavelength range of
SL1. Thus, for the majority of the observed ULIRGs, LL2 spectra are
essential to measure the optical depth of the 9.7 $\mu$m silicate
absorption feature with a small ambiguity.

Appropriate spectral binning with 2 or 4 pixels was applied to
reduce the scatter of data points at SL2 (5.2--7.7 $\mu$m) for some
faint ULIRGs, and at $\lambda_{\rm obs}$ $\sim$ 10 $\mu$m for ULIRGs
displaying very strong 9.7 $\mu$m silicate absorption features.

\section{Result}

Figures 3 and 4 present low-resolution 5--35 $\mu$m spectra of ULIRGs
classified optically as LINERs and HII-regions, respectively.  For
most of the sources, full 5--35 $\mu$m spectra are shown here for the
first time.  Although spectra of the five ULIRGs in the GTO program
(Arp 220, IRAS 08572+3915, 12112+0305, 14348$-$1447, and 22491$-$1808)
were presented elsewhere \citep{spo06,arm07}, we include their spectra
to comprise a complete sample, to which our detailed and uniform
analysis is applied. 
For these ULIRGs, their and our spectra are, in overall, consistent 
to each other.

The spectra in Figures 3 and 4 are suitable for displaying the
properties of the 9.7 $\mu$m and 18 $\mu$m silicate dust absorption
features. However, they are not very useful for PAH emission features. 
Figure 5 presents zoom-in spectra at $\lambda_{\rm obs}$ = 5.2--14.5
$\mu$m of optical LINER and HII-region ULIRGs, to better examine 
the properties of the PAH emission features.

\subsection{PAH emission}

The majority of ULIRGs in Figure 5 show clearly detectable PAH
emission features at $\lambda_{\rm rest}$ = 6.2 $\mu$m, 7.7 $\mu$m,
and 11.3 $\mu$m.  A fraction of the ULIRGs also display the PAH
emission feature at $\lambda_{\rm rest}$ = 8.6 $\mu$m.  Emission at
$\lambda_{\rm rest}$ = 12.8 $\mu$m is also detected (Figures 3 and 4),
but it can be attributed to both the 12.8 $\mu$m PAH and [NeII]
emission lines, which are not clearly resolved in our low-resolution
spectra.  We thus will not discuss the 12.8 $\mu$m PAH emission
feature further.

The 6.2 $\mu$m PAH emission feature is attributed to an aromatic C-C
stretching fundamental \citep{all89}.  It is relatively weak, but is
isolated both from other PAH emission and from the strong 9.7 $\mu$m
silicate dust absorption feature.  Although absorption at 6.0 $\mu$m
due to H$_{2}$O ice and at 6.85 $\mu$m by hydrogenated amorphous
carbon (HAC) are present \citep{spo02}, their relatively weak nature,
compared to the 9.7 $\mu$m silicate absorption, makes it possible to
measure the 6.2 $\mu$m PAH emission strength with little ambiguity. For
this reason, the 6.2 $\mu$m feature has often been used to measure the
PAH emission strength in a ULIRG \citep{fis00,spo07}. To estimate the strength
of the 6.2 $\mu$m PAH emission feature, we adopt a linear continuum
determined from data at $\lambda_{\rm rest}$ = 6.1 $\mu$m and 6.45
$\mu$m, which is shown as solid lines in Figure 5. 

The 11.3 $\mu$m PAH emission is due to an out-of-plane aromatic C-H
bending vibration \citep{all89}.  Since it is situated close to the
peak of the strong 9.7 $\mu$m silicate absorption, its flux
attenuation can be severe if the PAH-emitting regions are highly dust
obscured.  Thus, its flux relative to other PAH features can be a good
indicator of the dust obscuration of the PAH-emitting regions.  To
estimate the 11.3 $\mu$m PAH strength, we adopt a linear continuum
connecting data points at $\lambda_{\rm rest}$ = 11.0 $\mu$m and 11.6
$\mu$m, which is shown as solid lines in Figure 5.

The 7.7 $\mu$m PAH feature is due to aromatic C-C stretching
vibrations \citep{all89}. Although it is the strongest PAH emission
feature, its substantial wavelength overlap with the broad, strong 9.7
$\mu$m silicate dust absorption feature makes it very difficult to
distinguish between 7.7 $\mu$m PAH emission and 9.7 $\mu$m silicate
absorption, particularly for sources showing strong silicate
absorption, such as the majority of ULIRGs. To reduce the effects of
the 9.7 $\mu$m absorption feature, we adopt an excess above a
linear continuum determined from data at $\lambda_{\rm rest}$ = 7.3
and 8.1 $\mu$m (shown as solid lines in Figure 5) as the 7.7 $\mu$m
PAH emission. Since many previous papers (e.g., Genzel et al. 1998)
measured the 7.7 $\mu$m PAH fluxes based on different definitions,
readers must be careful when our value is compared with other estimates
in the literature. In this paper, although the 7.7 $\mu$m PAH
strength is shown for reference, it will not be used when possible
uncertainty coming from the effects of the strong 9.7 $\mu$m silicate
feature may complicate quantitative discussions.

The 8.6 $\mu$m PAH emission due to in-plane aromatic C-H bending
vibration \citep{all89} is not used in this paper, because it is
weak and overlaps with the 9.7 $\mu$m silicate absorption, 
which make its strength highly uncertain. Other weak PAH emission
features may be present (e.g., $\lambda_{\rm rest}$ = 5.7 $\mu$m;
Spoon et al. 2002), but are not discussed in detail because no new
meaningful information is expected to be obtained.

In summary, the 6.2 $\mu$m and 11.3 $\mu$m PAH emission features are
the main focus of our discussion, and the strength of the 7.7 $\mu$m PAH
feature based on our own definition is shown just for reference.  
We fit the PAH emission features with Gaussian profiles, which reproduce
the observed data reasonably well.  
The observed rest-frame equivalent widths and luminosities of the 6.2
$\mu$m, 7.7 $\mu$m, and 11.3 $\mu$m PAH emission features, based on 
our adopted continuum levels, are summarized in Table 3. 
The uncertainties coming from the fittings are unlikely to exceed 30\%,
so once the continuum levels are fixed, the total uncertainties are of
this order.

\subsection{Silicate absorption}

Based on our adopted continuum levels, the 9.7 $\mu$m silicate
absorption feature is found in almost all ULIRGs, and 18 $\mu$m
silicate absorption is also seen, particularly in ULIRGs with strong
9.7 $\mu$m absorption (Figures 3 and 4).  Silicate dust grains show
two strong absorption peaks centered at $\lambda_{\rm rest}$ = 9.7
$\mu$m and 18.5 $\mu$m.  Outside these peaks, a weak absorption tail
extends from $\lambda_{\rm rest}$ = 8 $\mu$m to $>$30 $\mu$m
\citep{chi06} (see also Figure 6a in this paper).  Thus, all
wavelengths between $\lambda_{\rm rest}$ = 8--30 $\mu$m show excess
absorption, compared to the extrapolation of dust extinction curve
from wavelengths outside $\lambda_{\rm rest}$ = 8--30 $\mu$m.  Hence,
to accurately estimate the optical depths of the 9.7 $\mu$m
($\tau_{9.7}$) and 18 $\mu$m ($\tau_{18}$) silicate absorption
features, a continuum level must be determined using data at
$\lambda_{\rm rest}$ $<$ 8 $\mu$m and $>$30 $\mu$m.  At $\lambda_{\rm
  rest}$ = 6--8 $\mu$m, PAH emission can contaminate the continuum
level, so that in practice data points at $\lambda_{\rm rest}$ $\sim$
5.6 $\mu$m are used at the shorter wavelength side.  Data points at
$\lambda_{\rm rest}$ $\sim$ 7.1 $\mu$m are also less affected by PAH
emission and absorption features \citep{spo06}, and so can be used as
additional continuum points. Table 4 summarizes the optical depths of
the 9.7 $\mu$m silicate absorption ($\tau_{9.7}$), against power-law
continuum levels determined from data points at $\lambda_{\rm rest}$ =
5.6 $\mu$m, 7.1 $\mu$m, and $\lambda_{\rm obs}$ = 34--35 $\mu$m (the
longest part of the IRS spectra) shown as the dashed lines in Figures
3 and 4.  A power-law continuum is assumed here, because the intrinsic
unabsorbed 5--35 $\mu$m spectrum of AGN-heated hot dust emission is
roughly approximated by a power-law form \citep{wee05}.

The wavelengths of these data points used for the continuum
determination are, however, far away from the 9.7 $\mu$m and 18 $\mu$m
silicate peaks.  The optical depth values of the 9.7 $\mu$m and 18
$\mu$m silicate absorption features can vary depending on the
assumptions of the continuum data points as well as the continuum
shape (power-law, linear, spline, or polynomial etc).  
Additionally, the power-low
continua do not necessarily provide perfect fits to observed data
points outside the silicate features.  In the spectrum of the weakly
obscured ULIRG IRAS 01004$-$2237 (Figure 4, third plot), data points
at $\lambda_{\rm obs}$ = 13--18 $\mu$m are clearly above the power-law
continuum (dashed line), indicating that the power-law continuum only
{\it roughly} represents the overall continuum.  The relative
ambiguity of the $\tau_{9.7}$ value is small because the 9.7 $\mu$m
silicate feature is intrinsically strong.  However, the optical depth
of the much weaker 18 $\mu$m silicate absorption feature can be highly
uncertain.  Consequently, discussions of the presence or absence of
strong dust temperature gradients, based on the relative strength of
the 9.7 $\mu$m and 18 $\mu$m silicate dust absorption features (Figure
2), could also be highly uncertain, if we use $\tau_{9.7}$ and
$\tau_{18}$ as defined above.

For this reason, we define $\tau_{9.7}'$ and $\tau_{18}'$.  The
$\tau_{9.7}'$ value is the optical depth of the 9.7 $\mu$m silicate
feature against a power-law continuum determined from data points at
$\lambda_{\rm rest}$ = 7.1 $\mu$m and 14.2 $\mu$m.  The $\tau_{18}'$
value is the optical depth of the 18 $\mu$m silicate feature against a
power-law continuum determined from data points at $\lambda_{\rm
  rest}$ = 14.2 $\mu$m and 24 $\mu$m.  These continua are shown as
dotted lines in Figures 3 and 4.  With these new definitions, the
ambiguities are reduced, because the continuum levels are determined
using data points just outside the 9.7 $\mu$m and 18 $\mu$m features,
closer to the absorption peaks.  The $\tau_{9.7}'$ values are shown
for all ULIRGs in Table 4 (column 3).  The $\tau_{18}'$ values are
also shown in Table 4 (column 4) for ULIRGs where the 18 $\mu$m
feature is clearly seen in absorption.

The $\tau_{18}'$ values are appropriate to estimate the strength of
the 18 $\mu$m silicate feature, if the 18 $\mu$m profile is similar to
that observed in the ULIRG IRAS 08572+3915 \citep{spo06}. However,
the 18 $\mu$m silicate profile of the Galactic, highly obscured star
GCS3-I extends far beyond $\lambda_{\rm rest}$ $\sim$ 24 $\mu$m, and
reaches to $\lambda_{\rm rest}$ $\sim$ 29 $\mu$m \citep{chi06} (see
also Figure 6a in this paper).  Thus, we also define $\tau_{18}''$,
which is the optical depth of the 18 $\mu$m feature against a
power-law continuum determined from data points at $\lambda_{\rm
  rest}$ = 14.2 $\mu$m and 29 $\mu$m (shown as dashed-dotted lines in
Figure 3 and 4).  The $\tau_{18}''$ values are also summarized in
Table 4 (column 5) for ULIRGs with clearly detectable 18 $\mu$m
silicate absorption.

\subsection{Ice absorption}

A significant fraction of ULIRGs display dips at the shorter
wavelength part of the 6.2 $\mu$m PAH emission features. To better
visualize the dips, Figure 7 presents zoom-in spectra at $\lambda_{\rm
  obs}$ = 5.2--9 $\mu$m for ULIRGs showing clear or possible dips.
Spectra of two ULIRGs without discernible dips (IRAS 17044+6720 and
00456$-$2904) are also shown as examples of no clear detection.  We
ascribe the dips to the 6.0 $\mu$m H$_{2}$O absorption feature
(bending mode), as has been found in both Galactic obscured stars
\citep{gib00,kea01} and obscured ULIRGs \citep{spo02}. Since the 6.0
$\mu$m ice absorption feature is sandwiched by PAH emission features
at $\lambda_{\rm rest}$ = 5.7 and 6.2 $\mu$m, it is sometimes
difficult to distinguish whether the dip is due to ice absorption or
to PAH emission features. For ULIRGs with {\it clearly} detectable
6.0 $\mu$m H$_{2}$O ice absorption features, their observed optical
depths ($\tau_{6.0}$) are summarized in Table 5.

For ULIRGs showing clear ice (and HAC) absorption features, the
continuum flux levels just outside of the 6.2 $\mu$m PAH emission are
reduced. 
For these ULIRGs, approximate EW$_{\rm 6.2PAH}$ values after these
corrections are also shown in Table 3.

\section{Discussion}

\subsection{Magnitudes of detected starbursts}

PAH emission luminosities can roughly probe the absolute magnitudes of
PAH-emitting normal starbursts, if the extinction of the starbursts is
say less than about 20--30 mag in A$_{\rm V}$, because this amount of
A$_{\rm V}$ produces little flux attenuation ($<$1 mag) at 
$\lambda_{\rm rest}$ $>$ 5 $\mu$m, aside from the strong 9.7 $\mu$m
silicate absorption feature \citep{rie85,lut96}.

Table 3 (columns 9 and 10) tabulates the 6.2 $\mu$m PAH to infrared
luminosity ratio, L$_{\rm 6.2PAH}$/L$_{\rm IR}$, and the 11.3 $\mu$m
PAH to infrared luminosity ratio, L$_{\rm 11.3PAH}$/L$_{\rm IR}$. The
ratios in normal starburst galaxies with modest dust obscuration
(A$_{\rm V}$ $<$ 20--30 mag) are estimated to be L$_{\rm
  6.2PAH}$/L$_{\rm IR}$ $\sim$ 3.4 $\times$ 10$^{-3}$ \citep{pee04}
and L$_{\rm 11.3PAH}$/L$_{\rm IR}$ $\sim$ 1.4 $\times$ 10$^{-3}$
\citep{soi02}.  Figure 8a and 8b display the PAH-to-infrared luminosity 
ratio (ordinate) and PAH equivalent width (abscissa) for the 
6.2 $\mu$m and 11.3 $\mu$m
PAH features, respectively. The observed L$_{\rm 6.2PAH}$/L$_{\rm IR}$
and L$_{\rm 11.3PAH}$/L$_{\rm IR}$ ratios in the majority of the
observed ULIRGs are systematically lower than those seen in 
modestly-obscured normal starburst galaxies, suggesting that the 
detected starburst activity as measured by the PAH emission in ULIRGs 
is energetically insignificant. 
The dominant energy sources of ULIRGs must therefore be
either (1) highly-obscured (A$_{\rm V}$ $>>$ 30 mag) starbursts, where
the emitted PAH luminosities are substantially attenuated by dust
extinction, or (2) buried AGNs, which produce intrinsically weak PAH
emission.

\subsection{Candidates of ULIRGs possessing powerful buried AGNs}

\subsubsection{Low equivalent widths of PAH emission features}

Whether the dominant energy sources of ULIRGs' cores are 
highly-obscured normal starbursts with mixed dust/source geometry 
(Figure 1a), 
or buried AGN with centrally-concentrated energy source
geometry (Figure 1b,c), can be distinguished using the
equivalent width of the PAH emission.  Since the PAH equivalent width
(EW$_{\rm PAH}$) must always be large in a normal starburst regardless
of the amount of dust extinction, a small EW$_{\rm PAH}$ value
requires a contribution from a PAH-free continuum-emitting energy
source ($\S$1).

Figure 9 (a), (b), and (c) compare the equivalent widths of the 6.2
$\mu$m, 7.7 $\mu$m, and 11.3 $\mu$m PAH emission features.  Positive
correlations are found in all plots, suggesting that (1) variation of
the relative PAH emission strength is modest, with some level of 
scatter,  and (2) PAH-emitting starbursts usually show strong PAH 
features in all the emission modes at 6.2 $\mu$m, 7.7 $\mu$m, and 
11.3 $\mu$m. 

\citet{bra06} have investigated the 6.2 $\mu$m and 11.3 $\mu$m PAH
emission equivalent widths, based on similar definitions of PAH
strength to ours, in nearby, normal starburst galaxies without obvious
Seyfert signatures in the optical (18 sources), and found that
EW$_{\rm 6.2PAH}$ = 561$\pm$73 nm (median 553 nm) and EW$_{\rm
  11.3PAH}$ = 617$\pm$187 nm (median 597 nm). The scatter is larger
for the 11.3 $\mu$m PAH emission feature than the 6.2 $\mu$m PAH
feature.  No starburst galaxies show values smaller than EW$_{\rm
  6.2PAH}$ = 459 nm and EW$_{\rm 11.3PAH}$ = 316 nm. Since the PAH
equivalent widths are independent of the infrared luminosities of the
starburst galaxies in the range L$_{\rm IR}$ =10$^{9.7-11.6}$L$_{\odot}$ 
\citep{bra06}, we apply these values to our ULIRG sample (L$_{\rm IR}$
$>$ 10$^{12}$L$_{\odot}$).

The median values for the 6.2 $\mu$m PAH emission equivalent widths
(rest-frame) are EW$_{\rm 6.2PAH}$ = 230 nm for the whole of our ULIRG
sample (LINER + HII-region), 225 nm for LINER ULIRGs, and 300 nm for
HII-region ULIRGs.  The median 11.3 $\mu$m PAH equivalent widths are
EW$_{\rm 11.3PAH}$ = 390 nm (LINER + HII-region), 350 nm (LINER) and
460 nm (HII-region).  The median equivalent widths of both PAH
features are smaller in ULIRGs than in starburst galaxies
\citep{bra06}, suggesting that PAH-free continuum emission,
originating in AGN-heated hot dust, contributes more strongly to the
observed fluxes in ULIRGs.  We assume that if a PAH equivalent width
is more than a factor of $\sim$3 smaller than the median value in
normal starburst galaxies, then an AGN-originating PAH-free continuum
contributes importantly to the observed 5--12 $\mu$m flux and dilutes
the PAH emission.  With this criterion, ULIRGs with EW$_{\rm 6.2PAH}$
$<$ 180 nm and EW$_{\rm 11.3PAH}$ $<$ 200 nm are classified as sources
displaying clear signatures of luminous AGNs.  Since these adopted
values are clearly below the minimum values observed in normal
starburst galaxies, we believe the criterion is reasonable and
conservative to pick up luminous buried AGN candidates.

For the 7.7 $\mu$m PAH equivalent widths, based on our
definition of the PAH strength ($\S$4.1), the median 7.7 $\mu$m PAH
equivalent widths in our ULIRG sample are EW$_{\rm 7.7PAH}$ = 490 nm
(LINER) and 695 nm (HII-region). In our ULIRG sample, the median PAH
equivalent widths are smaller in LINER ULIRGs than in HII-region
ULIRGs for all of the 6.2 $\mu$m, 7.7 $\mu$m, and 11.3 $\mu$m features.
We classify ULIRGs with EW$_{\rm 7.7PAH}$ $<$ 230 nm, one-third of the
median value for HII-region ULIRGs, as those showing luminous buried
AGN signatures. Since some fraction of HII-region ULIRGs may contain
luminous buried AGNs, this EW$_{\rm 7.7PAH}$ threshold is even more
conservative than those for the 6.2 $\mu$m and 11.3 $\mu$m PAH
emission features, as determined above.

Table 6 (columns 2--4) presents detection or non-detection of buried
AGN signatures based on the PAH equivalent width threshold. The
fraction of ULIRGs with clearly detectable buried AGN signatures is
16/48 (33\%; LINER + HII-region), 10/28 (36\%; LINER) and 6/20 (30\%;
HII-region) based on observed EW$_{\rm 6.2PAH}$, or 
20/48 (42\%; LINER + HII-region), 13/28 (46\%; LINER) and 7/20 (35\%;
HII-region) based on EW$_{\rm 6.2PAH}$ after correction for the 
ice and HAC absorption feature.
The fraction is 4/48 (8\%; LINER + HII-region), 3/28 (11\%; LINER) and 
1/20 (5\%; HII-region) based on observed EW$_{\rm 7.7PAH}$, 
12/48 (25\%; LINER + HII-region), 8/28 (29\%; LINER) and 
4/20 (20\%; HII-region) based on observed EW$_{\rm 11.3PAH}$. 
The fraction of ULIRGs where very low PAH equivalent widths are found 
in any of the features is 19/48 (40\%; LINER + HII-region), 13/28 
(46\%; LINER) and 6/20 (30\%; HII-region). 
If the corrected EW$_{\rm 6.2PAH}$ values are adopted, it is 
22/48 (46\%; LINER + HII-region), 15/28 (54\%; LINER) and 7/20 (35\%; 
HII-region).

Since we adopted a conservative threshold, this method provides only
candidate ULIRGs where PAH-free continuum emission from buried AGNs
contribute very importantly to the observed IRS spectra. In
starburst/buried-AGN composite ULIRGs (Figure 1c), PAH emission from 
starbursts surrounding the central buried AGNs is less obscured and 
less flux-attenuated, making the observed PAH equivalent widths 
relatively large, particularly when AGNs are deeply buried. 
Such buried AGNs should be searched for using other methods.

\subsubsection{Absolute optical depths of dust absorption features}

In a normal starburst with mixed dust/source geometry (Figure 1a), 
the foreground, less-obscured, less-attenuated emission 
(which shows only
weak dust absorption features) dominates the observed flux, and
therefore the observed optical depths of dust absorption features cannot
exceed a certain threshold, unless very unusual dust composition
patterns are assumed \citep{im03,idm06}. In a buried AGN, with 
centrally-concentrated source geometry (Figure 1b), the so-called 
foreground screen dust model is applicable, and the observed optical 
depths can be arbitrarily large. Hence, detection of strong dust absorption
features, with optical depths substantially larger than the upper limit
set by the mixed dust/source geometry, argues for a foreground screen
dust geometry, as expected from a buried AGN \citep{im03,idm06}.  

To differentiate the geometry using the optical depths of dust
absorption features, the 9.7 $\mu$m silicate dust absorption is
particularly useful, because (1) it is 
much stronger than the 18 $\mu$m silicate feature (Figure 6a), and (2)
its intrinsic strength has been widely investigated in Galactic
obscured objects \citep{roc84,roc85}.  Assuming the Galactic dust
extinction curve, the upper limit for the optical depth of the 9.7
$\mu$m silicate absorption feature in the mixed dust/source geometry
can be derived. Appendix A presents detailed
calculations and we obtain a maximum value of $\tau_{9.7}'$ $<$ 1.7 for
mixed dust/source geometry. In fact, none of the starburst
galaxies (L$_{\rm IR}$ $<$ 10$^{11.6}$L$_{\odot}$) 
lacking optical Seyfert signatures so far observed with 
{\it Spitzer} IRS \citep{bra06} show values that exceed the threshold
($\tau_{9.7}'$ $=$ 1.7). Thus, it is reasonable to assume that ULIRGs
with $\tau_{9.7}'$ substantially larger than 1.7 possess buried AGNs
with centrally-concentrated energy source geometry.
 
In ULIRGs, it is possible that the dust extinction curve is very
different from the Galactic one, because dust coagulation in the
high density environment in ULIRGs' cores could increase typical dust
grain size up to a few microns \citep{lao93,mai00a,mai00b,ima01}. However,
we believe this possibility has minimal effect on our conclusions
about the geometry, as long as we adopt $\tau_{9.7}'$ (peak optical
depth of the 9.7 $\mu$m feature, relative to the nearby continuum just
outside the feature; see $\S$4.2), rather than $\tau_{9.7}$ (peak
optical depth of the 9.7 $\mu$m feature and continuum, relative to
continuum outside 8--30 $\mu$m; see $\S$4.2).  Figure 6b shows the
absorption optical depth of astronomical silicate dust grains of three
different sizes. The absolute absorption optical depth at the peak of
the 9.7 $\mu$m feature becomes stronger with increasing dust grain size.
However, both the absorption optical depth inside the 9.7 $\mu$m
feature and the nearby continuum just outside the feature increase in a
very similar way with increasing dust size, so that the
strength of 9.7 $\mu$m feature relative to the nearby continuum
($\tau_{9.7}'$), is virtually unchanged with varying dust grain size.
The argument of the geometry based on $\tau_{9.7}'$ values are
therefore robust to possible increase of dust grain size in ULIRGs'
cores. We conservatively take ULIRGs with $\tau_{9.7}'$ $>$ 2 as
candidates for harboring luminous centrally-concentrated buried AGNs.

Individual ULIRGs with clearly detectable buried AGN signatures based
on large $\tau_{9.7}'$ values ($>$2; Table 4) are marked with open
circles in Table 6 (column 5).  The median $\tau_{9.7}'$ values are
1.8 (LINER + HII-region), 2.1 (LINER), and 1.6 (HII-region).  The
fraction of ULIRGs with $\tau_{9.7}'$ $>$ 2, indicative of buried AGNs
with centrally-concentrated energy source geometry, is 20/48 (42\%;
LINER + HII-region), 14/28 (50\%; LINER), and 6/20 (30\%; HII-region).
The fraction of ULIRGs showing buried AGN signatures is again higher
in LINER ULIRGs than HII-region ULIRGs, as was previously found based
on the PAH-equivalent-width method ($\S5.2.1$).

We must raise three cautions regarding this method. First, a
$\tau_{9.7}'$ value smaller than 1.7 does not necessarily preclude the
presence of a luminous AGN, because a weakly obscured AGN, of course,
shows a small $\tau_{9.7}'$ value. The best example is IRAS
01004$-$2237 ($\tau_{9.7}$ $\sim$ 0.4), which shows very low PAH
emission equivalent widths, as expected from an AGN-dominated source.

Next, we used the threshold of $\tau_{9.7}'$ $>$ 2, based on the 
{\it observed} values. For ULIRGs with detectable PAH emission from
modestly obscured starbursts, the observed fluxes are a
superposition of the emission from the highly obscured cores and
the starburst emission (Figure 1c), and so the {\it observed}
$\tau_{9.7}'$ values are reduced compared to the $\tau_{9.7}'$ values
{\it only toward the highly obscured energy sources at ULIRGs' cores}.
Since subtraction of the modestly obscured starburst component,
using a starburst template spectrum, introduces an additional
uncertainty, we do not attempt it. Thus, we may miss ULIRGs which
show large $\tau_{9.7}'$ values ($>$2) {\it toward the cores} but
small {\it observed} $\tau_{9.7}'$ values ($<$2) due to the starburst
dilution.

Finally, a starburst nucleus obscured by foreground absorbing dust 
(Figure 1d) can also show a $\tau_{9.7}'$ value larger
than the threshold determined by the mixed dust/source geometry,
because in this case, a foreground screen dust model is applicable.
Assuming Galactic dust properties, $\tau_{9.7}'$ $>$ 2 corresponds to
A$_{\rm V}$ $>$ 20--40 mag \citep{roc84,roc85}. 
If ULIRGs are the mergers of gas-rich spiral galaxies \citep{san88a}, 
gas and dust in the original galaxies can be quickly transferred to 
the nuclear regions \citep{bar96,mih96}.
The bulk of possibly newborn stars and resulting dust are also expected 
to distribute in the nuclear regions.
In fact, most of high density gas, dust, and energy sources 
(starbursts and/or AGNs) in ULIRGs are observed to be concentrated 
to nuclear cores with $<$kpc scale \citep{dow98,soi00,ink06}.
The amount of gas/dust in ULIRGs' hosts at $>$kpc scale 
(which are morphologically disturbed) is unlikely to substantially exceed 
that of the original merging gas/dust-rich galaxies. 
In our Galaxy, a gas-rich spiral, A$_{\rm V}$ value is estimated to be 
$\gtrsim$30 mag in an equatorial direction \citep{rie89}, and should be
much smaller in other directions.  
It is thus not easy to produce such a large A$_{\rm V}$ value
($>$20--40 mag) with dust in a host galaxy, unless the host is viewed 
from a direction with particularly large dust column density.
Although a small fraction of ULIRGs with $\tau_{9.7}'$ $>$ 2 may have 
this geometry (Figure 1d), it is very unlikely that the majority do. 

\subsubsection{Strong dust temperature gradients}

A buried AGN with centrally-concentrated energy source geometry
(Figure 1b) should show a strong dust temperature gradient (Figure 2),
whereas a normal starburst nucleus with mixed dust/source geometry
does not, whether it is obscured by foreground dust (Figure 1d) or not
(Figure 1a).  The presence of a strong dust temperature gradient thus
supports the buried AGN scenario, and it can be detected by comparing
the optical depths of dust absorption features at different infrared
wavelengths \citep{dud97,ima00,im03,idm06}.  As Figure 2 illustrates,
in a buried AGN, the dust column density estimated from observations
at $\lambda_{\rm rest}$ $>$ 3 $\mu$m probes deeper inside of the
obscuring material as the wavelength is decreased \{i.e., Av(3$\mu$m)
$>$ Av(10$\mu$m) $>$ Av(20$\mu$m)\}.
In this geometry, dust with a fixed temperature range from a fixed layer
contributes strongly to the observed continuum emission at each wavelength,
with a small contribution from other components (Figure 2).  Thus, in
a starburst/buried AGN composite (Figure 1c), PAH-free continuum
emission from the AGN-heated dust should be detectable even if its
optical depth is larger than unity at a particular observed
wavelength, provided the buried AGN is intrinsically sufficiently
luminous that the flux-attenuated (e$^{-\tau}$; $\tau >$ 1) AGN-heated
dust emission contributes significantly to the observed flux. This has
been demonstrated in the 3--4 $\mu$m spectra of the ULIRGs IRAS
08572+3915 and UGC 5101 ($\tau_{3-4 um}$ $>$ 5; Imanishi et al. 2000;
2001; Imanishi \& Maloney 2003).

On the other hand, in a normal starburst (Figure 1a), no significant
dust temperature gradient is expected, and so the estimated dust
extinction should be similar at different wavelengths 
\{Av(3$\mu$m) $\sim$ Av(10$\mu$m) $\sim$ Av(20$\mu$m)\}.
Since in a normal starburst, emission arises uniformly within an 
emitting volume, we can typically probe emission with optical depth 
less than about unity. 
If the extinction is large enough that the short
wavelengths do not probe the whole emitting region, 
the estimated dust extinction would decrease with decreasing wavelength 
\{Av(3$\mu$m) $<$ Av(10$\mu$m) $<$ Av(20$\mu$m)\}.
Hence, the relations among Av(3$\mu$m), Av(10$\mu$m), and Av(20$\mu$m) 
are totally different between a buried AGN (Figure 1b) and a normal 
starburst (Figure 1a).

For a buried AGN, the
$\tau_{18}'$/$\tau_{9.7}'$ or $\tau_{18}''$/$\tau_{9.7}'$ ratio 
in the {\it Spitzer} IRS spectrum becomes
small compared to the value in the absence of a strong dust temperature
gradient.  The $\tau_{18}'$/$\tau_{9.7}'$ ($\tau_{18}''$/$\tau_{9.7}'$)
ratio is also insensitive to possible changes in dust grain size in
ULIRGs, because the relative strength of $\tau_{18}'$ ($\tau_{18}''$)
and $\tau_{9.7}'$ varies little with varying dust grain size (Figure 6b).
Furthermore, since both absorption features are due to silicate dust
grains, the ratio is unaffected by possibly different dust composition
patterns in individual galaxies.

For ULIRGs with large contributions from modestly obscured
PAH-emitting starbursts to observed spectra, the 18 $\mu$m silicate
dust absorption features are diluted by the PAH complex at
$\lambda_{\rm rest}$ = 16.4 $\mu$m and 17.12 $\mu$m, as well as
H$_{\rm 2}$ emission at $\lambda_{\rm rest}$ = 17.04 $\mu$m and [SIII]
emission at $\lambda_{\rm rest}$ = 18.71 $\mu$m \citep{smi04}.
Subtraction of a template spectrum of the PAH-emitting starburst
introduces an additional ambiguity in the estimate of the intrinsic
$\tau_{18}'$ ($\tau_{18}''$) value toward the ULIRG's core.  To
minimize the uncertainties, we investigate the
$\tau_{18}'$/$\tau_{9.7}'$ ($\tau_{18}''$/$\tau_{9.7}'$) ratios only
for ULIRGs where the contributions from PAH-emitting normal starbursts
are small and the 18 $\mu$m silicate features are clearly seen in
absorption in the observed spectra.  The main aim of applying this
$\tau_{18}'$/$\tau_{9.7}'$ ($\tau_{18}''$/$\tau_{9.7}'$) method to
these ULIRGs is to further strengthen the case for buried AGN
previously suggested by the large $\tau_{9.7}'$ values, rather than to
discover new buried AGN candidates.

To investigate the presence of a strong dust temperature gradient in a
ULIRG, based on a small observed $\tau_{18}'$/$\tau_{9.7}'$
($\tau_{18}''$/$\tau_{9.7}'$) ratio, we first have to know its
intrinsic ratio in the absence of a temperature gradient.  Figure 6a
presents the observed profiles of two Galactic obscured stars, GCS3-I
and WR98a \citep{chi06}.  They are thought to be mostly obscured by
foreground {\it interstellar} medium far away from the central
illuminating stars, rather than circumstellar material, so no
significant temperature gradient is expected.  To conservatively argue
that a strong dust temperature gradient is found, based on a small
observed $\tau_{18}'$/$\tau_{9.7}'$ ($\tau_{18}''$/$\tau_{9.7}'$)
ratio, we have adopted the profile showing the smallest intrinsic
$\tau_{18}'$/$\tau_{9.7}'$ ($\tau_{18}''$/$\tau_{9.7}'$) ratio as a
template.  The intrinsic ratio of GCS3-I is estimated to be
$\tau_{18}'$/$\tau_{9.7}'$ = 0.3 and $\tau_{18}''$/$\tau_{9.7}'$ =
0.35.  The ratios of WR98a, and other Galactic obscured stars for
which silicate profiles are investigated in detail (OH-IR127.8+0.0,
WR118, and WR112; Chiar \& Tielens 2006), are larger than that of
GCS3-I, so that adopting the observed GCS3-I profile as a template is
appropriate here.  If an observed $\tau_{18}'$/$\tau_{9.7}'$ or
$\tau_{18}''$/$\tau_{9.7}'$ ratio is substantially smaller than the
adopted intrinsic value ($\tau_{18}'$/$\tau_{9.7}'$ = 0.3 and
$\tau_{18}''$/$\tau_{9.7}'$ = 0.35), then the most reasonable
explanation is the presence of a strong dust temperature gradient.

Table 4 (columns 6 and 7) summarizes the $\tau_{18}'$/$\tau_{9.7}'$
and $\tau_{18}''$/$\tau_{9.7}'$ ratios for ULIRGs showing clear 18
$\mu$m silicate absorption (mostly $\tau_{9.7}'$ $>$ 2). For a
significant fraction of such ULIRGs, the observed ratios are smaller
than the conservatively assumed intrinsic one, suggesting the presence
of strong dust temperature gradients. This provides additional
evidence for these ULIRGs for centrally-concentrated buried AGNs,
previously suggested by low PAH equivalent widths and/or large
$\tau_{9.7}'$ values ($>$2). 

ULIRGs showing detectable signatures of strong dust temperature
gradients are marked as open circles in Table 6 (column 6).  The
fraction of ULIRGs with detectable signatures of strong dust
temperature gradients is 7/48 (15\%; LINER + HII-region), 6/28 (21\%;
LINER), and 1/20 (5\%; HII-region). The fraction of ULIRGs with
detectable buried AGN signatures is again higher in LINER ULIRGs than
HII-region ULIRGs.

Again, there are several caveats. First, we adopted the observed
profile of GCS3-I as the template, since the
$\tau_{18}'$/$\tau_{9.7}'$ ratio in GCS3-I is the lowest among five
Galactic sources so far investigated in detail. A source with an even
smaller intrinsic $\tau_{18}'$/$\tau_{9.7}'$ ratio may be discovered
among Galactic sources in future observations.  
If ULIRGs showed intrinsically smaller $\tau_{18}'$/$\tau_{9.7}'$ ratios 
than the adopted ratio from GCS3-I, then the small observed ratios 
in ULIRGs would not necessarily require the strong dust
temperature gradients. However, for IRAS
08572+3915, a bright ULIRG with a small $\tau_{18}'$/$\tau_{9.7}'$
ratio (Table 4), \citet{spo06} independently found the signature of a
strong dust temperature gradient, based on the fact that the observed
optical depths of crystalline silicate dust absorption at
$\lambda_{\rm rest}$ $\sim$ 11 $\mu$m, 16 $\mu$m, 18.5 $\mu$m, 23
$\mu$m, and 28 $\mu$m are systematically weaker at the longer
wavelengths than expected (their Figure 2 bottom).  A similar trend
was found also in the average spectrum of ULIRGs with $\tau_{9.7}'$
$>$ 2 (Spoon et al. 2006; their Figure 5).  We thus believe that the
small $\tau_{18}'$/$\tau_{9.7}'$ ratios observed in several ULIRGs are
most naturally explained by strong dust temperature gradients.

Next, although we investigated the $\tau_{18}'$/$\tau_{9.7}'$ ratios
only for ULIRGs with small PAH contributions from modestly obscured
starbursts, these small contributions could reduce both $\tau_{18}'$
and $\tau_{9.7}'$ ($\S$5.2.2), and slightly alter the
$\tau_{18}'$/$\tau_{9.7}'$ ratios, compared to a pure spectrum of the
highly obscured energy sources in the cores.  However, the average
spectrum of starburst galaxies shows the 30 $\mu$m to 5 $\mu$m
(rest-frame) flux ratio to be $\sim$900 \citep{bra06}, while the
majority of the ULIRGs with small $\tau_{18}'$/$\tau_{9.7}'$ show flux
ratios larger than $>$900 (Figures 1 and 2).  That is, the average
starburst spectral energy distribution is bluer at $\lambda_{\rm
  rest}$ = 5--30 $\mu$m than those ULIRGs with small
$\tau_{18}'$/$\tau_{9.7}'$.  Assuming that the average spectrum is
representative of the spectral energy distribution of the modestly
obscured starbursts in these ULIRGs, this possible effect increases (not
decreases) the observed $\tau_{18}'$/$\tau_{9.7}'$ ratios, compared to
an uncontaminated spectrum of the highly obscured core.  Thus, the
observed small $\tau_{18}'$/$\tau_{9.7}'$ ratios in these ULIRGs are
still best explained by strong dust temperature gradients around
buried AGNs, with centrally concentrated energy source geometry.

Finally, of course, silicate dust grains in the cores of ULIRGs may
have intrinsically larger $\tau_{18}'$/$\tau_{9.7}'$ ratios than the
conservatively assumed one. In this case, if a ULIRG shows a decreased
$\tau_{18}'$/$\tau_{9.7}'$ ratio compared to the intrinsic one, due to
a dust temperature gradient, but still has an {\it observed} ratio
similar to the assumed one, then such a ULIRG would not be classified
as a buried AGN. 

\subsubsection{Comparison of the strength of silicate absorption and PAH
emission features}

Figure 10a compares the optical depth of the 9.7 $\mu$m silicate dust
absorption ($\tau_{9.7}'$) and the rest-frame equivalent widths of the
6.2 $\mu$m PAH emission feature (EW$_{\rm 6.2PAH}$). With the
exception of the two outliers with both small EW$_{\rm 6.2PAH}$ and
small $\tau_{9.7}'$ (IRAS 01004$-$2237 and 17044+6720, both weakly obscured
AGNs), there is a weak anti-correlation, and ULIRGs with
$\tau_{9.7}'$ $>$ 2 (i.e., buried AGN candidates) tend to show small
EW$_{\rm 6.2PAH}$ compared to starburst-like ULIRGs with small
$\tau_{9.7}'$ ($<$2).

A highly-obscured (A$_{\rm V}$ $>>$ 30 mag) normal starburst galaxy
with mixed dust/source geometry (scenario A) predicts a relatively
flat distribution (i.e., relatively constant EW$_{\rm 6.2PAH}$ values
with varying $\tau_{9.7}'$), because the PAH equivalent width is
unchanged by dust extinction ($\S$1).  On the other hand, scenario B,
where (1) the PAH emission comes from modestly obscured (A$_{\rm V}$
$<$ 20--30 mag) starbursts exterior to the deeply obscured ULIRG
cores, and (2) the large $\tau_{9.7}'$ values reflect dust column
density toward the buried AGNs in the cores, predicts an
anti-correlation between $\tau_{9.7}'$ and EW$_{\rm 6.2PAH}$.  As the
contribution from the AGN-produced PAH-free continuum to the observed
flux increases, the the EW$_{\rm 6.2PAH}$ value decreases.  At the
same time, the increasing contribution from the buried AGN emission
(with centrally-concentrated energy source geometry) results in the
increase of the observed $\tau_{9.7}'$ value, compared to a pure
starburst with mixed source/dust geometry.  The trend in Figure 10a 
(see also Spoon et al. 2007) supports the scenario B for ULIRGs with
$\tau_{9.7}'$ $>$ 2. 

Figure 10b compares $\tau_{9.7}'$ and the rest-frame equivalent widths
of the 11.3 $\mu$m PAH emission feature (EW$_{\rm 11.3PAH}$).
Compared to Figure 10a, the decreasing trend of EW$_{\rm 11.3PAH}$ at
$\tau_{9.7}'$ $>$ 2 appears slight, at best.  In scenario B, the flux
depression of the AGN-produced continuum flux at $\lambda_{\rm rest}$
= 11.3 $\mu$m is particularly large due to the strong 9.7 $\mu$m
silicate absorption.  However, the 11.3 $\mu$m PAH emission flux is
less affected by this absorption because the PAH emission comes from
starburst regions outside the core.  Consequently, the decrease of the
EW$_{\rm 11.3PAH}$ value with increasing $\tau_{9.7}'$ value is small
compared to EW$_{\rm 6.2PAH}$, explaining the trends in both Figures
10a and 10b.

Figure 11 compares the $\tau_{9.7}'$ values with the 11.3 $\mu$m to
6.2 $\mu$m PAH luminosity ratio. In scenario A, the 11.3 $\mu$m to
6.2 $\mu$m flux ratio should decrease with increasing $\tau_{9.7}'$,
as actually seen in normal starburst galaxies \citep{bra06}, because
the flux depression of the 11.3 $\mu$m PAH emission inside the strong
9.7 $\mu$m absorption is larger. On the other hand, in scenario B, no
correlation is expected because both the 6.2 $\mu$m and 11.3 $\mu$m
PAH emission come from the modestly obscured starbursts, unrelated to
the observed $\tau_{9.7}'$ values. Figure 11 shows no significant
correlation, supporting scenario B.

\subsubsection{Is the buried AGN scenario unique?}

ULIRGs with large $\tau_{9.7}'$ ($>$2) and small
$\tau_{18}'$/$\tau_{9.7}'$ ratios are naturally explained by the
presence of buried AGN (scenario B). We here briefly comment on two
alternative scenarios which are possible but we believe are unlikely.

First, in a normal starburst with mixed dust/source geometry, if each
star is deeply buried in a large amount of {\it circumstellar} dust
cocoon, such as a protostar, then a strong dust absorption feature
($\tau_{9.7}'$ $>$2) could be produced. However, the time scale of
such a protostar phase is only $<$10$^{6}$ yr \citep{lad87}, which is
much less than the typical time scale of a normal starburst
($\sim10^{7-8}$ yr; Moorwood 1996) or even the maximum starburst
($\sim10^{7-8}$ yr; Elmegreen 1999), or a ULIRG time scale
($\sim10^{8-9}$ yr; Murphy et al. 1996). It is therefore very
unlikely that all dust-heating stars are protostars in a significant
fraction of ULIRGs.
Furthermore, this scenario produces no strong dust temperature
gradient, and so cannot reduce the observed
$\tau_{18}'$/$\tau_{9.7}'$ ratio. 

Second, it may be argued that exceptionally centrally-concentrated
starbursts remain a possibility (Figure 1e).  However, 
while starburst galaxies with L$_{\rm IR}$ =10$^{9.7-11.6}$L$_{\odot}$ 
are well explained by the mixed dust/source geometry (Brandl et al. 2006; 
see $\S$5.2.4 of our paper), there is no reason why starbursts 
at the cores of ULIRGs (L$_{\rm IR}$ $>$ 10$^{12}$L$_{\odot}$) suddenly
change their properties to exceptionally centrally-concentrated nature 
(Figure 1e).
Furthermore, \citet{soi00} studied the cores of seven nearby ULIRGs
and found that the observed emission surface brightnesses in the
majority (six out of seven) of these ULIRGs' cores are very high
($\sim$10$^{13}$ L$_{\odot}$ kpc$^{-2}$).  Even if the energy sources
are {\it uniformly} distributed over the cores ($<$500 pc in size),
the observed surface brightnesses are close to the maximum values seen
in the cores of Galactic HII regions \citep{soi00}.  If an {\it
  exceptionally centrally-concentrated} starburst were the luminosity
source in the core of a ULIRG, the emission surface brightness of the
starburst would have to be extremely high, $>$10$^{14}$ L$_{\odot}$
kpc$^{-2}$.  Among known star-forming regions, such a high surface
brightness could only barely be achieved with a super star
cluster, where many young stars are formed within a area of less than
a few pc \citep{gor01}.  
The {\it absolute luminosity} of such super star clusters is, 
however, only of order 10$^{9-10}$L$_{\odot}$, and so the super star 
cluster scenario may work for the centrally-concentrated energy sources 
of low luminosity (L$_{\rm IR}$ $\sim$ 10$^{10}$L$_{\odot}$) galaxy 
nuclei \citep{in06}.
For the centrally-concentrated energy sources at ULIRG cores, 
both {\it high emission surface brightness} and {\it large absolute 
luminosity} ($\sim$10$^{12}$ L$_{\odot}$) are required, 
making the buried AGN scenario the most plausible choice 
(Soifer et al. 2003; their $\S$6).
In fact, known AGN populations (i.e. quasars) meet these requirements 
without difficulty, while for the starburst scenario, of order 100--1000 
super star clusters must be concentrated in a very compact region and 
such a phenomenon has not been clearly confirmed from observations.
We thus believe that buried AGNs are the most natural explanation for
the centrally-concentrated energy sources at ULIRGs' cores.
The presence of luminous buried AGNs in many nearby ULIRGs 
agrees with the suggestion by \citet{tak03} that the observed lump of 
the local infrared 60 $\mu$m luminosity function at the highest luminosity 
end is due to the AGN contribution.

\subsubsection{Combination of energy diagnostic methods}

Table 6 (column 7) summarizes the strengths of the detected buried AGN
signatures in {\it Spitzer} IRS 5--35 $\mu$m spectra based on the
three methods: (1) low PAH equivalent width; (2) large $\tau_{9.7}'$
value; and (3) small $\tau_{18}'$/$\tau_{9.7}'$ and
$\tau_{18}''$/$\tau_{9.7}'$ ratio.  When buried AGN signatures in
individual ULIRGs are consistently found using all or most of
the methods, the ULIRGs are classified as {\it very strong} buried AGN
candidates, marked with open double circles.  When buried AGNs
signatures are seen in the first method (particularly based on the 
observed EW$_{\rm 6.2PAH}$ value) and/or in the third method, then the ULIRGs are
classified as {\it strong} AGN candidates (open circles).  When the
signatures are detected only in the second method and/or low observed 
EW$_{\rm 11.3PAH}$ or corrected EW$_{\rm 6.2PAH}$ values (first method),
the ULIRGs are classified as {\it possible} buried AGN candidates (open
triangles), because (1) a normal starburst nucleus obscured by
foreground dust in the host galaxy (Figure 1d) cannot be ruled out
completely and (2) the 11.3 $\mu$m PAH emission strengths in normal
starbursts have intrinsically larger scatter ($\S$5.2.1).  The fraction
of ULIRGs with {\it strong} buried AGN signatures is 16/48 (33\%; LINER
+ HII-region), 10/28 (36\%; LINER), and 6/20 (30\%; HII-regions).  When
ULIRGs with {\it possible} buried AGN signatures are included, the
fraction increases to 26/48 (54\%; LINER + HII-region), 19/28 (68\%;
LINER), and 7/20 (35\%; HII-region).
Our careful look at the infrared 5--35 $\mu$m spectra suggests 
that luminous buried AGNs reside in a much higher fraction of 
optical non-Seyfert ULIRGs than previous estimates \citep{tan99,lut99}.

We stress that the above fraction of luminous buried AGNs is only for
ULIRGs classified optically as LINERs and HII-regions (i.e.
non-Seyferts), which comprise 70\% of ULIRGs.  Given that 30\% of
ULIRGs show optical Seyfert signatures, indicative of luminous AGNs
surrounded by dusty tori, and that the AGNs are estimated to be
energetically dominant in these optical Seyfert ULIRGs
\citep{vei99b}, the total fraction of ULIRGs possessing luminous AGNs
is at least 50\%, and possibly $>$65\%, in the whole ULIRG sample at
$z <$ 0.15, including both optical Seyfert and non-Seyfert objects.

Finally, in all of the methods employed by us ($\S$5.2.1--5.2.3 and
5.2.6), ULIRGs classified optically as LINERs tend to display
signatures of luminous buried AGN more frequently than those
classified optically as HII-regions.  For starburst/buried AGN
composite ULIRGs (Figure 1c), it is likely that the optical LINER or
HII-region classifications are largely affected by the properties of
the modestly-obscured starbursts at the exteriors of the ULIRG cores,
rather than buried AGN-related emission, because optical observations
can probe only the surfaces of dusty objects.  
If shocks or superwinds are important in 
the modestly-obscured starbursts, the optical classification will be
LINERs.  If emission from HII-regions is dominant, the ULIRGs will be
classed as HII-regions.  In this case, it is not obvious why two
phenomena, a luminous buried AGN at the center and shocks in the
surface starburst, are correlated.  \citet{vei95,vei99a} found that
the emission probed in the optical is dustier in LINER ULIRGs than in
HII-region ULIRGs.  In a dusty starburst, shock-related emission can
be relatively important in the optical compared to the emission from
HII-regions themselves, resulting in optical LINER classification.
When a luminous AGN is placed at the center of a less dusty starburst,
classified optically as an HII-region, the AGN emission is more easily
detectable in the optical spectrum, making such an object an optical
Seyfert.  On the other hand, when a luminous AGN is placed at the
center of a dusty starburst classified optically as a LINER, the AGN
emission is more elusive in the optical spectrum, so that such an
object is classified as an optical non-Seyfert.  The observed higher
fraction of optically elusive {\it buried} AGN in optical LINER ULIRGs
compared to HII-region ULIRGs can qualitatively be explained by this
scenario.

\subsubsection{Absorption-corrected luminosities of buried AGNs}

For ULIRGs which show buried AGN signatures and also small PAH
equivalent widths, the observed fluxes are largely ascribed to
AGN-heated dust continuum emission.  The resulting luminosity is
conserved at each temperature (Figure 2).  We can {\it quantitatively}
estimate the absorption-corrected intrinsic dust luminosity heated by
the AGN (whole infrared), or the absorption-corrected AGN energetic
radiation luminosity (X-ray -- UV -- optical), based on the observed
fluxes at $\lambda_{\rm rest}$ $\sim$ 10 $\mu$m and the dust column
density toward the 10 $\mu$m continuum emitting regions inferred from
$\tau_{9.7}'$.  Since in a buried AGN, the 10 $\mu$m continuum
emission regions are dominated by fixed layers ($\S$5.2.3; Figure 2),
the foreground screen dust absorption model is applicable, and the
dust extinction corrections are straightforward.

To do this, we assume that $\tau_{9.7}'$ and the extinction at
$\lambda_{\rm rest}$ = 8 or 13 $\mu$m continuum just outside the 9.7
$\mu$m silicate feature (A$_{\rm cont}$) are related with
$\tau_{9.7}'$/A$_{\rm cont}$ $\sim$ 2.3 \citep{rie85}.  Table 7
summarizes the absorption-corrected AGN luminosities for selected
ULIRGs with both strong buried AGN signatures and low PAH equivalent
widths.  The flux attenuation of the 8 or 13 $\mu$m continuum outside
the 9.7 $\mu$m silicate feature ranges from a factor of 1.2 (IRAS
01004$-$2237; $\tau_{9.7}'$ $\sim$ 0.4) to 5 (IRAS 01298$-$0744;
$\tau_{9.7}'$ $\sim$ 4).  The absorption-corrected AGN luminosities
are 1--5.5 $\times$ 10$^{45}$ ergs s$^{-1}$, which are of order
$\sim$10$^{12}$L$_{\odot}$ (= 3.8 $\times$ 10$^{45}$ ergs s$^{-1}$).

The ambiguity of the derived absorption-corrected AGN luminosities
come from (1) absolute flux calibration error of the
{\it Spitzer} IRS spectra and slit loss of the compact AGN emission; 
(2) possibly different dust extinction 
curves from the Galactic one that we assume; and (3) a small
contribution from PAH-emitting, modestly-obscured starbursts, which
decreases the $\tau_{9.7}'$ values and increases the continuum flux
level, compared to pure emission from the buried AGN.  The first error
is not significant, approximately 20\% or so.  The second
uncertainty is more difficult to quantify, but our adoption of
$\tau_{9.7}'$ reduces the effects of dust grain size change
($\S$5.2.2).  For the third source of uncertainty, the effect of the
decreased $\tau_{9.7}'$ values is more important than that of the
increased continuum flux level, because the AGN-produced flux at
$\lambda_{\rm rest}$ $\sim$ 9.7 $\mu$m is more highly attenuated.
Assume for example that the modestly obscured starburst emission
contributes 10\% of the observed flux at $\lambda_{\rm rest}$ = 8 and
13 $\mu$m, outside the 9.7 $\mu$m silicate feature.  The buried
AGN-produced continuum flux is then only 10\% smaller than the
observed flux.  For ULIRGs with $\tau_{9.7}'$ $>$ 2 (i.e., buried AGN
candidates we are now discussing), since the attenuation of the
observed flux at $\lambda_{\rm rest}$ = 9.7 $\mu$m is a factor of $>$7
compared to the nearby continuum, more than 50\% of the continuum at
$\lambda_{\rm rest}$ = 9.7 $\mu$m is produced by the starburst
emission.  The true, uncontaminated $\tau_{9.7}'$ value {\it toward
  the buried AGN} is larger by a factor of $>$0.7 than the observed
value, so that the correction factor of dust extinction for the
AGN-heated dust continuum is larger by $>$30\% ($\tau_{9.7}'$/A$_{\rm
  cont}$ $\sim$ 2.3).  Therefore, the small starburst contamination
generally results in the underestimate of the intrinsic
absorption-corrected AGN luminosity.  Despite these uncertainties, the
rough agreement between the absorption-corrected buried AGN
luminosities and the observed infrared luminosities of these ULIRGs
(1--3 $\times$ 10$^{12}$L$_{\odot}$) makes it reasonable for us to
argue that the putative AGNs could quantitatively account for the bulk
of the observed infrared luminosities, at least for ULIRGs with strong
buried AGN signatures in {\it Spitzer} IRS spectra.

\subsection{Comparison with infrared 3--4 $\mu$m spectra}

\subsubsection{Ice absorption feature} 

H$_{2}$O ice absorption is also present at 3.1 $\mu$m (due to a
stretching mode) and has been detected in 3--4 $\mu$m spectra of many
ULIRGs in this sample \citep{im03,idm06,ris06}. Comparison of the 3.1
$\mu$m and 6.0 $\mu$m ice absorption features can provide a better
understanding of the properties of ice in ULIRGs.

Column 4 of Table 5 summarizes the detection or non-detection of the
3.1 $\mu$m ice feature. For the detections its optical depth is shown
in column 5. Many ULIRGs have clear detections of both the 3.1 $\mu$m
and 6.0 $\mu$m ice absorption features, reinforcing the picture that
the energy sources in many ULIRGs are obscured by a large amount of
ice-covered dust grains.

Figure 12 compares the optical depths of the 3.1 $\mu$m ($\tau_{3.1}$)
and 6.0 $\mu$m ($\tau_{6.0}$) H$_{2}$O ice absorption features, for
ULIRGs clearly showing both features.  The $\tau_{6.0}$/$\tau_{3.1}$
ratios are several times larger than expected from laboratory data.
Systematically larger $\tau_{6.0}$/$\tau_{3.1}$ ratios compared to the
laboratory prediction have previously been found in Galactic obscured
stars, but the reason is unclear \citep{gib00,kea01}.  This trend may
be common in astronomical obscured sources.  Our poor understanding of
ice absorption features hampers the search for buried AGNs with
centrally-concentrated energy source geometry based on small
$\tau_{6.0}$/$\tau_{3.1}$ ratios, as was done based on the small
$\tau_{18}'$/$\tau_{9.7}'$ ratio ($\S$5.2.3).

\subsubsection{PAH emission features} 

The 3.3 $\mu$m PAH emission feature assigned to the aromatic CH
stretch \citep{all89} was detected in the majority of this ULIRG
sample \citep{idm06}.  Figure 9 (d), (e) and (f) compare the
rest-frame equivalent widths of the 3.3 $\mu$m PAH feature (EW$_{\rm
  3.3PAH}$) with those of the 6.2 $\mu$m, 7.7 $\mu$m, and 11.3 $\mu$m
PAH features.

In Figure 9, the scatter is generally large in plots comparing PAH
features at widely separated wavelengths. 
Contribution from PAH-free AGN continuum to an observed flux can vary
to a larger degree with increasing wavelength separation, explaining
the trend in Figure 9.  

\subsubsection{Support for the strong dust temperature gradient}

In a buried AGN with a strong dust temperature gradient, dust
extinction toward the 3--4 $\mu$m continuum emission regions,
estimated using 3--4 $\mu$m data, A$_{\rm V}$(3$\mu$m), should be
larger than that toward 10 $\mu$m continuum emission region estimated
from the $\tau_{9.7}'$ value (Figure 2).  For ULIRGs with small
contamination from PAH-emitting starbursts to their 3--4 $\mu$m
spectra (small EW$_{\rm 3.3PAH}$ values), A$_{\rm V}$(3$\mu$m) can be
estimated with little ambiguity, from absorption optical depths at 3.4
$\mu$m ($\tau_{3.4}$) by bare carbonaceous dust and at 3.1 $\mu$m
($\tau_{3.1}$) by ice-covered dust deep inside molecular clouds.

IRAS 08572+3915, 12127$-$1412, 17044+6720 show 3--4 $\mu$m spectra with
very small PAH contaminations.  The $\tau_{3.4}$ values are 0.8, 0.35,
and 0.15 for IRAS 08572+3915, 12127$-$1412, 17044+6720, respectively
\citep{idm06}. 
The $\tau_{3.4}$ value reflects the column density of {\it bare}
carbonaceous dust (without an ice mantle) \citep{pen94,ima96,raw03}.
In the Galaxy, the $\tau_{9.7}'$/$\tau_{3.4}$ ratio is surprisingly
similar in different directions ($\tau_{9.7}'$/$\tau_{3.4}$ = 14--17;
Roche \& Aitken 1984, 1985; Pendleton et al. 1994).
Thus, a $\tau_{9.7}'$/$\tau_{3.4}$
ratio substantially smaller than 14--17 can be a signature of a strong
dust temperature gradient.  Adopting the $\tau_{9.7}'$ values shown in
Table 4, we obtain $\tau_{9.7}'$/$\tau_{3.4}$ = 5, 7, and 12 for IRAS
08572+3915, 12127$-$1412, 17044+6720, respectively.  These three
ULIRGs have previously been classified as very strong buried AGN
candidates, based on {\it Spitzer} IRS 5--35 $\mu$m spectra
($\S$5.2.6).  The small $\tau_{9.7}'$/$\tau_{3.4}$ ratios support the
buried AGN hypothesis.

For IRAS 08572+3915 and 17044+6720, no 3.1 $\mu$m absorption by
ice-covered dust is detected \citep{idm06}, suggesting that most of
the dust toward the 3--4 $\mu$m continuum emission sources is bare,
without an ice mantle.  However, for IRAS 12127$-$1412, the 3.1 $\mu$m
absorption feature is found with $\tau_{3.1}$ $\sim$ 0.4, suggesting a
large amount of ice-covered dust, in addition to bare dust grains.
Since the 9.7 $\mu$m absorption is seen whether the silicate dust is
bare or ice-covered, adding the column density of ice-covered dust to
that of bare carbonaceous dust grains, will raise A$_{\rm V}$(3$\mu$m)
even further, strengthening the argument for a strong dust temperature
gradient in IRAS 12127$-$1412.

\subsubsection{Classification -- AGN or starburst?} 

Table 6 (column 8) summarizes the detection or non-detection, and the
detection significance in the case of detection, of buried AGN
signatures based on ground-based 3--4 $\mu$m spectra \citep{idm06}.
The trend of buried AGN detection is generally consistent in
individual ULIRGs, in that (1) very strong buried AGN candidates at
3--4 $\mu$m are also very strong buried AGN candidates in {\it
  Spitzer} IRS 5--35 $\mu$m spectra, and (2) ULIRGs with no AGN
signatures at 3--4 $\mu$m usually show no AGN signatures at 5--35
$\mu$m.  There are some ULIRGs which have detectable AGN signatures at
5--35 $\mu$m, but not at 3--4 $\mu$m, which are explained by a larger
contamination from modestly obscured starburst emission at a shorter
wavelength \citep{ima04,ima06}. 

Finally, \citet{idm06} have previously argued, based on 3--4 $\mu$m
spectra, that the fraction of ULIRGs showing luminous buried AGN
signatures is higher in LINER ULIRGs than HII-region ULIRGs in this
sample.  This trend is exactly what we have seen in the {\it Spitzer}
IRS 5--35 $\mu$m spectra discussed here ($\S$5.2.1--5.2.6).

\subsection{Buried AGNs and infrared colors}

Figure 13 compares the strength of the 9.7 $\mu$m silicate absorption
feature ($\tau_{9.7}'$) and infrared colors, {\it Spitzer} IRS 5.4
$\mu$m to 35 $\mu$m flux ratio (Figure 13a) and {\it IRAS} 25 $\mu$m
to 60 $\mu$m flux ratio (Figure 13b).  The infrared colors are
independent of $\tau_{9.7}'$ values, with no clear positive or
negative correlation found.

The {\it IRAS} 25 $\mu$m to 60 $\mu$m flux ratio is often used to
determine whether ULIRGs show cool (f$_{25}$/f$_{60}$ $<$ 0.2) or warm
($>$ 0.2) far-infrared colors \citep{san88b}.  Although many classical
Seyfert-type AGNs surrounded by dusty tori show warm far-infrared
colors \citep{deg87}, the bulk of ULIRGs possessing luminous buried
AGN candidates ($\tau_{9.7}'$ $>$ 2) belong to cool sources (Figure
13b).  As already argued by \citet{idm06}, it is quite reasonable that
buried AGNs tend to show cooler infrared colors than optical Seyfert
AGNs.  A larger amount of dust around a central AGN in a buried AGN
compared to a Seyfert AGN surrounded by a torus ($\S$1) can naturally
produce a cooler infrared color, because the contribution from an
outer, cooler component becomes important, and makes the optical
Seyfert signatures fainter or even vanish, because even the torus axis
direction can be opaque to the bulk of the AGN's ionizing radiation.

\subsection{Future prospects}

Although infrared spectroscopic energy diagnostics enable us to
discuss the overall properties of ULIRGs statistically, follow-up
observations of infrared-selected buried AGN candidates are very
important to cross-check the reliability of these methods.  For
example, we argue an exceptionally centrally-concentrated extreme
starburst is unlikely, based on the emission surface brightness and
absolute luminosity ($\S$5.2.5). X-ray observations can be
particularly powerful in distinguishing a buried AGN from an extreme
starburst, because a luminous AGN is a much stronger X-ray emitter
than any kind of starburst. Since the column densities in the
cores of ULIRGs can be large enough to be Compton thick (X-ray
absorption N$_{\rm H}$ $>$ 10$^{24}$ cm$^{-2}$), high sensitivity
X-ray observations at E $>$ 10 keV are particularly desirable.

In a strongly X-ray emitting buried AGN surrounded by dense gas and
dust, X-ray dissociation regions (XDRs; Maloney et al. 1996) should
develop, rather than the photo-dissociation regions (PDRs) usually
seen in a starburst (a strong UV emitter). Finding XDR signatures can
thus be another powerful way to test the buried AGN scenario.
\citet{ima04,ink06} found that several infrared-selected buried AGN
candidates tend to show millimeter molecular line flux ratios expected
from XDRs.  For one of the strongest infrared-selected buried AGN
candidates, IRAS 08572+3915, \citet{sir07} detected CO (v=1--0)
absorption features up to J = 17 in ground-based infrared 4.5--5
$\mu$m observations. It is argued that such high-J lines can be
explained only by XDRs \citep{spa06}, further supporting the buried
AGN scenario.  A further search for XDR signatures in ULIRGs may
reinforce the buried AGN picture suggested from the infrared
spectroscopy.

\section{Summary}

We presented {\it Spitzer} IRS 5--35 $\mu$m low-resolution (R $\sim$
100) spectra of ULIRGs at $z <$ 0.15 in the {\it IRAS} 1 Jy sample.
ULIRGs classified optically as non-Seyferts (LINERs and HII-regions),
which comprise 70\% of the 1 Jy sample at $z <$ 0.15, were our prime
targets to 
search for signatures of optically undetected AGNs deeply buried in
dust along virtually all lines-of-sight, without well-developed narrow
line regions.  In total, 28 LINER and 20 HII-region ULIRGs (a complete
sample) were observed.  Three methods were mainly utilized: (1) the
equivalent widths of the PAH emission features at 6.2 $\mu$m, 7.7
$\mu$m, and 11.3 $\mu$m, (2) the absolute optical depth of the 9.7
$\mu$m silicate dust absorption feature, and (3) the 9.7 $\mu$m to 18
$\mu$m silicate dust absorption optical depth ratio.  The first method
traces the hardness of the radiation field (starburst-like soft or
AGN-like hard).  The latter two methods probe the geometry, i.e.
whether the dust and energy sources are spatially well mixed (a normal
starburst) or the energy source is more centrally concentrated than
dust (a buried AGN).  We found the following main conclusions.

\begin{enumerate}
\item For the majority of ULIRGs, PAH emission features were detected,
      suggesting the presence of PAH-emitting normal starburst activity.
      However, the observed PAH to infrared luminosity ratios were
      systematically smaller than in modestly obscured 
      (A$_{\rm V}$ $<$ 20--30 mag) starburst galaxies. 
      Taken at face value, the PAH-emitting starbursts with 
      modest dust obscuration are energetically insignificant in ULIRGs.  
      The dominant energy sources are (1) very highly obscured 
      (A$_{\rm V}$ $>>$ 30 mag) PAH-emitting starbursts with
      mixed dust/source geometry, whose PAH flux is attenuated, and/or
      (2) non-PAH-emitting AGNs.
\item In 19 out of the 48 (40\%) ULIRGs we studied, the observed PAH
      equivalent widths were found to be substantially lower than normal
      starburst galaxies at either of the 6.2 $\mu$m, 7.7 $\mu$m, or
      11.3 $\mu$m PAH features.  
      This fraction increased to 22/48 (46\%) if we adopted the 
      6.2 $\mu$m PAH equivalent widths after correction for the ice 
      and HAC absorption features. 
      The highly obscured energy sources of
      these ULIRGs' cores were argued to be dominated by
      non-PAH-emitting AGNs, rather than highly-obscured PAH-emitting
      starbursts.
\item In 20/48 (42\%) of ULIRGs, the observed optical depths of
      the 9.7 $\mu$m silicate dust absorption feature ($\tau_{9.7}'$)
      were substantially larger than the upper limit determined by
      mixed dust/source geometry in normal starbursts.  A
      non-PAH-emitting buried AGN with centrally-concentrated energy
      source geometry is the most natural explanation for the energy
      sources at the cores of these ULIRGs.  Comparisons of the
      $\tau_{9.7}'$ value with the PAH equivalent widths, and with the
      11.3 $\mu$m to 6.2 $\mu$m PAH luminosity ratio, supported the
      buried AGN scenario in these ULIRGs.
\item For selected ULIRGs with small PAH contamination (mostly
      ULIRGs with large $\tau_{9.7}'$), the 18 $\mu$m to 9.7 $\mu$m
      silicate absorption optical depth ratios were often significantly
      smaller than the intrinsic ratio.  A strong dust temperature
      gradient around a centrally-concentrated energy source geometry
      was strongly suggested, further supporting the buried AGN
      scenario.
\item 16/48 (33\%) of optical non-Seyfert ULIRGs showed {\it
        strong} signatures of luminous buried AGNs, and 26/48 (54\%)
      ULIRGs displayed some signatures.  Given that 30\% of ULIRGs showed
      optical Seyfert signatures and were known to possess luminous
      AGNs surrounded by dusty tori, it was suggested that luminous AGNs
      are present in at least 50\%, and possibly $>$65\%, of the
      complete ULIRG sample at $z <$ 0.15.
\item The fraction of ULIRGs showing luminous buried AGN
      signatures was always higher in optical LINER ULIRGs than
      HII-region ULIRGs using any of the above methods.  This was
      qualitatively explained by the scenario that the starburst
      regions probed in the optical are dustier in LINER ULIRGs than
      in HII-region ULIRGs.
\item For ULIRGs with strong buried AGN signatures and small PAH
      contamination, the absorption-corrected intrinsic luminosities
      of the buried AGNs were found to be quantitatively comparable to
      10$^{12}$L$_{\odot}$, accounting for the bulk of the observed
      infrared luminosities of ULIRGs.
\item Most of the luminous buried AGN candidates were found in ULIRGs
      with cool far-infrared colors. 
\item The above conclusions drawn from the {\it Spitzer} IRS 5--35
      $\mu$m spectra were generally consistent with those based on 3--4
      $\mu$m spectra obtained by \citet{idm06}. Our main conclusions
      were that optically undetected luminous buried AGNs are surely
      present in a significant fraction of non-Seyfert ULIRGs. It is
      therefore very important to carefully and quantitatively
      investigate the energetic roles of optically elusive luminous
      buried AGNs, if we are to understand the true nature of the nearby
      ULIRG population.
\end{enumerate}

\acknowledgments

We are grateful to J. E. Chiar for providing M.I. with silicate dust
absorption profiles of the Galactic sources in an electronic format,
and H. Suto for valuable discussions about the properties of silicate
dust grains. We thank V. Charmandaris for the advice about the 
{\it Spitzer} IRS data analysis and the anonymous referee for valuable 
comments.
M.I. is supported by Grants-in-Aid for Scientific
Research (16740117).  This work is based on observations made with the
Spitzer Space Telescope, which is operated by the Jet Propulsion
Laboratory, California Institute of Technology under a contract with
NASA. Support for this work was provided by NASA and also by an award
issued by JPL/Caltech.  Research in Infrared Astronomy at the Naval
Research Laboratory is supported by the Office of Naval Research
(USA).  This research has made use of the SIMBAD database, operated at
CDS, Strasbourg, France, and of the NASA/IPAC Extragalactic Database
(NED) which is operated by the Jet Propulsion Laboratory, California
Institute of Technology, under contract with the National Aeronautics
and Space Administration.

\clearpage

\appendix
\section{The maximum $\tau_{9.7}'$ value in a mixed dust/source geometry}

Figure 1a shows a schematic diagram of a typical mixed dust/source
geometry. In this geometry, the observed flux I($\tau_{\nu}$) is given by
\begin{eqnarray}
I(\tau_{\lambda}) & = & I_{0}(\lambda) \times \frac{1 -
e^{-\tau_{\lambda}}}{\tau_{\lambda}}, 
\end{eqnarray} 
where $I_{0}$($\lambda$) is an unattenuated intrinsic flux and
$\tau_{\lambda}$ is the optical depth at each wavelength, which takes
different values inside absorption features and nearby continua outside
the features.  

We here consider radiation with an intrinsically smooth spectral energy
distribution, namely I$_{0}$(8 $\mu$m) $\sim$ I$_{0}$(9.7 $\mu$m) $\sim$
I$_{0}$(13 $\mu$m), such as expected for dust emission in ULIRGs.
The peak absorption optical depth of the 9.7 $\mu$m silicate feature,
relative to the nearby continuum at 8 $\mu$m and 13 $\mu$m  
(which corresponds to $\tau_{9.7}'$ and not $\tau_{9.7}$ in our
definition in $\S$4.2), is
\begin{eqnarray}
\tau_{9.7}' & \equiv & \ln [\frac{1 - e^{-\tau_{cont}}}{\tau_{cont}} 
\times \frac{\tau_{97sil}}{1 - e^{-\tau_{97sil}}}] \\
& = & \ln [\frac{1 - e^{-\tau_{cont}}}{1 - e^{-\tau_{97sil}}}
\times \frac{\tau_{97sil}}{\tau_{cont}}], 
\end{eqnarray}
where $\tau_{97sil}$ and $\tau_{cont}$ refer to the optical depths at the
peak of the 9.7 $\mu$m silicate absorption feature and nearby continuum
at 8 $\mu$m and 13 $\mu$m just outside the feature, respectively.
The $\tau_{97sil}$ value is always larger than the $\tau_{cont}$ value,
because $\tau_{97sil}$ is the superposition of the optical depth for
continuum emission and the 9.7 $\mu$m silicate feature.

In the case of the Galactic dust extinction curve derived by \citet{rie85}, 
\begin{eqnarray}
\tau_{cont}  & = & 0.025/1.08 \times A_{\rm V} \\
\tau_{97sil} & = & 0.087/1.08 \times A_{\rm V} 
\end{eqnarray} 
where, by definition, $A_{\rm V} \equiv 1.08 \times
\tau_{V}$.  

The absorption strength of the 9.7 $\mu$m silicate feature, relative to 
A$_{\rm V}$, is different by a factor of $\sim$2 in different directions
in the Galaxy \citep{roc84,roc85}. If we take this into account, 
\begin{eqnarray}
\tau_{cont}  & = & 0.025/1.08 \times A_{\rm V} \\
\tau_{97sil} & = & (0.05 \sim 0.1) \times A_{\rm V} + 
0.025/1.08 \times A_{\rm V}. 
\end{eqnarray}

The $\tau_{cont}$ and $\tau_{97sil}$ values can be arbitrarily large
with increasing amount of dust ($A_{\rm V}$) in the mixed dust/source
geometry.  
The $\tau_{9.7}'$ value is an increasing function of A$_{\rm V}$, 
and the first term in the equation A3 ($\frac{1 - e^{-\tau_{cont}}}{1 -
e^{-\tau_{97sil}}}$) becomes unity for a very large A$_{\rm V}$ value.    
Thus, the upper limit to $\tau_{9.7}'$ is 
\begin{eqnarray}
\tau_{9.7}'(upper-limit) & = & \ln [\frac{\tau_{97sil}}{\tau_{cont}}].
\end{eqnarray}
We obtain a stringent upper limit of $\tau_{9.7}'$ $\sim$ 1.7 
in this mixed dust/source geometry for any amount of dust.

The $\tau_{cont}$/A$_{\rm V}$ and $\tau_{97sil}$/A$_{\rm V}$ ratios
can vary substantially with varying dust grain size, because the
wavelength difference between the optical V-band and infrared $\sim$10
$\mu$m is large. However, the upper limit of $\tau_{9.7}'$ in the
mixed dust/source geometry is dependent only on
$\tau_{cont}$/$\tau_{97sil}$, which should be less sensitive to
varying dust grain size (see Figure 6b), because their wavelengths are
very close.  Therefore, the above conclusion about the upper limit of
$\tau_{9.7}'$ in the mixed dust/source geometry should be robust to
possible differences in dust grain size between the cores of ULIRGs
and the Galactic diffuse interstellar medium.

\clearpage

\begin{deluxetable}{lcrrrrcrc}
\tabletypesize{\scriptsize}
\tablecaption{Observed ULIRGs at $z <$ 0.15 and their {\it IRAS}-based
infrared emission properties
\label{tbl-1}}
\tablewidth{0pt}
\tablehead{
\colhead{Object} & \colhead{Redshift}   & 
\colhead{f$_{\rm 12}$}   & 
\colhead{f$_{\rm 25}$}   & 
\colhead{f$_{\rm 60}$}   & 
\colhead{f$_{\rm 100}$}  & 
\colhead{log L$_{\rm IR}$} & 
\colhead{f$_{25}$/f$_{60}$} & 
\colhead{Optical}   \\
\colhead{} & \colhead{}   & \colhead{(Jy)} & \colhead{(Jy)} 
& \colhead{(Jy)} & \colhead{(Jy)}  & \colhead{L$_{\odot}$} & \colhead{}
& \colhead{Class}   \\
\colhead{(1)} & \colhead{(2)} & \colhead{(3)} & \colhead{(4)} & 
\colhead{(5)} & \colhead{(6)} & \colhead{(7)} & \colhead{(8)} & 
\colhead{(9)}
}
\startdata
IRAS 00188$-$0856 & 0.128 & $<$0.12 & 0.37 & 2.59 & 3.40 & 12.3 & 0.14 (C) & LINER \\  
IRAS 00482$-$2721 & 0.129 & $<$0.10 & $<$0.18 & 1.13 & 1.84 & 12.0 & $<$0.16 (C) & LINER \\
IRAS 03250+1606 & 0.129 & $<$0.10 & $<$0.15 & 1.38 & 1.77 & 12.1 & $<$0.11 (C) & LINER \\  
IRAS 04103$-$2838 & 0.118 & 0.08 & 0.54 & 1.82 & 1.71 & 12.2 & 0.30 (W) & LINER \\
IRAS 08572+3915 & 0.058 & 0.32 & 1.70 & 7.43 & 4.59 & 12.1 & 0.23 (W) & LINER \\
IRAS 09039+0503 & 0.125 & 0.07 & 0.12 & 1.48 & 2.06 & 12.1 & 0.08 (C) &LINER \\ 
IRAS 09116+0334 & 0.146 & $<$0.09 & $<$0.14 & 1.09 & 1.82 & 12.1 &$<$0.13 (C) & LINER \\  
IRAS 09539+0857 & 0.129 & $<$0.15 & $<$0.15 & 1.44 & 1.04 & 12.0 & $<$0.11 (C) &LINER \\  
IRAS 10378+1108 & 0.136 & $<$0.11 & 0.24 & 2.28 & 1.82 & 12.3 & 0.11 (C) &LINER \\  
IRAS 10485$-$1447 & 0.133 & $<$0.11 & 0.25 & 1.73 & 1.66 & 12.2 & 0.14 (C)& LINER \\  
IRAS 10494+4424 & 0.092 & $<$0.12 & 0.16 & 3.53 & 5.41 & 12.1 & 0.05 (C) &LINER \\  
IRAS 11095$-$0238 & 0.106 & 0.06 & 0.42 & 3.25 & 2.53 & 12.2 & 0.13 (C)& LINER \\  
IRAS 11130$-$2659 & 0.136 & $<$0.09 & 0.20 & 1.21 & 1.24 & 12.1 & 0.17 (C) & LINER \\
IRAS 12112+0305 & 0.073 & 0.12 & 0.51 & 8.50 & 9.98 & 12.3 & 0.06 (C) &LINER \\  
IRAS 12127$-$1412 & 0.133 & $<$0.13 & 0.24 & 1.54 & 1.13 & 12.1 & 0.16 (C)& LINER \\  
IRAS 12359$-$0725 & 0.138 & 0.09 & 0.15 & 1.33 & 1.12 & 12.1 & 0.11 (C)& LINER \\  
IRAS 13335$-$2612 & 0.125 & $<$0.13 & $<$0.14 & 1.40 & 2.10 & 12.1 & $<$0.10 (C) & LINER \\
IRAS 14252$-$1550 & 0.149 & $<$0.09 & $<$0.23 & 1.15 & 1.86 & 12.2 & $<$0.20 (C)& LINER \\  
IRAS 14348$-$1447 & 0.083 & 0.07 & 0.49 & 6.87 & 7.07 & 12.3 & 0.07 (C)& LINER \\  
IRAS 15327+2340 (Arp 220) & 0.018 & 0.48 & 7.92 & 103.33 & 112.40 & 12.1 & 0.08 (C) & LINER \\  
IRAS 16090$-$0139 & 0.134 & 0.09 & 0.26 & 3.61 & 4.87 & 12.5 & 0.07 (C)& LINER \\  
IRAS 16468+5200   & 0.150 & $<$0.06 & 0.10 & 1.01 & 1.04 & 12.1 & 0.10 (C)& LINER \\  
IRAS 16487+5447 & 0.104 & $<$0.07 & 0.20 & 2.88 & 3.07 & 12.1 & 0.07 (C) &LINER \\ 
IRAS 17028+5817   & 0.106 & $<$0.06 & 0.10 & 2.43 & 3.91 & 12.1 & 0.04 (C)& LINER \\  
IRAS 17044+6720   & 0.135 & $<$0.07 & 0.36 & 1.28 & 0.98 & 12.1 & 0.28 (W)& LINER \\  
IRAS 21329$-$2346 & 0.125 & 0.05 & 0.12 & 1.65 & 2.22 & 12.1 & 0.07 (C)& LINER \\  
IRAS 23234+0946 & 0.128 & $<$0.06 & 0.08 & 1.56 & 2.11 & 12.1 & 0.05 (C) &LINER \\  
IRAS 23327+2913 & 0.107 & $<$0.06 & 0.22 & 2.10 & 2.81 & 12.1 & 0.10 (C) &LINER \\ \hline  
IRAS 00091$-$0738 & 0.118 & $<$0.07 & 0.22 & 2.63 & 2.52 & 12.2 & 0.08 (C) & HII \\
IRAS 00456$-$2904 & 0.110 & $<$0.08 & 0.14 & 2.60 & 3.38 & 12.2 & 0.05 (C) & HII \\
IRAS 01004$-$2237 & 0.118 & 0.11 & 0.66 & 2.29 & 1.79 & 12.3 & 0.29 (W) & HII \\
IRAS 01166$-$0844 & 0.118 & 0.07 & 0.17 & 1.74 & 1.42 & 12.1 & 0.10 (C) & HII \\
IRAS 01298$-$0744 & 0.136 & $<$0.12 & 0.19 & 2.47 & 2.08 & 12.3 & 0.08 (C) & HII \\
IRAS 01569$-$2939 & 0.141 & $<$0.11 & 0.14 & 1.73 & 1.51 & 12.2 & 0.08 (C) & HII \\
IRAS 02411+0353 & 0.144 & $<$0.08 & 0.22 & 1.37 & 1.95 & 12.2 & 0.16 (C) & HII \\
IRAS 10190+1322 & 0.077 & $<$0.07 & 0.38 & 3.33 & 5.57 & 12.0 & 0.11 (C) &HII \\  
IRAS 11387+4116 & 0.149 & 0.12 & $<$0.14 & 1.02 & 1.51 & 12.2 & $<$0.14 (C) &HII \\  
IRAS 11506+1331 & 0.127 & $<$0.10 & 0.19 & 2.58 & 3.32 & 12.3 & 0.07 (C) &HII \\ 
IRAS 13509+0442 & 0.136 & 0.10 & $<$0.23 & 1.56 & 2.53 & 12.2 & $<$0.15 (C) &HII \\ 
IRAS 13539+2920 & 0.108 & $<$0.09 & 0.12 & 1.83 & 2.73 & 12.0 & 0.07 (C) &HII \\  
IRAS 14060+2919 & 0.117 & $<$0.10 & 0.14 & 1.61 & 2.42 & 12.1 & 0.09 (C) &HII \\ 
IRAS 15206+3342 & 0.125 & 0.08 & 0.35 & 1.77 & 1.89 & 12.2 & 0.20 (C) &HII \\  
IRAS 15225+2350 & 0.139 & $<$0.07 & 0.18 & 1.30 & 1.48 & 12.1 & 0.14 (C) &HII \\  
IRAS 16474+3430 & 0.111 & $<$0.13 & 0.20 & 2.27 & 2.88 & 12.1 & 0.09 (C) &HII \\ 
IRAS 20414$-$1651 & 0.086 & $<$0.65 & 0.35 & 4.36 & 5.25 & 12.2 & 0.08 (C) & HII \\  
IRAS 21208$-$0519 & 0.130 & $<$0.09 & $<$0.15 & 1.17 & 1.66 & 12.0 & $<$0.13 (C) & HII \\ 
IRAS 22206$-$2715 & 0.132 & $<$0.10 & $<$0.16 & 1.75 & 2.33 & 12.1 & $<$0.09 (C) & HII \\
IRAS 22491$-$1808 & 0.076 & 0.05 & 0.55 & 5.44 & 4.45 & 12.1 & 0.10 (C) & HII \\ \hline  
\enddata

\tablecomments{
Col.(1): Object name.  
Col.(2): Redshift.
Col.(3)--(6): f$_{12}$, f$_{25}$, f$_{60}$, and f$_{100}$ are 
{\it IRAS} fluxes at 12 $\mu$m, 25 $\mu$m, 60 $\mu$m, and 100 $\mu$m,
respectively, taken from \citet{kim98}.
Col.(7): Decimal logarithm of infrared (8$-$1000 $\mu$m) luminosity
in units of solar luminosity (L$_{\odot}$), calculated with
$L_{\rm IR} = 2.1 \times 10^{39} \times$ D(Mpc)$^{2}$
$\times$ (13.48 $\times$ $f_{12}$ + 5.16 $\times$ $f_{25}$ +
$2.58 \times f_{60} + f_{100}$) ergs s$^{-1}$ \citep{sam96}.
Since the calculation is based on our adopted cosmology, the infrared
luminosities slightly ($<$10\%) differ from the values shown in Kim \&
Sanders (1998, their Table 1, column 15).  
For sources that have upper limits in some {\it IRAS} bands, 
we can derive upper and lower limits of the infrared luminosity, by
assuming that the actual flux is the {\it IRAS}-upper limit and zero
value, respectively.  
The difference of the upper and lower values is usually very small, less
than 0.2 dex.
We assume that the infrared luminosity is the average of these values. 
Col.(8): {\it IRAS} 25 $\mu$m to 60 $\mu$m flux ratio.
ULIRGs with f$_{25}$/f$_{60}$ $<$ 0.2 and $>$ 0.2 are
classified as cool and warm sources (denoted as ``C'' and ``W''),
respectively \citep{san88b}.
Col.(9): Optical spectral classification by \citet{vei99a}.
}

\end{deluxetable}

\clearpage

\begin{deluxetable}{lcccccc}
\tabletypesize{\scriptsize}
\tablecaption{{\it Spitzer} IRS observing log
\label{tbl-2}}
\tablewidth{0pt}
\tablehead{
\colhead{Object} & \colhead{PID} &\colhead{Date} & \multicolumn{4}{c}
{Integration time [sec]} \\ 
\colhead{} & \colhead{} & \colhead{[UT]} & \colhead{SL2} & \colhead{SL1} &
\colhead{LL2} & \colhead{LL1} \\ 
\colhead{(1)} & \colhead{(2)} & \colhead{(3)} & \colhead{(4)} & 
\colhead{(5)} & \colhead{(6)} & \colhead{(7)} 
}
\startdata 
IRAS 00188$-$0856 & 105  & 2003 Dec 17 & 240 & 240 & 180 & 180 \\
IRAS 00482$-$2721 & 3187 + 2306 & 2005 Jul 7 + 2004 Jul 18 & 240 & 240 &
240 & 240 \\ 
IRAS 03250+1606   & 3187 + 2306 & 2005 Feb 11 & 240 & 240 & 240 & 240 \\ 
IRAS 04103$-$2838 & 3187 + 2306 & 2005 Feb 10 + 2004 Aug 10 & 240 & 240
& 240 & 240 \\  
IRAS 08572+3915   & 105 & 2004 Apr 15 & 84 & 84 & 84 & 84 \\ 
IRAS 09039+0503   & 3187 + 2306 & 2005 Apr 18 + 2004 Nov 17 & 240 & 240 &
240 & 240 \\  
IRAS 09116+0334   & 2306 & 2005 Apr 21 & 240 & 240 & 240 & 240 \\ 
IRAS 09539+0857   & 3187 + 2306 & 2005 Jun 5 + 2005 May 31 & 240 & 240 &
240 & 240 \\  
IRAS 10378+1108   & 105 & 2005 Jun 8 & 240 & 240 & 180 & 180 \\ 
IRAS 10485$-$1447 & 3187 + 2306 & 2005 May 23 + 2005 Jan
4 & 240 & 240 & 480 \tablenotemark{a} & 480 \tablenotemark{a} \\   
IRAS 10494+4424   & 2306 & 2004 Nov 17 & 240 & 240 & 240 & 240 \\ 
IRAS 11095$-$0238 & 105  & 2005 Jun 7 & 240 & 120 & 240 & 120 \\ 
IRAS 11130$-$2659 & 2306 & 2005 Jul 12 & 240 & 240 & 240 & 240 \\ 
IRAS 12112+0305   & 105 & 2004 Jan 4 & 84 & 84 & 120 & 120 \\ 
IRAS 12127$-$1412 & 3187 + 2306 & 2005 Jun 30 & 240 & 240 & 240 & 240 \\ 
IRAS 12359$-$0725 & 2306 & 2005 Jun 30 & 240 & 240 & 240 & 240 \\ 
IRAS 13335$-$2612 & 3187 + 2306 & 2005 Feb 15 + 2004 Jul 17 & 240 & 240
& 240 & 240 \\  
IRAS 14252$-$1550 & 2306 & 2004 Jul 17 & 240 & 240 & 240 & 240 \\ 
IRAS 14348$-$1447 & 105 & 2004 Feb 7 & 120 & 120 & 120 & 120 \\ 
Arp 220           & 105 & 2004 Feb 29 & 84 & 84 & 60 & 60 \\ 
IRAS 16090$-$0139 & 105 & 2005 Aug 6 & 120 & 120 & 180 & 180 \\ 
IRAS 16468+5200   & 2306 & 2004 Jul 14 & 480 \tablenotemark{b} & 
480 \tablenotemark{b} & 480 \tablenotemark{a} & 480 \tablenotemark{a} \\ 
IRAS 16487+5447   & 2306 & 2004 Jul 17 & 240 & 240 & 240 & 240 \\ 
IRAS 17028+5817   & 2306 & 2004 Jul 17 & 240 & 240 & 240 & 240 \\ 
IRAS 17044+6720   & 2306 & 2004 Jul 17 & 240 & 240 & 240 & 240 \\ 
IRAS 21329$-$2346 & 3187 + 2306 & 2004 Nov 16 + 2004 Oct 23 & 240 & 240
& 240 & 240 \\  
IRAS 23234+0946   & 3187 + 2306 & 2004 Dec 13 + 2004 Dec 8 & 240 & 240 &
240 & 240 \\  
IRAS 23327+2913   & 2306 & 2004 Dec 8 & 240 & 240 & 240 & 240 \\ \hline
IRAS 00091$-$0738 & 3187 + 2306 & 2005 Jun 30 + 2004 Dec 8 & 240 & 240 &
240 & 240 \\  
IRAS 00456$-$2904 & 3187 + 2306 & 2005 Jul 14 + 2004 Jul 17 & 240 & 240
& 240 & 240 \\  
IRAS 01004$-$2237 & 105 & 2004 Jan 4 & 120 & 120 & 120 & 120 \\ 
IRAS 01166$-$0844 & 3187 + 2306 & 2005 Jan 3 + 2005 Jan 15 + 2004 Jul 17
& 480 \tablenotemark{c} & 480 \tablenotemark{c} & 480 \tablenotemark{a}
& 480 \tablenotemark{a} \\     
IRAS 01298$-$0744 & 105 & 2005 Jul 14 & 240 & 240 & 120 & 120 \\ 
IRAS 01569$-$2939 & 2306 & 2004 Jul 18 & 240 & 240 & 240 & 240 \\ 
IRAS 02411+0353 & 2306 & 2005 Jan 14 & 240 & 240 & 240 & 240 \\ 
IRAS 10190+1322   & 3187 + 2306 & 2005 May 22 + 2004 Dec 12 & 
240 + 240 \tablenotemark{d} & 240 + 240 \tablenotemark{d} & 
480 \tablenotemark{a} & 480 \tablenotemark{a} \\    
IRAS 11387+4116   & 2306 & 2005 Jan 11 & 240 & 240 & 240 & 240 \\ 
IRAS 11506+1331   & 3187 + 2306 & 2005 May 25 + 2005 Jan 3 & 240 & 240 &
240 & 240 \\  
IRAS 13509+0442   & 2306 & 2004 Jul 17 & 240 & 240 & 240 & 240 \\ 
IRAS 13539+2920   & 2306 & 2005 Feb 7  & 240 & 240 & 240 & 240 \\ 
IRAS 14060+2919   & 2306 & 2004 Jul 16 & 240 & 240 & 240 & 240 \\ 
IRAS 15206+3342   & 105  & 2004 Jun 24 & 120 & 120 & 180 & 180 \\ 
IRAS 15225+2350   & 2306 & 2005 Feb 7  & 240 & 240 & 240 & 240 \\ 
IRAS 16474+3430   & 2306 & 2004 Jul 14 & 240 & 240 & 240 & 240 \\ 
IRAS 20414$-$1651 & 105  & 2004 May 14 & 120 & 120 & 120 & 120 \\ 
IRAS 21208$-$0519 & 3187 + 2306 & 2004 Nov 13 + 2004 Oct 22 & 
240 \tablenotemark{e} & 240 \tablenotemark{e}
& 480 \tablenotemark{a} & 480 \tablenotemark{a} \\   
IRAS 22206$-$2715 & 3187 + 2306 & 2004 Nov 15 + 2004 Oct 25 & 240 & 240
& 480 \tablenotemark{a} & 480 \tablenotemark{a} \\  
IRAS 22491$-$1808 & 105 & 2004 Jun 24 & 120 & 120 & 120 & 120 \\ 
\enddata

\tablenotetext{a}{Although separate spectroscopy of double nuclei
\citep{kim02} was proposed, the same coordinate was observed twice, 
probably because IRS blue peak up camera (13--18.5 $\mu$m) pointed to
the same position.} 

\tablenotetext{b}{We combined SL spectra of double nuclei with $\sim$3''
separation \citep{kim02}, because both nuclei are very faint.
The combined SL1 spectrum is smoothly connected to the LL2 spectrum.} 

\tablenotetext{c}{We combined SL spectra of double nuclei with $\sim$5''
separation \citep{kim02}, because the fainter northern nucleus is very
faint. 
The combined SL1 spectrum is smoothly connected to the LL2 spectrum.} 

\tablenotetext{d}{SL spectra of the eastern and western nuclei with 
$\sim$4'' separation \citep{kim02} are analyzed separately, because both
nuclei have similar flux levels. 
The combined SL1 spectrum is smoothly connected to the LL2 spectrum.
}  

\tablenotetext{e}{Only the brighter northern nucleus \citep{kim02} was
analyzed, because (1) the fainter southern nucleus with $\sim$7''
separation is very faint, (2) it may not be fully covered with the LL
spectrum, and (3) the SL1 spectrum of the northern nucleus is smoothly
connected to the LL2 spectrum.} 

\tablecomments{
Col.(1): Object name.
Col.(2): PID number: 105 (PI = J. Houck), 2306 (PI = M. Imanishi), and
3187 (PI = S. Veilleux). 
For sources denoted with ``3187 + 2306'', SL and LL spectra were taken
by observations with PID = 3187 and 2306, respectively. 
Col.(3): Observing date in UT. 
Col.(4): Net on-source integration time for SL2 spectroscopy in sec.
Col.(5): Net on-source integration time for SL1 spectroscopy in sec.
Col.(6): Net on-source integration time for LL2 spectroscopy in sec.
Col.(7): Net on-source integration time for LL1 spectroscopy in sec.
}

\end{deluxetable}

\clearpage

\begin{deluxetable}{lcccccccccc}
\rotate
\tabletypesize{\scriptsize}
\tablecaption{PAH emission
\label{tbl-3}}
\tablewidth{0pt}
\tablehead{
\colhead{Object} & \colhead{EW$_{\rm 6.2PAH}$} & 
\colhead{EW$_{\rm 7.7PAH}$ \tablenotemark{a}} & 
\colhead{EW$_{\rm 11.3PAH}$}  & \colhead{EW$_{\rm 3.3PAH}$}  & 
\colhead{L$_{\rm 6.2PAH}$} & 
\colhead{L$_{\rm 7.7PAH}$ \tablenotemark{a}} &
\colhead{L$_{\rm 11.3PAH}$} & \colhead{L$_{\rm 6.2PAH}$/L$_{\rm IR}$} & 
\colhead{L$_{\rm 11.3PAH}$/L$_{\rm IR}$} \\
\colhead{} & \colhead{[nm]} & \colhead{[nm]} & \colhead{[nm]} &
\colhead{[nm]} & \colhead{10$^{42}$ [ergs s$^{-1}$]}  &
\colhead{10$^{42}$ [ergs s$^{-1}$]} & 
\colhead{10$^{42}$ [ergs s$^{-1}$]} & \colhead{[$\times$ 10$^{-3}$]} & 
\colhead{[$\times$ 10$^{-3}$]} \\  
\colhead{(1)} & \colhead{(2)} & \colhead{(3)} & \colhead{(4)} &
\colhead{(5)} & \colhead{(6)} & \colhead{(7)} & \colhead{(8)} & 
\colhead{(9)} & \colhead{(10)}        
}
\startdata 
IRAS 00188$-$0856 & 85 ($\sim$40) & 305 & 290 & 50 & 3.5 & 17 & 3.4 & 0.4 & 0.4 \\
IRAS 00482$-$2721 & 260 & 710 & 465 & \nodata & 1.8 & 4.8 & 0.9 & 0.5 & 0.2 \\
IRAS 03250+1606 & 325 & 740 & 570 & 80 & 7.0 & 18 & 4.6 & 1.6 & 1.1 \\ 
IRAS 04103$-$2838 & 165 & 365 & 120 & \nodata & 8.4 & 21 & 6.6 & 1.4 & 1.1 \\ 
IRAS 08572+3915 & $<$15 & $<$100 & $<$30 & $<$5 & $<$1.9 & $<$18 &
$<$0.5 & $<$0.4 & $<$0.1 \\ 
IRAS 09039+0503 & 325 ($\sim$200) & 625 & 580 & 95 & 3.7 & 11 & 2.2 & 0.8 & 0.5 \\ 
IRAS 09116+0334 & 360 & 730 & 630 & 75 & 9.4 & 27 & 7.4 & 1.9 & 1.5 \\ 
IRAS 09539+0857 & 145 & 405 & 390 & 65 & 2.5 & 12 & 1.3 & 0.6 & 0.3 \\ 
IRAS 10378+1108 & 110 ($\sim$50) & 450 & 210 & 40 & 2.2 & 14 & 1.5 & 0.3 & 0.2 \\ 
IRAS 10485$-$1447 & 225 ($\sim$150) & 400 & 150 & 80 & 3.7 & 11 & 1.1 & 0.6 & 0.2 \\ 
IRAS 10494+4424 & 345 ($\sim$250) & 715 & 495 & 110 & 6.1 & 19 & 3.1 & 1.1 & 0.6 \\ 
IRAS 11095$-$0238 & 20 ($\sim$10) & 250 & 110 & 150 & 1.0 & 17 & 0.9 & 0.2 & 0.1 \\ 
IRAS 11130$-$2659 & 95 ($\sim$60) & 490 & 280 & \nodata & 1.9 & 13 & 1.0 & 0.4 & 0.2 \\ 
IRAS 12112+0305 & 390 & 565 & 425 & 100+160 & 6.1 & 15 & 3.2 & 0.8 & 0.4\\ 
IRAS 12127$-$1412 & 5 ($\sim$3) & 95 & 20 & 0 & 0.7 & 11 & 0.6 & 0.1 & 0.1 \\ 
IRAS 12359$-$0725 & 220 & 445 & 165 & 75 & 3.1 & 8.8 & 1.1 & 0.6 & 0.2 \\ 
IRAS 13335$-$2612 & 310 & 700 & 610 & \nodata & 6.6 & 17 & 5.1 & 1.5 & 1.2 \\ 
IRAS 14252$-$1550 & 225 & 685 & 320 & 60+60 & 3.6 & 12 & 2.3 & 0.6 & 0.4\\ 
IRAS 14348$-$1447 & 260 ($\sim$110) & 540 & 500 & 70+110 & 5.9 & 19 & 3.6 & 0.8 & 0.5 \\ 
Arp 220 & 215 ($\sim$120) & 410 & 315 & 80 & 1.5 & 5.1 & 0.8 & 0.3 & 0.2 \\ 
IRAS 16090$-$0139 & 110 ($\sim$70) & 265 & 325 & 75 & 8.1 & 30 & 5.6 & 0.7 & 0.5 \\ 
IRAS 16468+5200   & 230 & 470 & 435 & 120 & 2.9 & 11 & 2.0 & 0.7 & 0.5 \\ 
IRAS 16487+5447   & 250 & 580 & 570 & 70 & 2.8 & 9.6 & 2.2 & 0.5 & 0.4 \\ 
IRAS 17028+5817   & 375 & 800 & 595 & 120 & 4.6 & 14 & 2.8 & 0.9 & 0.6 \\ 
IRAS 17044+6720   & 55 & 165 & 30 & 10 & 3.9 & 13 & 1.0 & 0.8 & 0.2 \\ 
IRAS 21329$-$2346 & 230 & 635 & 370 & 50 & 3.0 & 12 & 1.3 & 0.6 & 0.3 \\ 
IRAS 23234+0946   & 220 & 695 & 445 & 75 & 3.1 & 11 & 2.0 & 0.7 & 0.4 \\ 
IRAS 23327+2913   & 255 & 485 & 175 & 45 & 2.5 & 6.9 & 1.4 & 0.6 & 0.3\\ 
\hline
IRAS 00091$-$0738 & 50 ($\sim$30) & 350 & 255 & \nodata & 1.3 & 15 & 1.5 & 0.2 & 0.3 \\ 
IRAS 00456$-$2904 & 385 & 750 & 545 & \nodata & 11 & 27 & 7.6 & 1.9 & 1.4 \\ 
IRAS 01004$-$2237 & 45 & 190 & 8 & \nodata & 3.1 & 19 & 0.5 & 0.4 & 0.1 \\ 
IRAS 01166$-$0844 & 30 & 430 & 190 & \nodata & 0.5 & 8.5 & 0.7 & 0.1 & 0.2 \\ 
IRAS 01298$-$0744 & 10 ($\sim$6) & 380 & 155 & \nodata & 0.7 & 37 & 1.1 & 0.1 & 0.1 \\ 
IRAS 01569$-$2939 & 115 & 555 & 255 & \nodata & 3.1 & 20 & 2.0 & 0.5 & 0.3\\ 
IRAS 02411+0353 & 300 & 695 & 375 & \nodata & 13 & 38 & 7.9 & 2.1 & 1.3 \\ 
IRAS 10190+1322E & 375 & 680 & 510 & 95 & 4.4 & 12 & 3.0 & 1.2 & 0.8\\ 
IRAS 10190+1322W & 375 & 745 & 685 & 50 & 4.6 & 11 & 3.6 & 1.2 & 0.9\\ 
IRAS 11387+4116 & 265 & 680 & 475 & 75 & 4.2 & 15 & 3.9 & 0.7 & 0.7\\ 
IRAS 11506+1331 & 195 ($\sim$120) & 540 & 305 & 95 & 8.5 & 26 & 3.4 & 1.1 & 0.4\\ 
IRAS 13509+0442 & 490 & 795 & 590 & 135 & 9.7 & 22 & 4.7 & 1.5 & 0.7\\ 
IRAS 13539+2920 & 355 & 755 & 600 & 85 & 7.6 & 22 & 4.4 & 1.9 & 1.1\\ 
IRAS 14060+2919 & 340 & 860 & 530 & 150 & 8.6 & 27 & 6.5 & 2.0 & 1.5\\ 
IRAS 15206+3342 & 225 & 525 & 230 & 55 & 12 & 32 & 7.7 & 2.0 & 1.3\\ 
IRAS 15225+2350 & 145 ($\sim$100) & 290 & 130 & 40 & 5.0 & 16 & 2.0 & 1.0 & 0.4\\ 
IRAS 16474+3430 & 390 & 820 & 575 & 105 & 9.2 & 27 & 5.8 & 1.7 & 1.1\\ 
IRAS 20414$-$1651 & 330 & 705 & 465 & 75 & 3.2 & 10 & 1.7 & 0.5 & 0.3\\ 
IRAS 21208$-$0519 & 505 & 810 & 420 & 100 & 6.0 & 14 & 2.7 & 1.5 & 0.7\\ 
IRAS 22206$-$2715 & 210 & 975 & 565 & \nodata & 3.1 & 13 & 2.7 & 0.6 & 0.5 \\ 
IRAS 22491$-$1808 & 325 & 695 & 460 & \nodata & 3.9 & 12 & 2.6 & 0.8 & 0.5 \\ 
\enddata

\tablenotetext{a}{We regard flux excess at $\lambda_{\rm rest}$ =
7.3--8.1 $\mu$m above an adopted continuum level as 7.7 $\mu$m PAH 
emission, to reduce the effects of the strong 9.7 $\mu$m silicate dust
absorption feature. 
Our definition is different from those in many previous literatures.}

\tablecomments{
Col.(1): Object name.  
Col.(2): Rest-frame equivalent width of the 6.2 $\mu$m PAH emission.
         The values in parentheses are approximate ones after
         correction for the ice and HAC absorption features. 
Col.(3): Rest-frame equivalent width of the 7.7 $\mu$m PAH emission.
Col.(4): Rest-frame equivalent width of the 11.3 $\mu$m PAH emission.
Col.(5): Rest-frame equivalent width of the 3.3 $\mu$m PAH emission,
taken from \citet{idm06}.  ---: no data.
Col.(6): Luminosity of the 6.2 $\mu$m PAH emission in units 
of 10$^{42}$ ergs s$^{-1}$.
Col.(7): Luminosity of the 7.7 $\mu$m PAH emission in units 
of 10$^{42}$ ergs s$^{-1}$.
Col.(8): Luminosity of the 11.3 $\mu$m PAH emission in units 
of 10$^{42}$ ergs s$^{-1}$.
Col.(9): The 6.2 $\mu$m PAH to infrared luminosity ratio in units of 
10$^{-3}$. 
The ratio for normal starbursts with modest dust obscuration 
(A$_{\rm V}$ $<$ 20--30 mag) is $\sim$3.4 $\times$ 10$^{-3}$ \citep{pee04}.
Col.(10): The 11.3 $\mu$m PAH to infrared luminosity ratio in units of 
10$^{-3}$.
The ratio for normal starbursts with modest dust obscuration 
(A$_{\rm V}$ $<$ 20--30 mag) is $\sim$1.4 $\times$ 10$^{-3}$ \citep{soi02}.
}

\end{deluxetable}

\clearpage

\begin{deluxetable}{lrccccc}
\tabletypesize{\scriptsize}
\tablecaption{Optical depth of the 9.7 $\mu$m and 18 $\mu$m silicate
dust absorption features
\label{tbl-4}}
\tablewidth{0pt}
\tablehead{
\colhead{Object} & \colhead{$\tau_{9.7}$} & \colhead{$\tau_{9.7}'$} &
\colhead{$\tau_{18}'$} & \colhead{$\tau_{18}''$} &
\colhead{$\tau_{18}'$/$\tau_{9.7}'$} & \colhead{$\tau_{18}''$/$\tau_{9.7}'$}   \\
\colhead{(1)} & \colhead{(2)} & \colhead{(3)} & \colhead{(4)} & 
\colhead{(5)} & \colhead{(6)} & \colhead{(7)} 
}
\startdata 
IRAS 00188$-$0856 & 3.1 & 2.5 & 0.55 & 0.65 & 0.22 & 0.26 \\
IRAS 00482$-$2721 & 2.6 & 2.1 & 0.6 & 0.75 & 0.29 & 0.36 \\ 
IRAS 03250+1606 & 1.8 & 1.6 & \nodata & \nodata & \nodata & \nodata \\
IRAS 04103$-$2838 & 0.5 & 0.6 & \nodata & \nodata & \nodata & \nodata \\
IRAS 08572+3915 & 4.1 & 3.8 & 0.9 & 1.0 & 0.24 & 0.26 \\
IRAS 09039+0503 & 2.6 & 2.2 & 0.9 & 1.0 & 0.41 & 0.45 \\
IRAS 09116+0334 & 1.5 & 1.1 & \nodata & \nodata & \nodata & \nodata \\
IRAS 09539+0857 & 3.8 & 3.5 & 1.3 & 1.6 & 0.37 & 0.46 \\
IRAS 10378+1108 & 2.6 & 2.4 & 0.45 & 0.55 & 0.19 & 0.23 \\
IRAS 10485$-$1447 & 3.0 & 3.0 & 0.8 & 0.95 & 0.27 & 0.32 \\
IRAS 10494+4424 & 2.2 & 1.7 & \nodata & \nodata & \nodata & \nodata \\
IRAS 11095$-$0238 & 3.5 & 3.4 & 0.8 & 0.9 & 0.23 & 0.26 \\
IRAS 11130$-$2659 & 3.5 & 3.1 & 1.0 & 1.2 & 0.32 & 0.39 \\
IRAS 12112+0305 & 2.3 & 1.7 & \nodata & \nodata & \nodata & \nodata \\
IRAS 12127$-$1412 & 3.0 & 2.5 & 0.6  & 0.7 & 0.24 & 0.28 \\
IRAS 12359$-$0725 & 2.2 & 1.8 & 0.5 & 0.55 & 0.28 & 0.31 \\
IRAS 13335$-$2612 & 1.6 & 1.4 & \nodata & \nodata & \nodata & \nodata \\
IRAS 14252$-$1550 & 1.6 & 1.4 & \nodata & \nodata & \nodata & \nodata \\
IRAS 14348$-$1447 & 2.8 & 2.0 & \nodata & \nodata & \nodata & \nodata \\
Arp 220           & 3.7 & 3.2 & 0.95 & 1.1 & 0.30 & 0.34 \\
IRAS 16090$-$0139 & 3.0 & 2.6 & 0.7 & 0.8 & 0.27 & 0.31 \\
IRAS 16468+5200   & 3.0 & 2.5 & 0.65 & 0.8 & 0.26 & 0.32 \\
IRAS 16487+5447 & 2.4 & 1.8 & \nodata & \nodata & \nodata & \nodata \\
IRAS 17028+5817   & 2.0 & 1.5 & \nodata & \nodata & \nodata & \nodata \\
IRAS 17044+6720   & 1.7 & 1.8 & 0.4 & 0.5 & 0.22 & 0.28 \\
IRAS 21329$-$2346 & 2.7 & 2.2 & \nodata & \nodata & \nodata & \nodata \\
IRAS 23234+0946 & 2.2 & 1.8 & \nodata & \nodata & \nodata & \nodata \\
IRAS 23327+2913 & 1.5 & 1.3 & \nodata & \nodata & \nodata & \nodata \\ \hline
IRAS 00091$-$0738 & 3.2 & 3.2 & 1.05 & 1.25 & 0.33 & 0.39 \\
IRAS 00456$-$2904 & 1.5 & 1.2 & \nodata & \nodata & \nodata & \nodata\\
IRAS 01004$-$2237 & 0.4 & 0.4 & \nodata & \nodata & \nodata & \nodata \\
IRAS 01166$-$0844 & 3.2 & 3.0 & 0.75 & 0.9 & 0.25 & 0.30 \\
IRAS 01298$-$0744 & 4.3 & 4.0 & 1.1 & 1.35 & 0.27 & 0.34 \\
IRAS 01569$-$2939 & 3.2 & 2.8 & 1.0 & 1.2 & 0.36 & 0.43 \\
IRAS 02411+0353 & 1.3 & 1.0 & \nodata & \nodata & \nodata & \nodata \\
IRAS 10190+1322 \tablenotemark{a} & 1.7 & 1.3 & \nodata & \nodata &
\nodata & \nodata \\ 
IRAS 11387+4116 & 1.6 & 1.1 & \nodata & \nodata & \nodata & \nodata \\
IRAS 11506+1331 & 2.6 & 2.3 & 0.7 & 0.85 & 0.30 & 0.37 \\
IRAS 13509+0442 & 2.0 & 1.6 & \nodata & \nodata & \nodata & \nodata \\
IRAS 13539+2920 & 1.8 & 1.6 & \nodata & \nodata & \nodata & \nodata \\
IRAS 14060+2919 & 1.5 & 1.0 & \nodata & \nodata & \nodata & \nodata \\
IRAS 15206+3342 & 0.9 & 0.6 & \nodata & \nodata & \nodata & \nodata \\
IRAS 15225+2350 & 2.5 & 2.2 & 0.65 & 0.75 & 0.29 & 0.34 \\
IRAS 16474+3430 & 1.8 & 1.5 & \nodata & \nodata & \nodata & \nodata \\
IRAS 20414$-$1651 & 2.4 & 1.9 & \nodata & \nodata & \nodata & \nodata \\
IRAS 21208$-$0519 & 1.5 & 1.4 & \nodata & \nodata & \nodata & \nodata \\
IRAS 22206$-$2715 & 2.4 & 1.7 & \nodata & \nodata & \nodata & \nodata \\
IRAS 22491$-$1808 & 2.0 & 1.5 & \nodata & \nodata & \nodata & \nodata \\ \hline  
\enddata

\tablenotetext{a}{To estimate the optical depths of silicate dust
absorption features, SL spectra of the eastern and western nuclei are
summed.
}

\tablecomments{
Col.(1): Object name.  
Col.(2): $\tau_{9.7}$ is an optical depth of the 9.7 $\mu$m silicate
dust absorption feature, against a power-law continuum determined from
data points at $\lambda_{\rm rest}$ = 5.6 $\mu$m, 7.1 $\mu$m and
$\lambda_{\rm obs}$ = 34--35 $\mu$m, shown as dashed lines in
Figures 3 and 4. 
Col.(3): $\tau_{9.7}'$ is an optical depth of the 9.7 $\mu$m silicate
dust absorption feature, against a power-law continuum determined from
data points at  $\lambda_{\rm rest}$ = 7.1 $\mu$m and 14.2 $\mu$m, shown
as dotted lines in Figures 3 and 4.  
Once the continuum levels are fixed, the uncertainty of $\tau_{9.7}'$ 
is $<$5\% for ULIRGs with large $\tau_{9.7}'$ values ($>$2) and 
can be $\sim$10\% for ULIRGs with small $\tau_{9.7}'$.
Col.(4): $\tau_{18}'$ is an optical depth of the 18 $\mu$m silicate dust
absorption feature, against a power-law continuum determined from data
points at $\lambda_{\rm rest}$ = 14.2 $\mu$m and 24 $\mu$m, shown as
dotted lines in Figures 3 and 4. 
Once the continuum is fixed, the statistic uncertainty of $\tau_{18}'$
is $<$10\%, because the value is estimated only for ULIRGs with clearly
detectable 18 $\mu$m silicate absorption features. 
Col.(5): $\tau_{18}''$ is an optical depth of the 18 $\mu$m silicate
dust absorption feature, against a power-law continuum determined from
data points at $\lambda_{\rm rest}$ = 14.2 $\mu$m and 29 $\mu$m, 
shown as dashed-dotted lines in Figures 3 and 4. 
Once the continuum is fixed, the statistic uncertainty is $<$10\%.
Col.(6): $\tau_{18}'$/$\tau_{9.7}'$ ratio, only for ULIRGs with clearly
detectable 18 $\mu$m silicate absorption.
The uncertainty is $<$10\%, once the continuum is fixed.
Col.(7): $\tau_{18}''$/$\tau_{9.7}'$ ratio, only for ULIRGs with clearly
detectable 18 $\mu$m silicate absorption.
}

\end{deluxetable}

\clearpage

\begin{deluxetable}{lcccc}
\tabletypesize{\scriptsize}
\tablecaption{Ice absorption
\label{tbl-5}}
\tablewidth{0pt}
\tablehead{
\colhead{Object} & \colhead{6.0 $\mu$m} & \colhead{$\tau_{6.0}$} &
\colhead{3.1 $\mu$m} & \colhead{$\tau_{3.1}$}  \\
\colhead{(1)} & \colhead{(2)} & \colhead{(3)} & \colhead{(4)} & 
\colhead{(5)}  
}
\startdata 
IRAS 00188$-$0856 & $\bigcirc$ & 0.9 & $\bigcirc$ & 1.8 \\
IRAS 00482$-$2721 & X & \nodata & \nodata & \nodata \\
IRAS 03250+1606 & X & \nodata & $\bigcirc$ & 0.8 \\ 
IRAS 04103$-$2838 & X & \nodata & \nodata & \nodata \\ 
IRAS 08572+3915 & X & \nodata & X & \nodata \\ 
IRAS 09039+0503 & $\bigcirc$ & 0.6 & $\bigcirc$ & 0.8 \\ 
IRAS 09116+0334 & X & \nodata & $\bigcirc$ & 0.4 \\ 
IRAS 09539+0857 & X & \nodata & X & \nodata \\ 
IRAS 10378+1108 & $\bigcirc$ & 0.8 & $\bigcirc$ & 0.6 \\ 
IRAS 10485$-$1447 & $\bigcirc$ & 0.5 & $\bigcirc$ & 0.8 \\ 
IRAS 10494+4424 & $\bigcirc$ & 0.4 & $\bigcirc$ & 1.0 \\ 
IRAS 11095$-$0238 & $\bigcirc$ & 0.5 & X & \nodata \\ 
IRAS 11130$-$2659 & $\bigcirc$ & 0.5 & \nodata & \nodata \\ 
IRAS 12112+0305   & $\bigtriangleup$ & \nodata & X & \nodata \\ 
IRAS 12127$-$1412 & $\bigcirc$ & 0.4 & $\bigcirc$ & 0.4 \\ 
IRAS 12359$-$0725 & X & \nodata  & $\bigcirc$ & 0.5 \\ 
IRAS 13335$-$2612 & X & \nodata & \nodata & \nodata \\ 
IRAS 14252$-$1550 & X & \nodata & $\bigcirc$ & 0.7 \\ 
IRAS 14348$-$1447 & $\bigcirc$ & 0.9 & $\bigcirc$ & 0.5 \\ 
Arp 220           & $\bigcirc$ & 0.6 & \nodata & \nodata \\ 
IRAS 16090$-$0139 & $\bigcirc$ & 0.5 & $\bigcirc$ & 0.8 \\ 
IRAS 16468+5200   & $\bigtriangleup$ & \nodata & X & \nodata \\ 
IRAS 16487+5447   & $\bigtriangleup$ & \nodata & $\bigcirc$ & 1.2 \\ 
IRAS 17028+5817   & $\bigtriangleup$ & \nodata & $\bigcirc$ & 0.6 \\ 
IRAS 17044+6720   & X & \nodata & X & \nodata \\ 
IRAS 21329$-$2346 & $\bigtriangleup$ & \nodata & $\bigcirc$ & 1.0 \\ 
IRAS 23234+0946   & X & \nodata & X & \nodata \\ 
IRAS 23327+2913   & X & \nodata & X & \nodata \\ \hline
IRAS 00091$-$0738 & $\bigcirc$ & 0.5 & \nodata & \nodata \\ 
IRAS 00456$-$2904 & X & \nodata & \nodata & \nodata \\ 
IRAS 01004$-$2237 & X & \nodata & \nodata & \nodata \\ 
IRAS 01166$-$0844 & X & \nodata & \nodata & \nodata \\ 
IRAS 01298$-$0744 & $\bigcirc$ & 0.5 & \nodata & \nodata \\ 
IRAS 01569$-$2939 & X & \nodata & \nodata & \nodata \\ 
IRAS 02411+0353 & X & \nodata  & \nodata & \nodata \\ 
IRAS 10190+1322E & X & \nodata & X & \nodata \\ 
IRAS 10190+1322W & X & \nodata & X & \nodata \\ 
IRAS 11387+4116 & X & \nodata & X & \nodata \\ 
IRAS 11506+1331 & $\bigcirc$ & 0.5 & $\bigcirc$ & 1.0 \\ 
IRAS 13509+0442 & X & \nodata & X & \nodata \\ 
IRAS 13539+2920 & X & \nodata & X & \nodata \\ 
IRAS 14060+2919 & X & \nodata & X & \nodata \\ 
IRAS 15206+3342 & X & \nodata & X & \nodata \\ 
IRAS 15225+2350 & $\bigcirc$ & 0.5 & $\bigcirc$ & 0.4 \\ 
IRAS 16474+3430 & X & \nodata  & $\bigcirc$ & 0.8 \\ 
IRAS 20414$-$1651 & X & \nodata & X & \nodata \\ 
IRAS 21208$-$0519 & X & \nodata & X & \nodata \\ 
IRAS 22206$-$2715 & X & \nodata & \nodata & \nodata \\ 
IRAS 22491$-$1808 & X & \nodata & \nodata & \nodata \\ 
\enddata

\tablecomments{
Col.(1): Object name.  
Col.(2): Detection or non-detection of the 6.0 $\mu$m ice absorption
feature. $\bigcirc$: clearly detected. 
$\bigtriangleup$: possibly detected.
X: non-detected. 
The non-detection includes two cases;  
(i) the absorption feature is indeed weak or absent in high
signal-to-noise ratios spectra, 
or (ii) the presence of ice absorption is not clear due to 
relatively large scatter in spectra or to the difficulty 
in separating the ice absorption and nearby PAH emission features.  
Col.(3): Optical depth of the 6.0 $\mu$m ice absorption feature. 
---: not measurable.  
Since data scatter in spectra is not the same for individual ULIRGs,
there is no fixed detection threshold.
The uncertainty is $\sim$0.1 for detected sources.
Col.(4): Detection or non-detection of the 3.1 $\mu$m ice absorption
feature in the 2.8--4.1 $\mu$m spectra of \citet{idm06}.
$\bigcirc$: clearly detected. X: non-detected. ---: no data.
Col.(5): Optical depth of the 3.1 $\mu$m ice absorption feature. 
---: not measurable because (i) it is weak, (ii) data scatter is large in
spectra, and (iii) 2.8--4.1 $\mu$m spectra are lacking.
Again there is no fixed detection threshold.
}

\end{deluxetable}

\clearpage 

\begin{deluxetable}{lccccccc}
\tabletypesize{\scriptsize}
\tablecaption{Buried AGN signatures \label{tbl-6}}
\tablewidth{0pt}
\tablehead{
\colhead{Object} & \multicolumn{6}{c}{{\it Spitzer} IRS 5--35 $\mu$m} & 
\colhead{3--4 $\mu$m} \\
\colhead{} & \colhead{EW$_{\rm 6.2PAH}$} & \colhead{EW$_{\rm 7.7PAH}$} & 
\colhead{EW$_{\rm 11.3PAH}$} & \colhead{$\tau_{9.7}'$} & 
\colhead{T-gradient} &\colhead{Total} & \colhead{} \\
\colhead{(1)} & \colhead{(2)} & \colhead{(3)} &  
\colhead{(4)} & \colhead{(5)} & \colhead{(6)} & 
\colhead{(7)} & \colhead{(8)} 
}
\startdata 
IRAS 00188$-$0856 & $\bigcirc$ & X & X & $\bigcirc$ & $\bigcirc$ &
{\large $\circledcirc$} & $\bigcirc$ \\ 
IRAS 00482$-$2721 & X & X & X & $\bigcirc$ & X & $\bigtriangleup$ & \nodata \\
IRAS 03250+1606   & X & X & X & X & X & X & $\bigcirc$ \\ 
IRAS 04103$-$2838 & $\bigcirc$ & X & $\bigcirc$ & X & X & $\bigcirc$ &
\nodata  \\  
IRAS 08572+3915   & $\bigcirc$ & $\bigcirc$ & $\bigcirc$ & $\bigcirc$ &
$\bigcirc$ & {\large $\circledcirc$} & {\large $\circledcirc$} \\  
IRAS 09039+0503   & X & X & X & $\bigcirc$ & X & $\bigtriangleup$ &
$\bigcirc$ \\  
IRAS 09116+0334   & X & X & X & X & X & X & $\bigcirc$ \\ 
IRAS 09539+0857   & $\bigcirc$ & X & X & $\bigcirc$ & X &
$\bigcirc$ & X \\  
IRAS 10378+1108   & $\bigcirc$ & X & X & $\bigcirc$ & $\bigcirc$ &
$\bigcirc$ & $\bigcirc$ \\   
IRAS 10485$-$1447 & X ($\bigcirc$) & X & $\bigcirc$ & $\bigcirc$ & $\bigtriangleup$ &
$\bigtriangleup$ & $\bigcirc$ \\  
IRAS 10494+4424   & X & X & X & X & X & X & $\bigcirc$ \\ 
IRAS 11095$-$0238 & $\bigcirc$ & X & $\bigcirc$ & $\bigcirc$ &
$\bigcirc$ & {\large $\circledcirc$} & X \\ 
IRAS 11130$-$2659 & $\bigcirc$ & X & X & $\bigcirc$ & X &
$\bigcirc$ & X \\ 
IRAS 12112+0305   & X & X & X & X & X & X & X \\ 
IRAS 12127$-$1412 & $\bigcirc$ & $\bigcirc$ & $\bigcirc$ & $\bigcirc$ &
$\bigcirc$ & {\large $\circledcirc$} & {\large $\circledcirc$} \\ 
IRAS 12359$-$0725 & X & X & $\bigcirc$ & X & $\bigtriangleup$ & $\bigtriangleup$
& $\bigcirc$ \\ 
IRAS 13335$-$2612 & X & X & X & X & X & X & \nodata  \\ 
IRAS 14252$-$1550 & X & X & X & X & X & X & $\bigcirc$ \\ 
IRAS 14348$-$1447 & X ($\bigcirc$) & X & X & X & X & $\bigtriangleup$ & $\bigcirc$ \\ 
Arp 220 & X ($\bigcirc$) & X & X & $\bigcirc$ & X &
$\bigtriangleup$ & X \\ 
IRAS 16090$-$0139 & $\bigcirc$ & X & X & $\bigcirc$ & $\bigtriangleup$ &
$\bigcirc$ & $\bigcirc$ \\ 
IRAS 16468+5200   & X & X & X & $\bigcirc$ & $\bigtriangleup$ & 
$\bigtriangleup$ & X \\ 
IRAS 16487+5447   & X & X & X & X & X & X & $\bigcirc$ \\ 
IRAS 17028+5817   & X & X & X & X & X & X & $\bigcirc$ \\ 
IRAS 17044+6720   & $\bigcirc$ & $\bigcirc$ & $\bigcirc$ & X (---)
\tablenotemark{a} & $\bigcirc$ & {\large $\circledcirc$} & 
{\large $\circledcirc$} \\   
IRAS 21329$-$2346 & X & X & X & $\bigcirc$ & X & $\bigtriangleup$ & 
$\bigcirc$ \\ 
IRAS 23234+0946   & X & X & X & X & X & X & X \\ 
IRAS 23327+2913   & X & X & $\bigcirc$ & X & X & $\bigtriangleup$ & X \\
\hline   
IRAS 00091$-$0738 & $\bigcirc$ & X & X & $\bigcirc$ & X &
$\bigcirc$ & \nodata \\ 
IRAS 00456$-$2904 & X & X & X & X & X & X & \nodata \\ 
IRAS 01004$-$2237 & $\bigcirc$ & $\bigcirc$ & $\bigcirc$ & 
X (---) \tablenotemark{a} & X & {\large $\circledcirc$} &
\nodata \\  
IRAS 01166$-$0844 & $\bigcirc$ & X & $\bigcirc$ & $\bigcirc$ & $\bigcirc$ &
{\large $\circledcirc$} & \nodata \\ 
IRAS 01298$-$0744 & $\bigcirc$ & X & $\bigcirc$ & $\bigcirc$ & X &
$\bigcirc$ & \nodata \\  
IRAS 01569$-$2939 & $\bigcirc$ & X & X & $\bigcirc$ & X &
$\bigcirc$ & \nodata \\  
IRAS 02411+0353 & X & X & X & X & X & X & \nodata \\ 
IRAS 10190+1322  & X \tablenotemark{b} & X \tablenotemark{b} & X
\tablenotemark{b} & X & X & X & X \\ 
IRAS 11387+4116   & X & X & X & X & X & X & X \\ 
IRAS 11506+1331   & X ($\bigcirc$) & X & X & $\bigcirc$ & X & 
$\bigtriangleup$ & $\bigcirc$ \\ 
IRAS 13509+0442   & X & X & X & X & X & X & X \\ 
IRAS 13539+2920   & X & X & X & X & X & X & X \\ 
IRAS 14060+2919   & X & X & X & X & X & X & X \\ 
IRAS 15206+3342   & X & X & X & X & X & X & X \\ 
IRAS 15225+2350   & $\bigcirc$ & X & $\bigcirc$ & $\bigcirc$ & X & 
$\bigcirc$ & $\bigcirc$ \\ 
IRAS 16474+3430   & X & X & X & X & X & X & $\bigcirc$ \\ 
IRAS 20414$-$1651 & X & X & X & X & X & X & X \\ 
IRAS 21208$-$0519 & X & X & X & X & X & X & X \\ 
IRAS 22206$-$2715 & X & X & X & X & X & X & \nodata \\ 
IRAS 22491$-$1808 & X & X & X & X & X & X & \nodata \\ 
\enddata

\tablenotetext{a}{The small $\tau_{9.7}'$ value is due to
weak obscuration of a luminous AGN, rather than the normal starburst
nature with mixed dust/source geometry.}

\tablenotetext{b}{PAH equivalent widths are derived for both nuclei
separately. No AGN signatures are found in both nuclei.}  

\tablecomments{
Col.(1): Object name.  
Col.(2): Buried AGN signatures based on the low equivalent width of the
         6.2 $\mu$m PAH emission (EW$_{\rm 6.2PAH}$ $<$ 180 nm) ($\S$5.2.1).
         $\bigcirc$: present.  X: none. 
         When the corrected EW$_{\rm 6.2PAH}$ value becomes less than 180 nm, 
         the symbol ``$\bigcirc$'' is shown in parentheses.
Col.(3): Buried AGN signatures based on the low equivalent width of the
         7.7 $\mu$m PAH emission (EW$_{\rm 7.7PAH}$ $<$ 230 nm) ($\S$5.2.1).
         $\bigcirc$: present.  X: none. 
Col.(4): Buried AGN signatures based on the low equivalent width of the
         11.3 $\mu$m PAH emission (EW$_{\rm 11.3PAH}$ $<$ 200 nm) ($\S$5.2.1).
         $\bigcirc$: present.  X: none. 
Col.(5): Buried AGN signatures based on the large $\tau_{9.7}'$ value 
         ($>$2) ($\S$5.2.2).
         $\bigcirc$: present.  X: none. 
Col.(6): Buried AGN signatures based on the small
         $\tau_{18}'$/$\tau_{9.7}'$ and $\tau_{18}''$/$\tau_{9.7}'$ 
         ratios. 
         $\bigcirc$: present. $\bigtriangleup$: possibly present. 
         X: none. 
         Considering the uncertainty of the ratio with $<$10\% (Table
         4), ULIRGs showing $\tau_{18}'$/$\tau_{9.7}'$ $<$ 0.27 and 
         $\tau_{18}''$/$\tau_{9.7}'$ $<$ 0.32 are taken as sources with
         detectable signatures of strong dust temperature gradients 
         ($\S$5.2.3).  
Col.(7): Buried AGN signatures from combined methods in Col. (2)--(6) 
         based on {\it Spitzer} IRS spectra.
         {\large $\circledcirc$}: very strong.
         $\bigcirc$: strong.
         $\bigtriangleup$: possibly present. 
         X: none.
         Please see $\S$5.2.6 for more details. 
Col.(7): Buried AGN signatures based on infrared 3--4 $\mu$m
         spectroscopic diagnostic by \citet{idm06}. 
         {\large $\circledcirc$}: very strong.
         $\bigcirc$: strong.
         X: none.  ---: no data.
}

\end{deluxetable}

\clearpage 

\begin{deluxetable}{lc}
\tablecaption{Absorption-corrected AGN luminosity \label{tbl-7}}
\tablewidth{0pt}
\tablehead{
\colhead{Object} & \colhead{L(AGN) [10$^{45}$ ergs s$^{-1}$]} \\
\colhead{(1)} & \colhead{(2)} 
}
\startdata 
IRAS 00188$-$0856 & 1.5 \\
IRAS 04103$-$2838 & 1.5 \\
IRAS 08572+3915   & 5.5 \\
IRAS 10378+1108   & 1.0 \\
IRAS 10485$-$1447 & 1.0 \\
IRAS 11095$-$0238 & 2.5 \\
IRAS 11130$-$2659 & 1.0 \\
IRAS 12127$-$1412 & 2.5 \\
IRAS 16090$-$0139 & 2.0 \\
IRAS 17044+6720   & 2.0 \\
IRAS 00091$-$0738 & 1.5 \\
IRAS 01004$-$2237 & 1.5 \\
IRAS 01166$-$0844 & 1.0 \\
IRAS 01298$-$0744 & 3.5 \\
IRAS 15225+2350   & 1.5 \\
\enddata

\tablecomments{
Col.(1): Object name.  
Col.(2): Absorption-corrected intrinsic luminosity of the AGN energetic
radiation (X-ray -- UV -- optical) in units of 10$^{45}$ ergs s$^{-1}$.
}

\end{deluxetable}

\clearpage 


\begin{figure}
\includegraphics[angle=0,scale=.5]{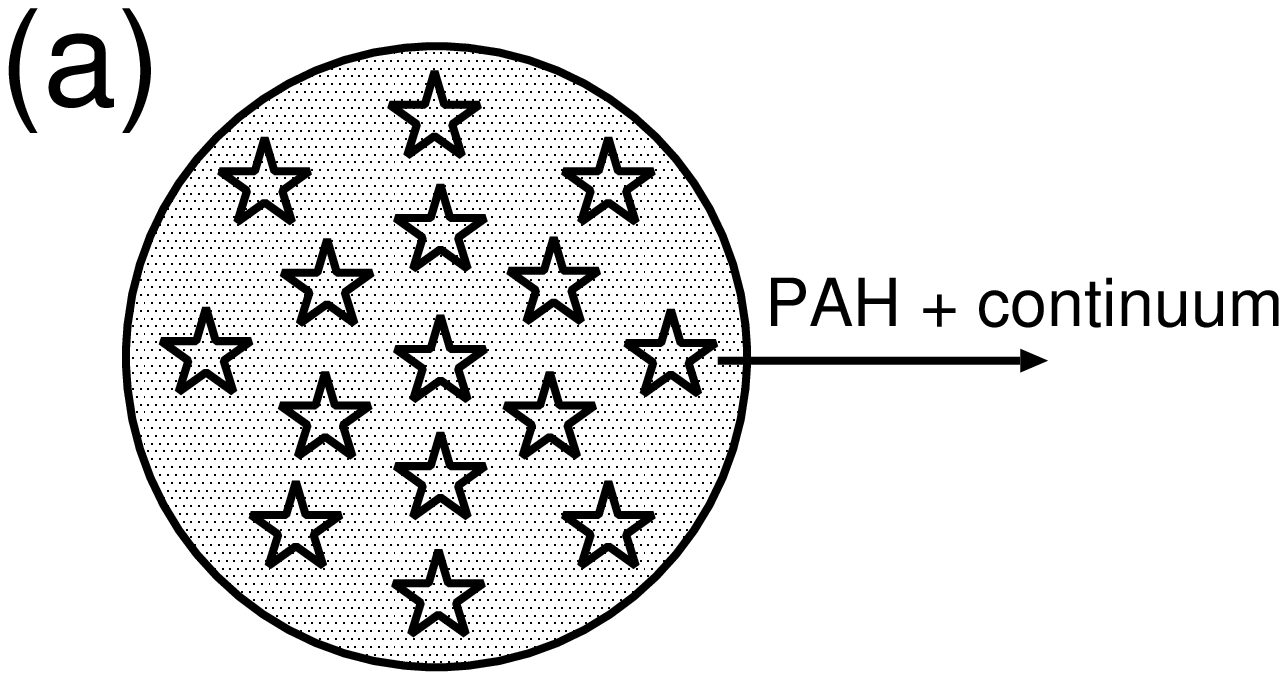} \hspace{0.2cm} 
\includegraphics[angle=0,scale=.5]{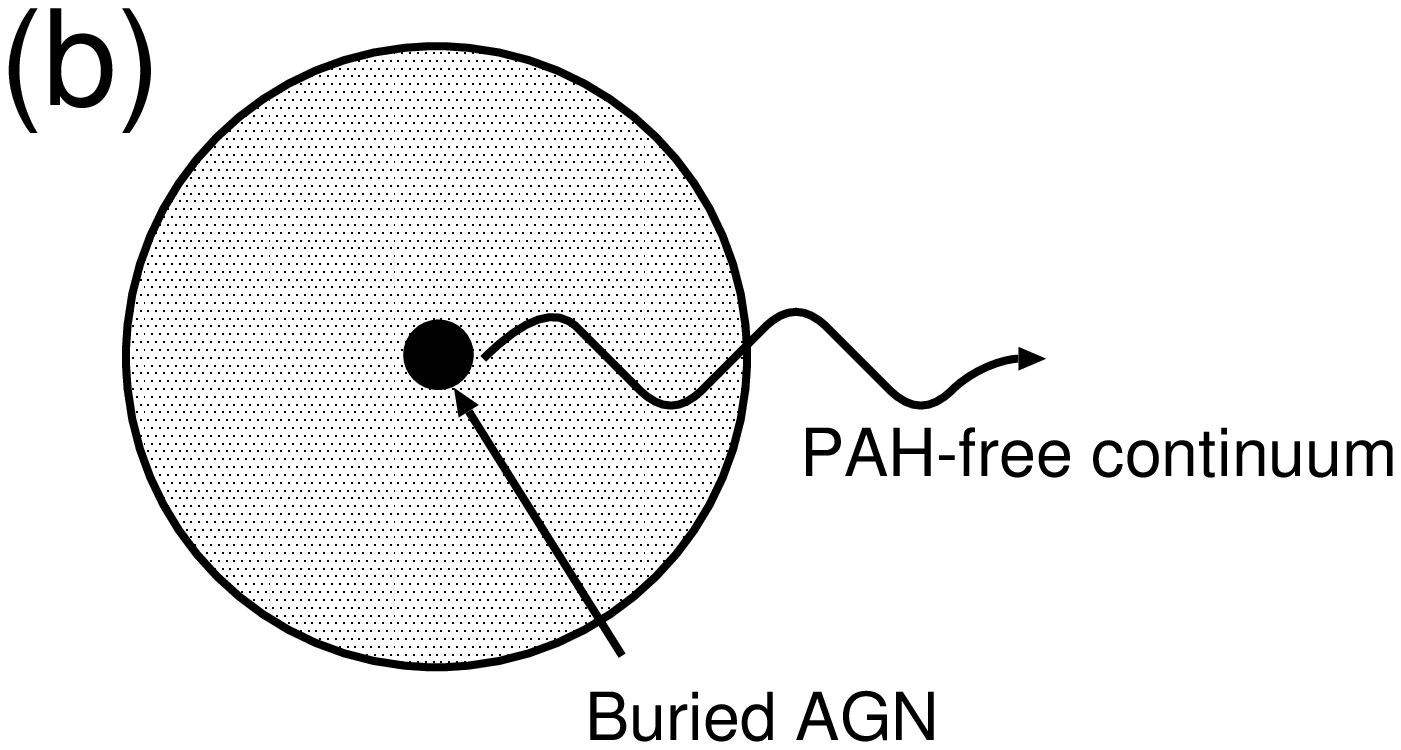} \\ 
\includegraphics[angle=0,scale=.5]{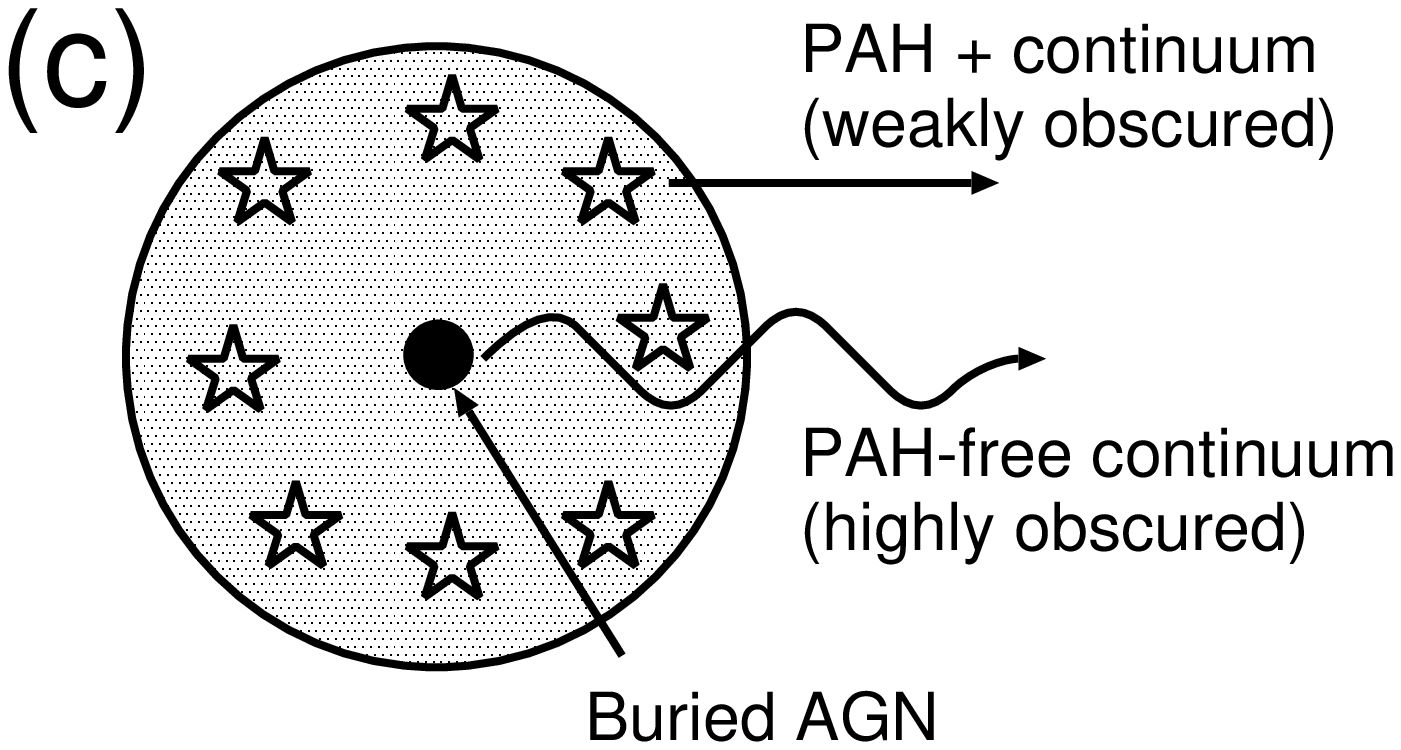} \hspace{0.2cm}
\includegraphics[angle=0,scale=.5]{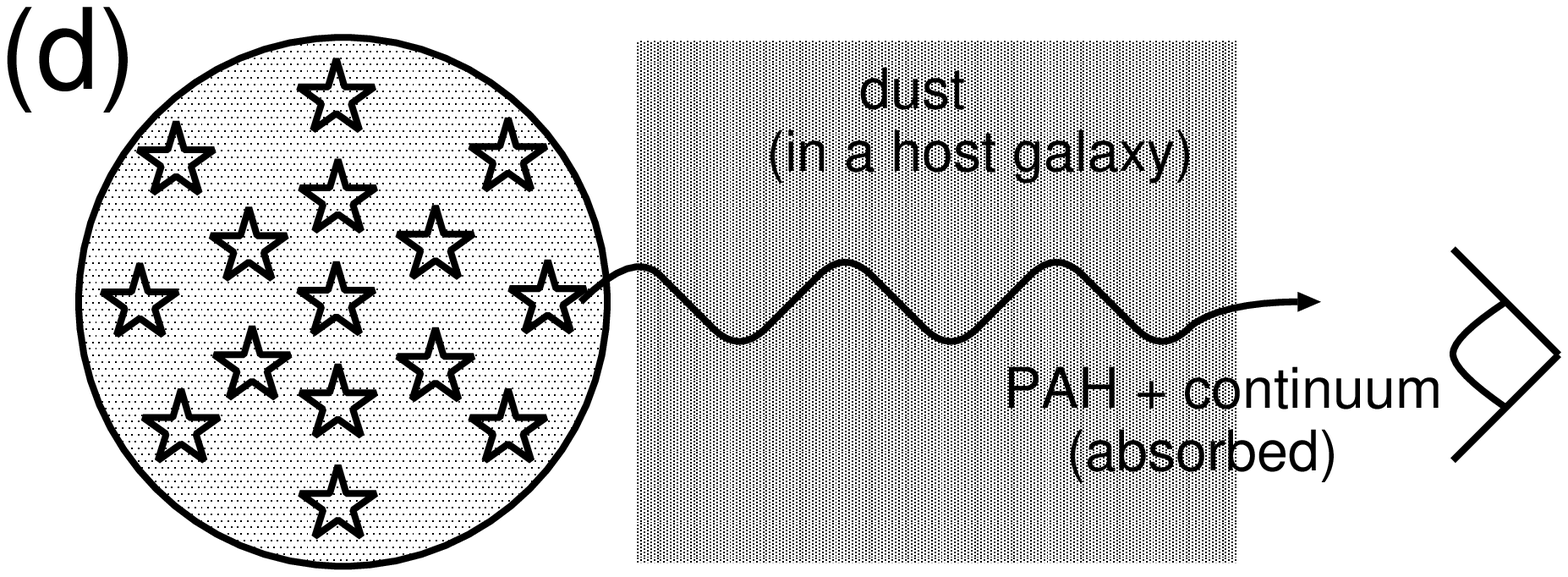} \\
\includegraphics[angle=0,scale=.5]{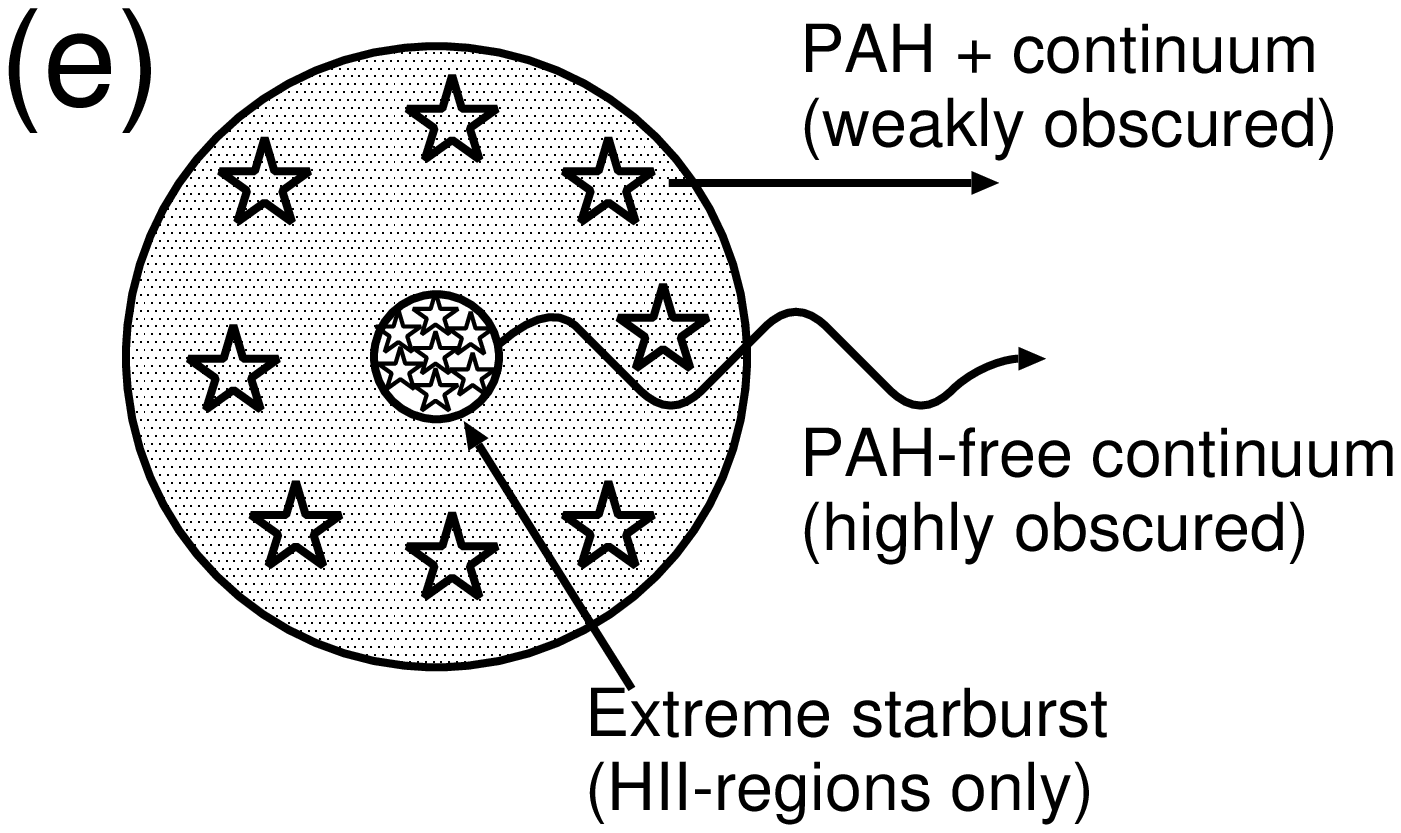} 
\caption{
{\it (a)}: Geometry of energy sources and dust in a normal starburst. 
The open star symbols indicate ``stars'' in a starburst.
The energy sources (stars) and dust are spatially well mixed. 
Large equivalent width PAH emission is observed. 
{\it (b)}:  Geometry of the energy source and dust in a
buried AGN. The energy source (a mass-accreting supermassive blackhole) is
more centrally concentrated than the surrounding dust. No PAH emission
is found.
{\it (c)}: A buried AGN/starburst composite.
The starburst surrounds the central buried AGN.
The observed spectrum is a superposition of PAH + continuum emission
from the starburst and PAH-free continuum from the AGN.
Since the buried AGN is more highly obscured than the surrounding
starbursts, the starburst emission generally makes a strong contribution
to an observed flux even if the starburst is energetically
insignificant. 
{\it (d)}: A normal starburst nucleus (mixed dust/source geometry)
obscured by a large amount of foreground dust in a host galaxy.
This geometry can produce a strong dust absorption feature whose optical
depth is larger than the maximum threshold obtained in the mixed
dust/source geometry. 
{\it (e)}: An exceptionally centrally-concentrated extreme starburst,
whose emitting volume is predominantly occupied by HII-regions.  
Such an extreme starburst produces no PAH emission, like a buried AGN.
}
\end{figure} 


\begin{figure}
\begin{center}
\includegraphics[angle=0,scale=.6]{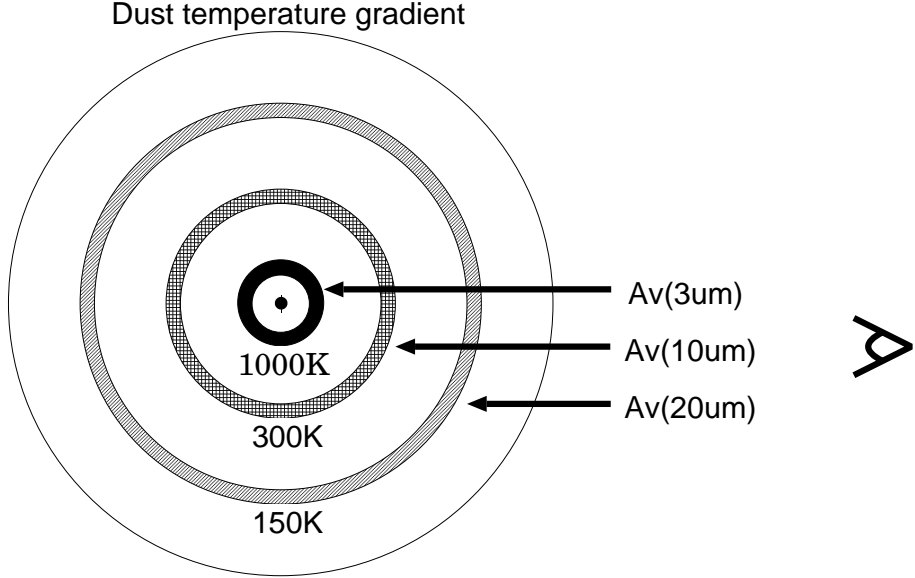} 
\end{center}
\caption{Schematic diagram of a strong dust temperature gradient
  around a centrally-concentrated energy source (the central small
  filled circle).  Inner dust close to the central energy source has
  higher temperature than dust far away from the energy source.
  Assuming approximately blackbody radiation, dust at 1000K dust,
  close to the innermost dust sublimation radius, shows an emission
  peak at $\lambda_{\rm rest}$ $ \sim$ 3 $\mu$m. More distant 300K
  dust and even 150K dust have emission peaks at $\lambda_{\rm rest}$
  $\sim$ 10 $\mu$m and $\sim$20 $\mu$m, respectively.  This is a very
  simplified picture, and in an actual case, dust with other
  temperatures can also contribute to the observed flux at each
  wavelength.  For any reasonable dust radial density distribution, it
  is always true that the contribution from inner, hotter dust
  emission is larger at a shorter wavelength in this wavelength range.
  When we look at the energy source from the right side, an observed
  flux is dominated by dust at the foreground side (right), rather
  than dust with the same temperature at the distant side (left), due
  to flux attenuation, particularly for outer cooler dust.  Thus, the
  relation A$_{\rm V}$(3$\mu$m) $>$ A$_{\rm V}$(10$\mu$m) $>$ A$_{\rm
    V}$(20$\mu$m) should hold.  Dust extinction toward the 3 $\mu$m
  continuum emitting region, A$_{\rm V}$(3$\mu$m), can be estimated
  from the optical depths of absorption features at 3.4 $\mu$m due to
  bare carbonaceous dust and at 3.1 $\mu$m due to ice-covered dust
  \citep{idm01,idm06,ima06}.  The extinction toward the 10 $\mu$m and
  20 $\mu$m continuum emitting regions, A$_{\rm V}$(10$\mu$m) and
  A$_{\rm V}$(20$\mu$m), can be estimated from the optical depths of
  9.7 $\mu$m and 18 $\mu$m silicate dust absorption features in {\it
    Spitzer} IRS spectra, respectively.  Although a spherical geometry
  is illustrated here for simplicity, the essence is unchanged as long
  as a centrally-concentrated compact energy source is obscured by
  dust along our line-of-sight.  }
\end{figure}

\clearpage

\begin{figure}
\includegraphics[angle=-90,scale=.24]{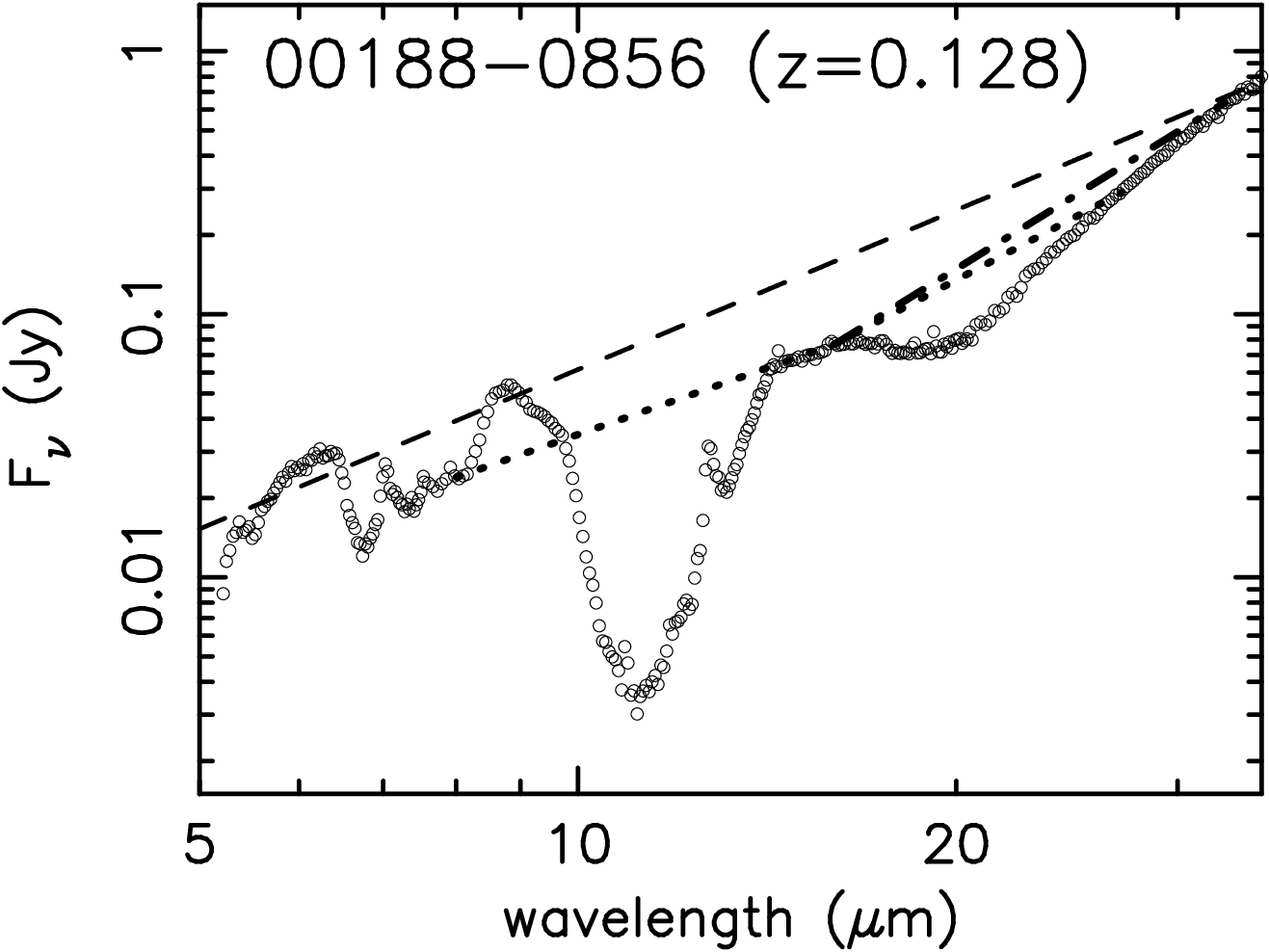} 
\includegraphics[angle=-90,scale=.24]{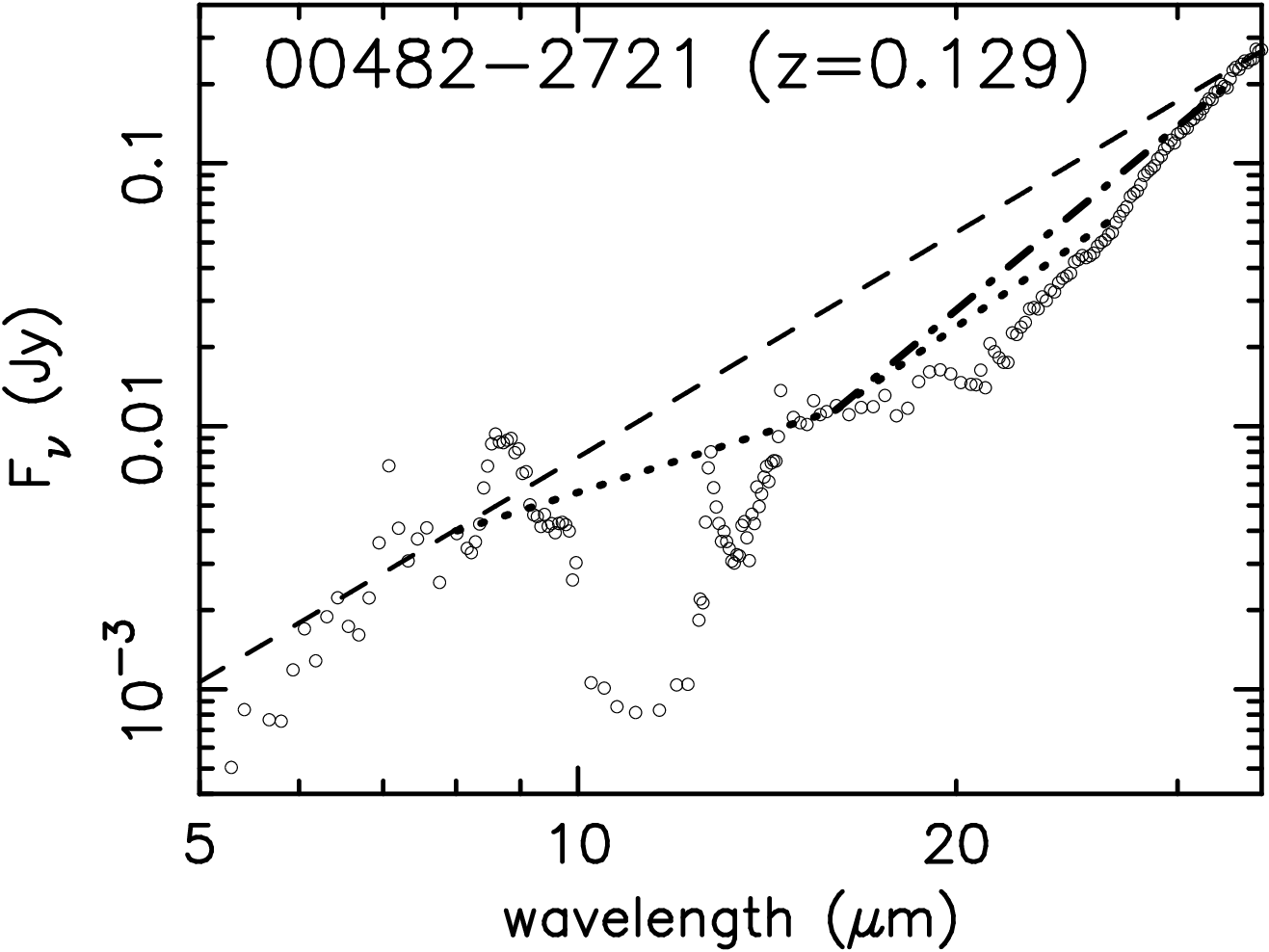} 
\includegraphics[angle=-90,scale=.24]{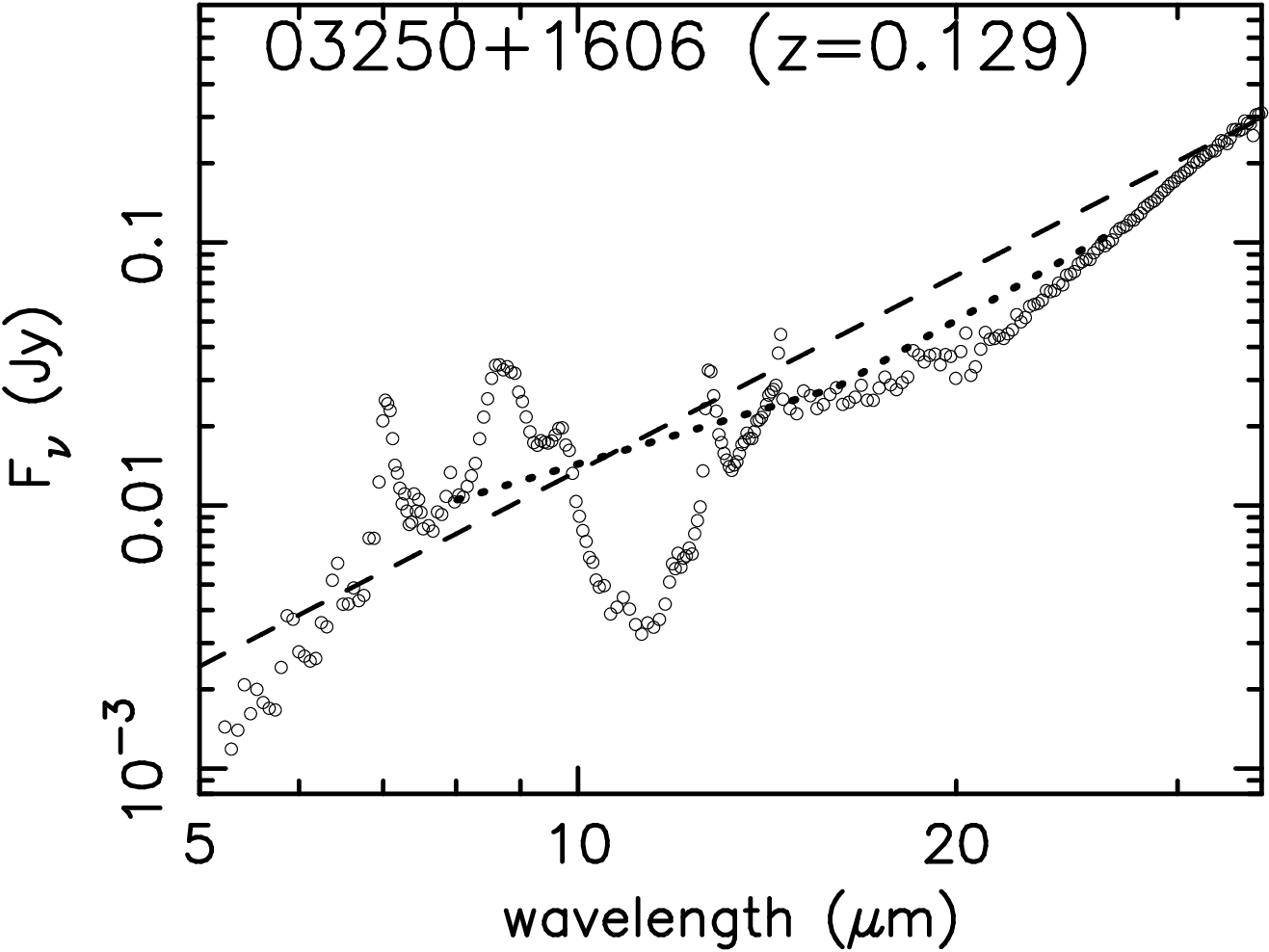} \\
\includegraphics[angle=-90,scale=.24]{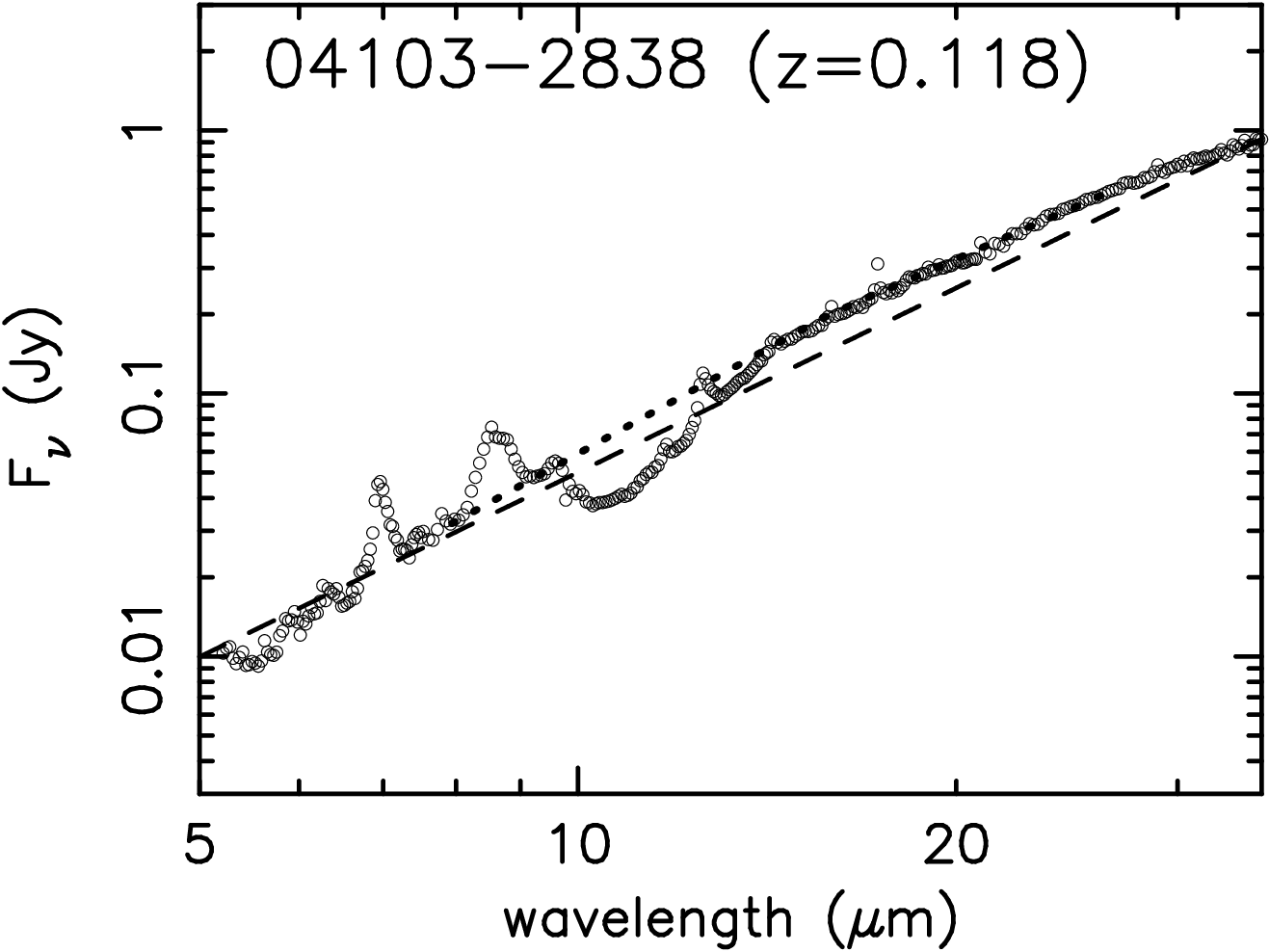}  
\includegraphics[angle=-90,scale=.24]{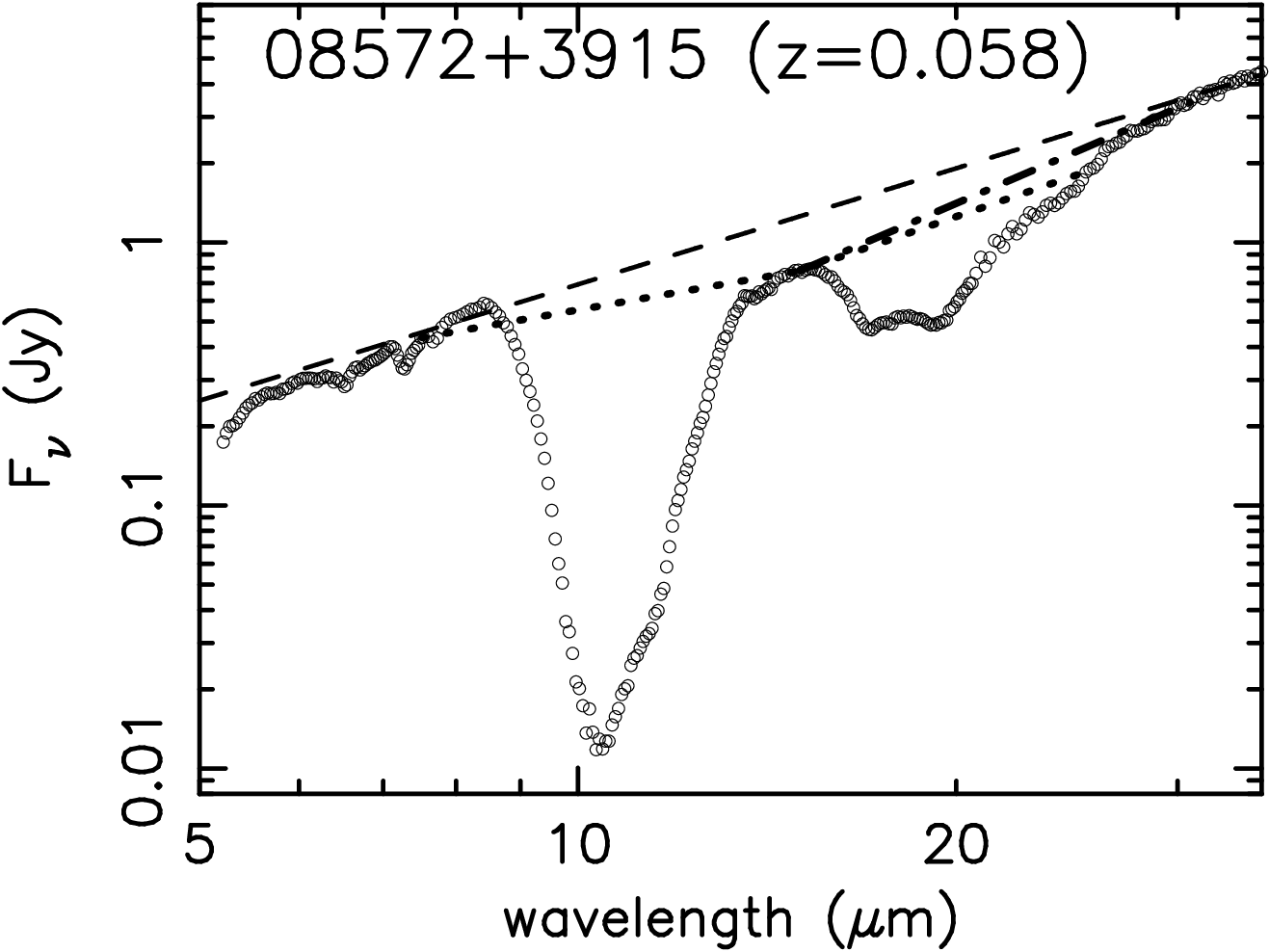} 
\includegraphics[angle=-90,scale=.24]{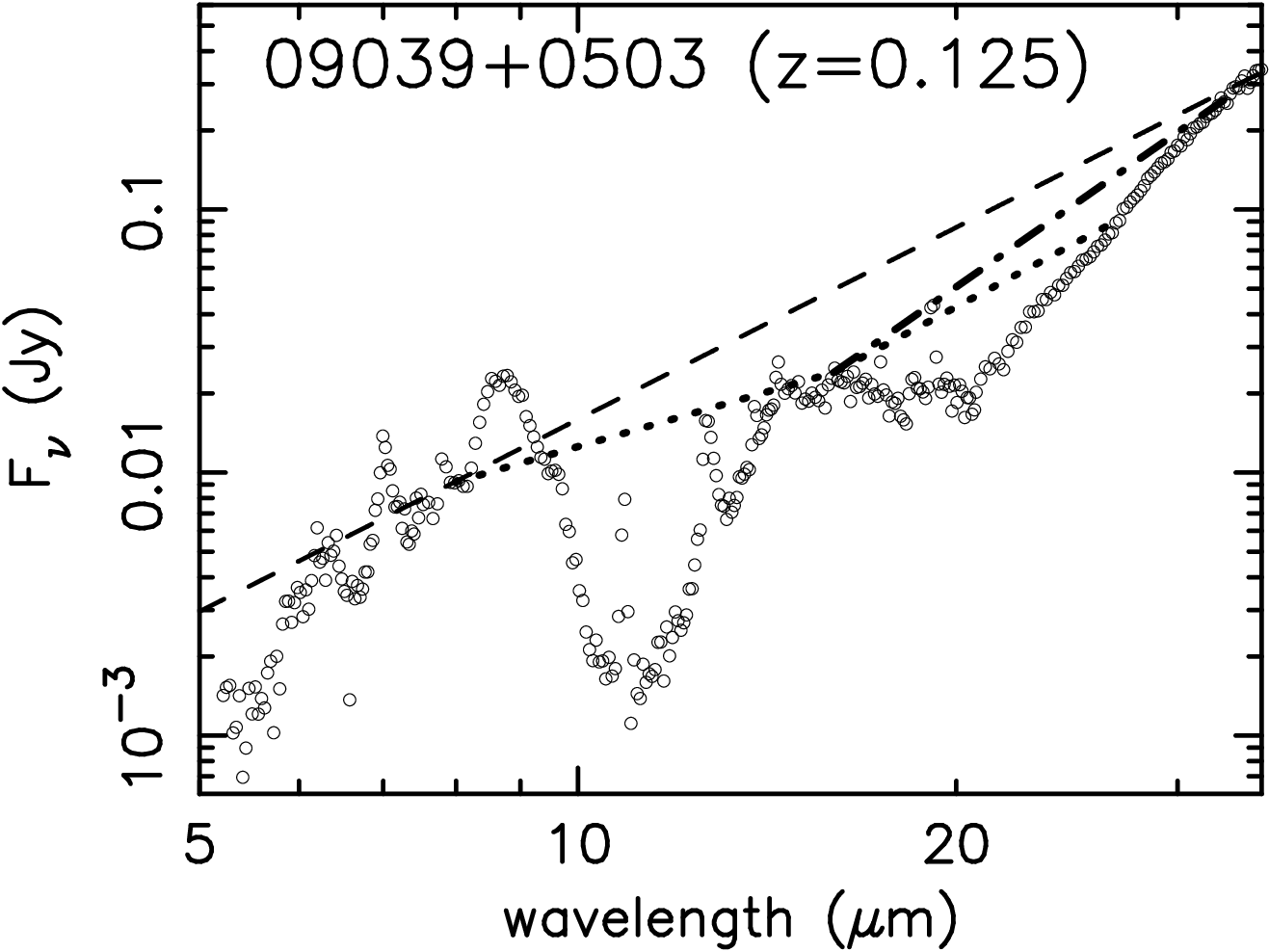} \\ 
\includegraphics[angle=-90,scale=.24]{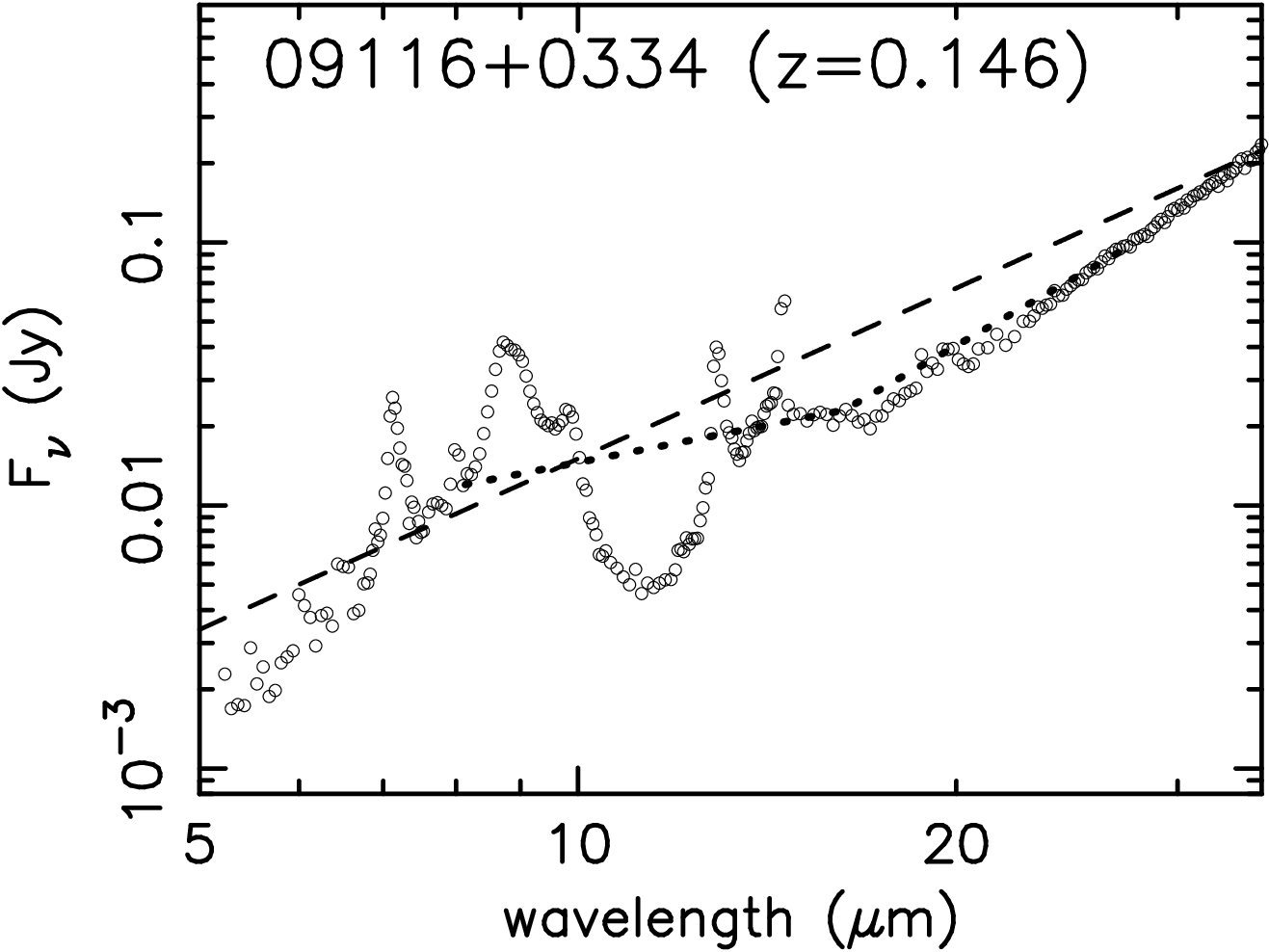} 
\includegraphics[angle=-90,scale=.24]{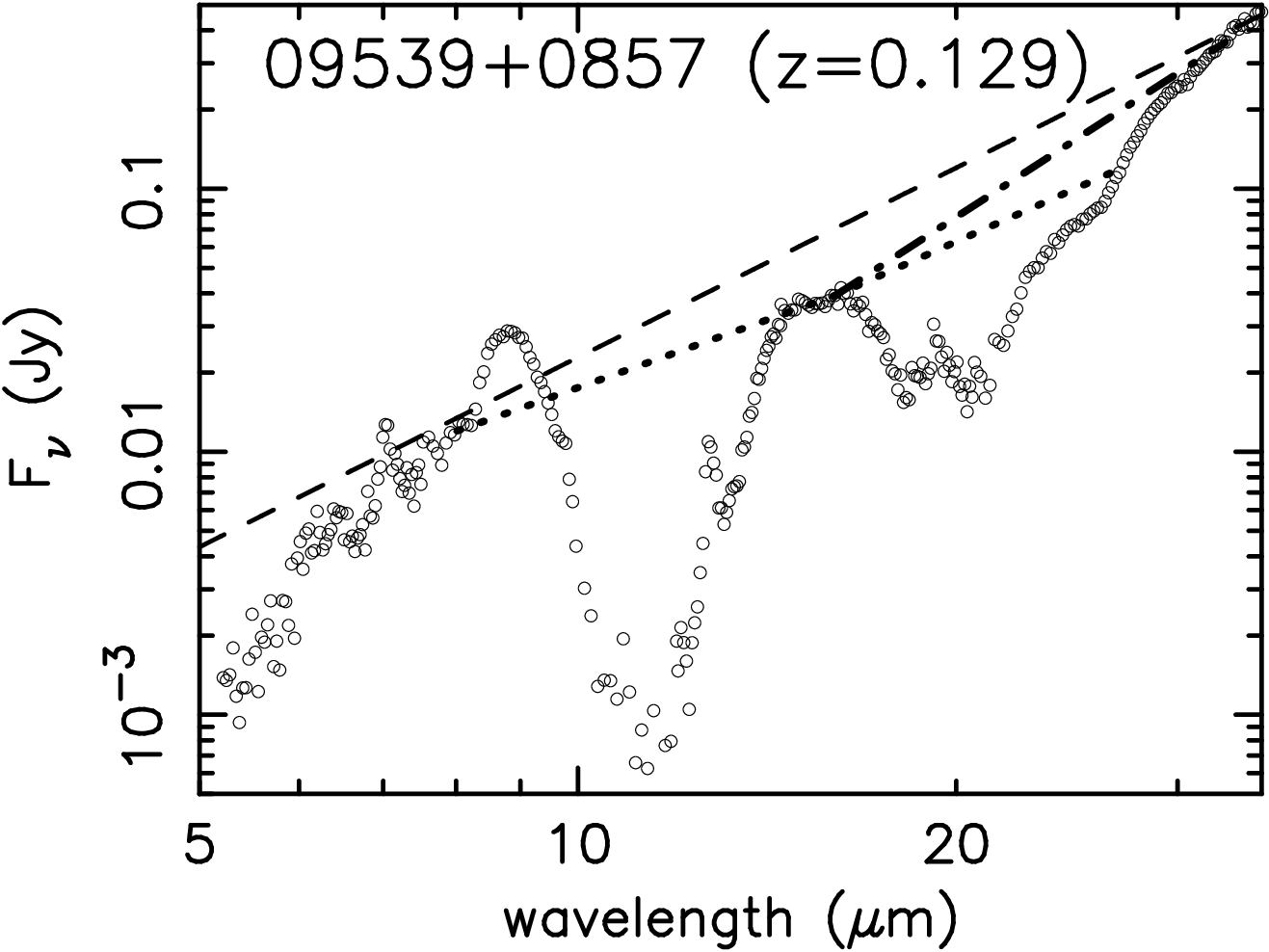} 
\includegraphics[angle=-90,scale=.24]{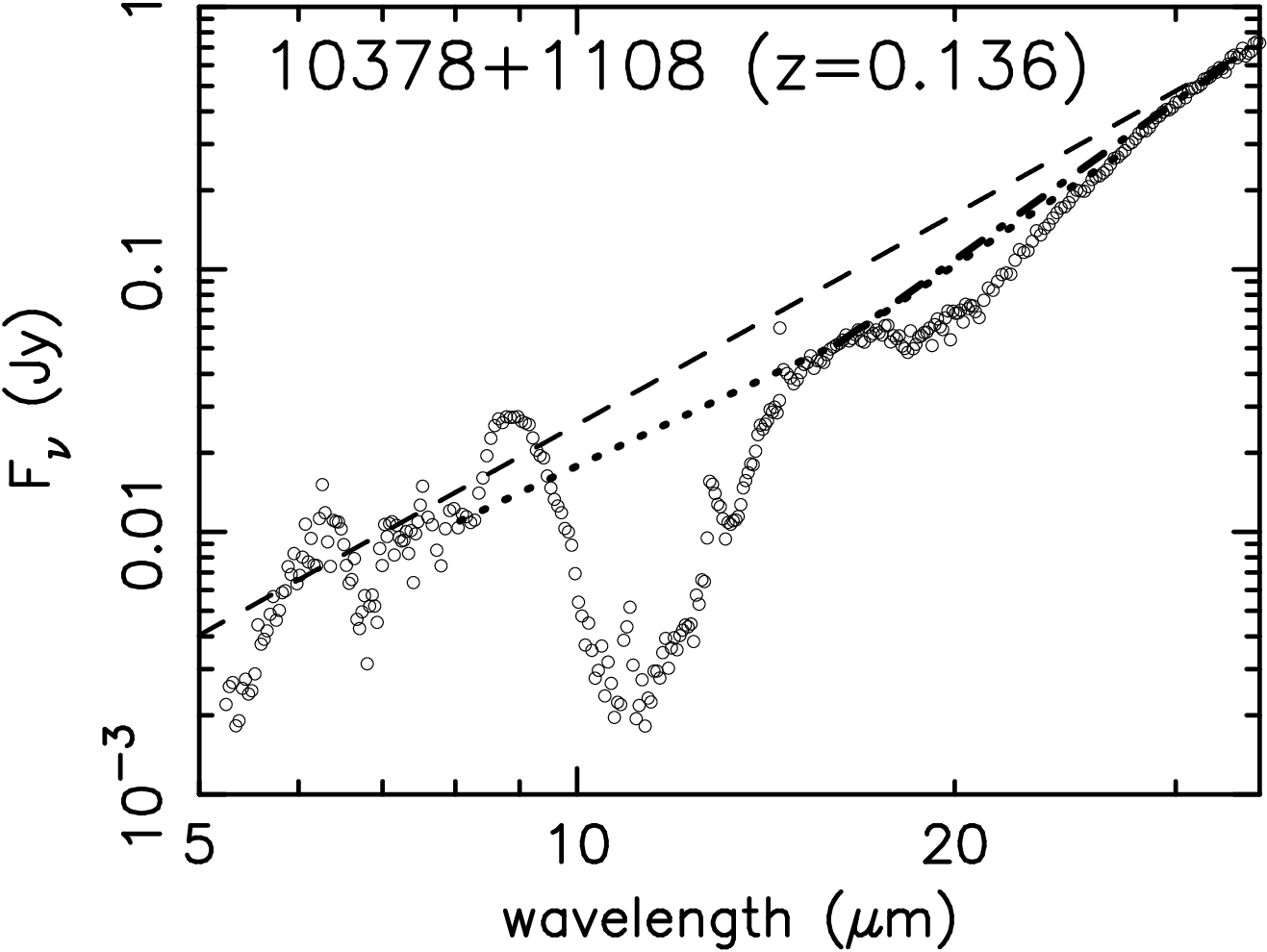} \\ 
\includegraphics[angle=-90,scale=.24]{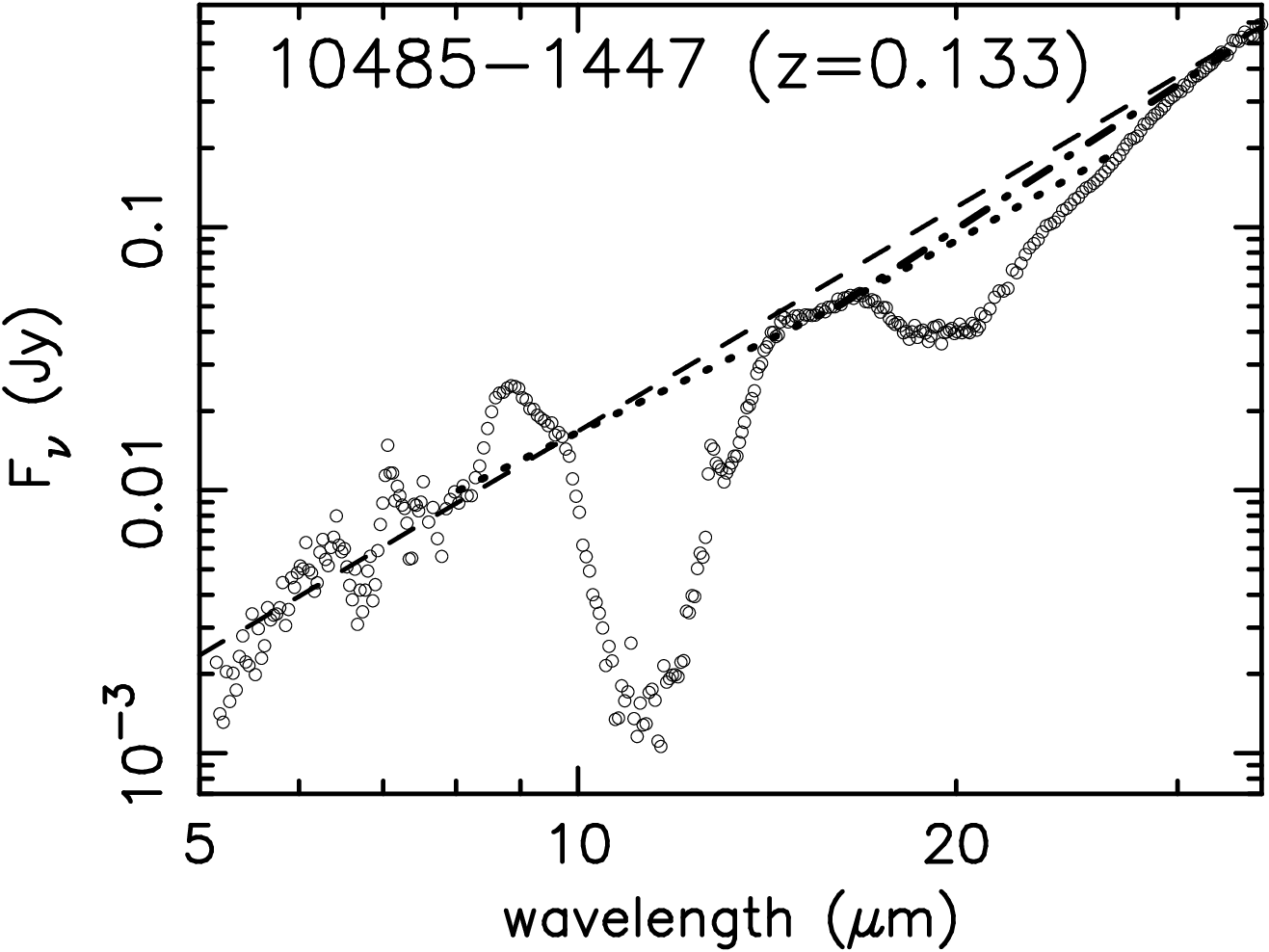} 
\includegraphics[angle=-90,scale=.24]{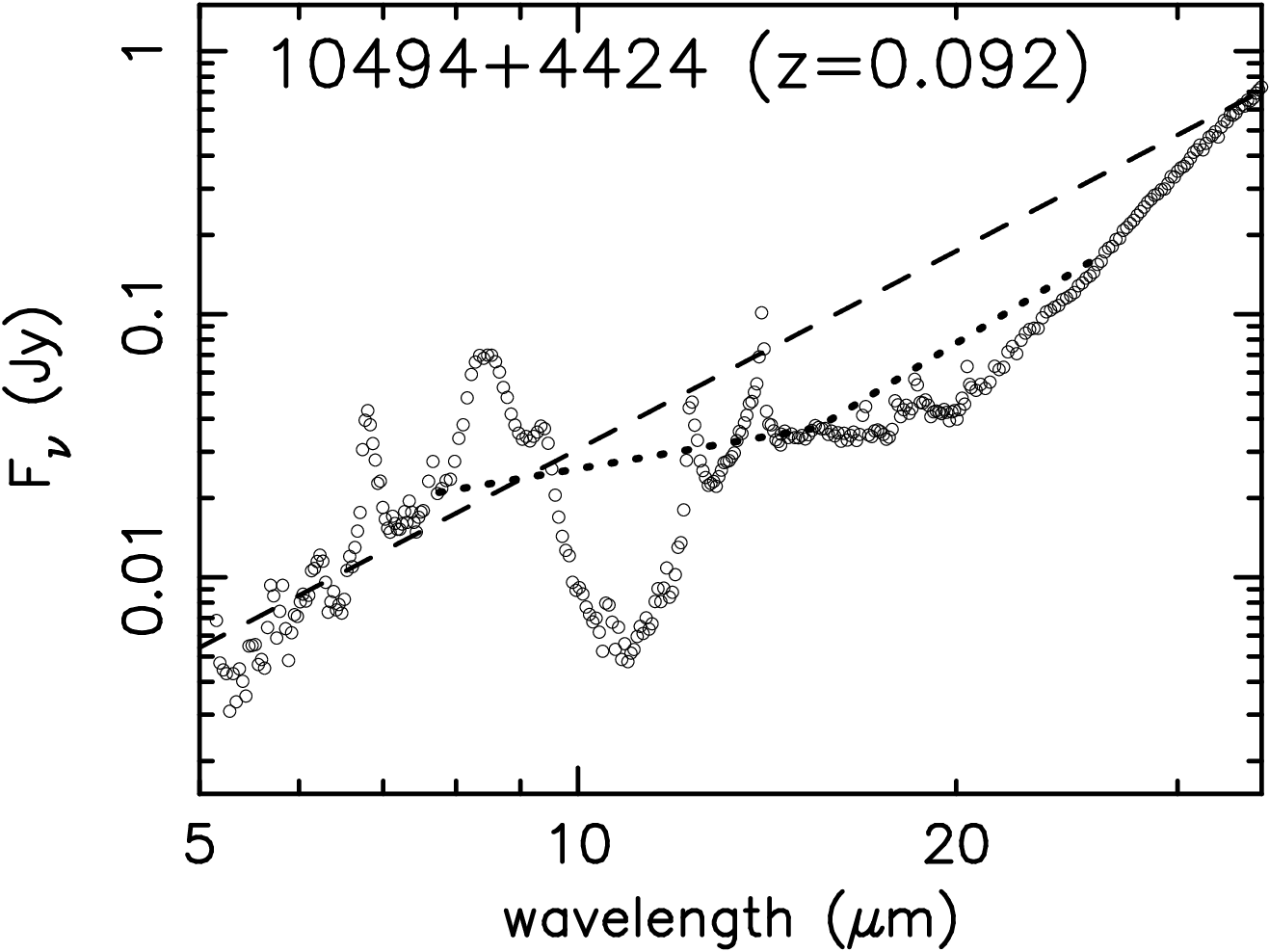}  
\includegraphics[angle=-90,scale=.24]{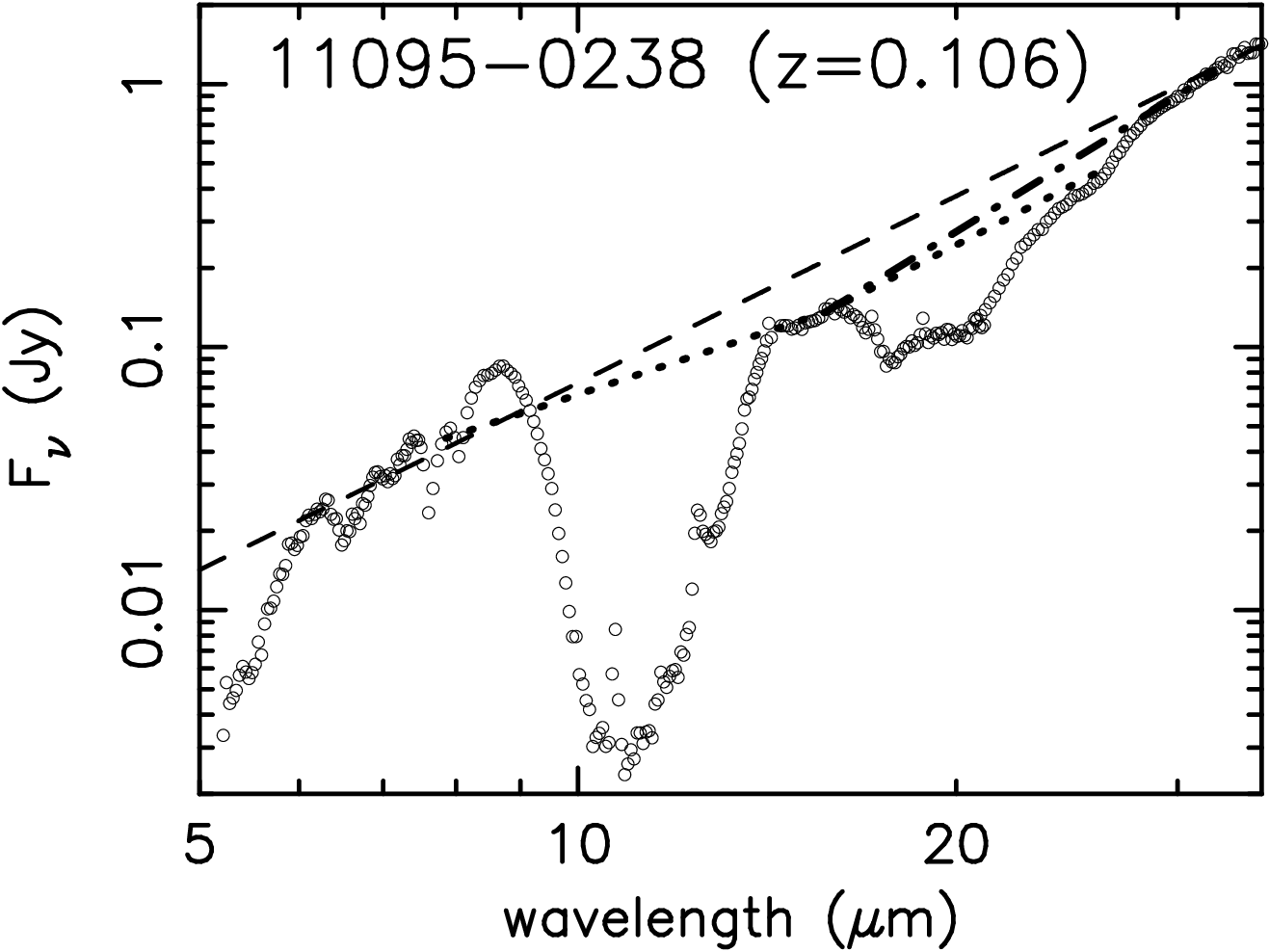} \\ 
\end{figure}

\clearpage

\begin{figure}
\includegraphics[angle=-90,scale=.24]{f3m.eps} 
\includegraphics[angle=-90,scale=.24]{f3n.eps}  
\includegraphics[angle=-90,scale=.24]{f3o.eps}  \\
\includegraphics[angle=-90,scale=.24]{f3p.eps} 
\includegraphics[angle=-90,scale=.24]{f3q.eps}  
\includegraphics[angle=-90,scale=.24]{f3r.eps}  \\
\includegraphics[angle=-90,scale=.24]{f3s.eps} 
\includegraphics[angle=-90,scale=.24]{f3t.eps} 
\includegraphics[angle=-90,scale=.24]{f3u.eps} \\
\includegraphics[angle=-90,scale=.24]{f3v.eps} 
\includegraphics[angle=-90,scale=.24]{f3w.eps} 
\includegraphics[angle=-90,scale=.24]{f3x.eps} \\
\end{figure}

\clearpage

\begin{figure}
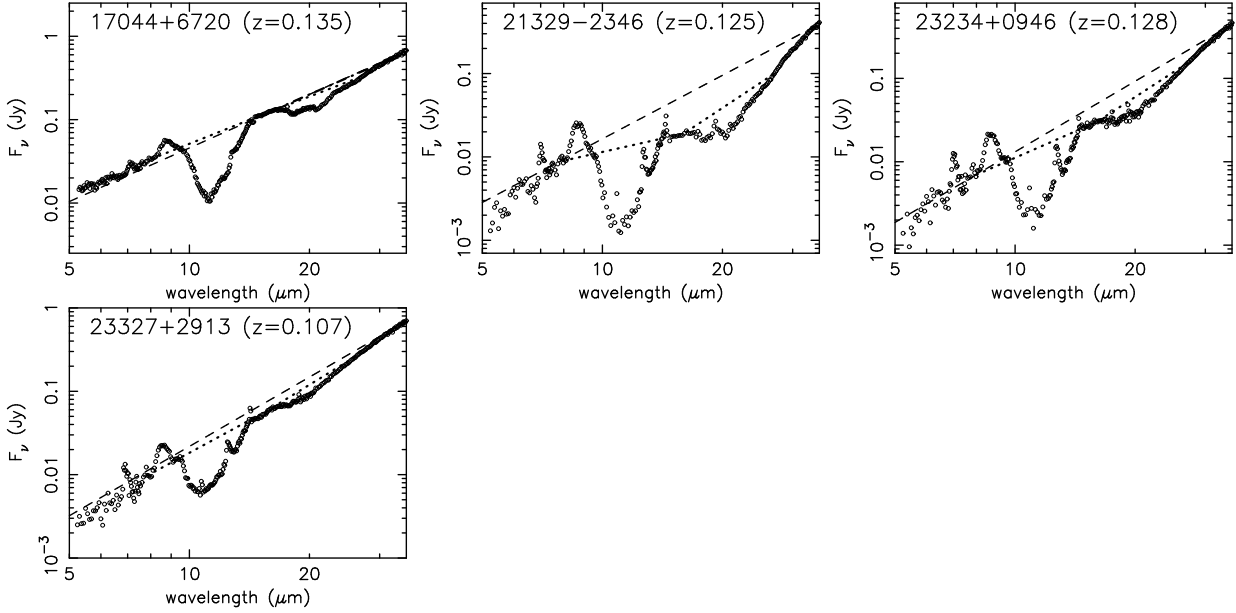

\includegraphics[angle=-90,scale=.24]{f3y.eps} 
\includegraphics[angle=-90,scale=.24]{f3z.eps} 
\includegraphics[angle=-90,scale=.24]{f3aa.eps} \\
\includegraphics[angle=-90,scale=.24]{f3ab.eps} \\
\caption{Spectra of ULIRGs classified optically as LINERs.  The
  abscissa and ordinate are, respectively, the observed wavelength in
  $\mu$m and flux F$_{\nu}$ in Jy, both shown as decimal logarithmic
  scale.  For all objects, the ratio of the uppermost to lowermost
  scale in the ordinate is a factor of 1000, to show the variation of
  the overall spectral energy distribution.  Dashed line: power-law
  continuum determined from data points at $\lambda_{\rm rest}$ = 5.6
  $\mu$m, 7.1 $\mu$m, and $\lambda_{\rm obs}$ = 34--35 $\mu$m.  Dotted
  line: power-law continuum determined from data points at
  $\lambda_{\rm rest}$ = 7.1 $\mu$m and 14.2 $\mu$m for 9.7 $\mu$m
  silicate absorption, and at $\lambda_{\rm rest}$ = 14.2 $\mu$m and
  24 $\mu$m for 18 $\mu$m silicate absorption.  Dashed-dotted line:
  power-law continuum determined from data points at $\lambda_{\rm
    rest}$ = 14.2 $\mu$m and 29 $\mu$m, used only for ULIRGs showing
  strong 18 $\mu$m silicate dust absorption features.  }
\end{figure}

\clearpage

\begin{figure}
\includegraphics[angle=-90,scale=.24]{f4a.eps} 
\includegraphics[angle=-90,scale=.24]{f4b.eps} 
\includegraphics[angle=-90,scale=.24]{f4c.eps} \\
\includegraphics[angle=-90,scale=.24]{f4d.eps} 
\includegraphics[angle=-90,scale=.24]{f4e.eps} 
\includegraphics[angle=-90,scale=.24]{f4f.eps} \\
\includegraphics[angle=-90,scale=.24]{f4g.eps} 
\includegraphics[angle=-90,scale=.24]{f4h.eps} 
\includegraphics[angle=-90,scale=.24]{f4i.eps} \\
\includegraphics[angle=-90,scale=.24]{f4j.eps} 
\includegraphics[angle=-90,scale=.24]{f4k.eps} 
\includegraphics[angle=-90,scale=.24]{f4l.eps} \\ 
\end{figure}

\clearpage

\begin{figure}
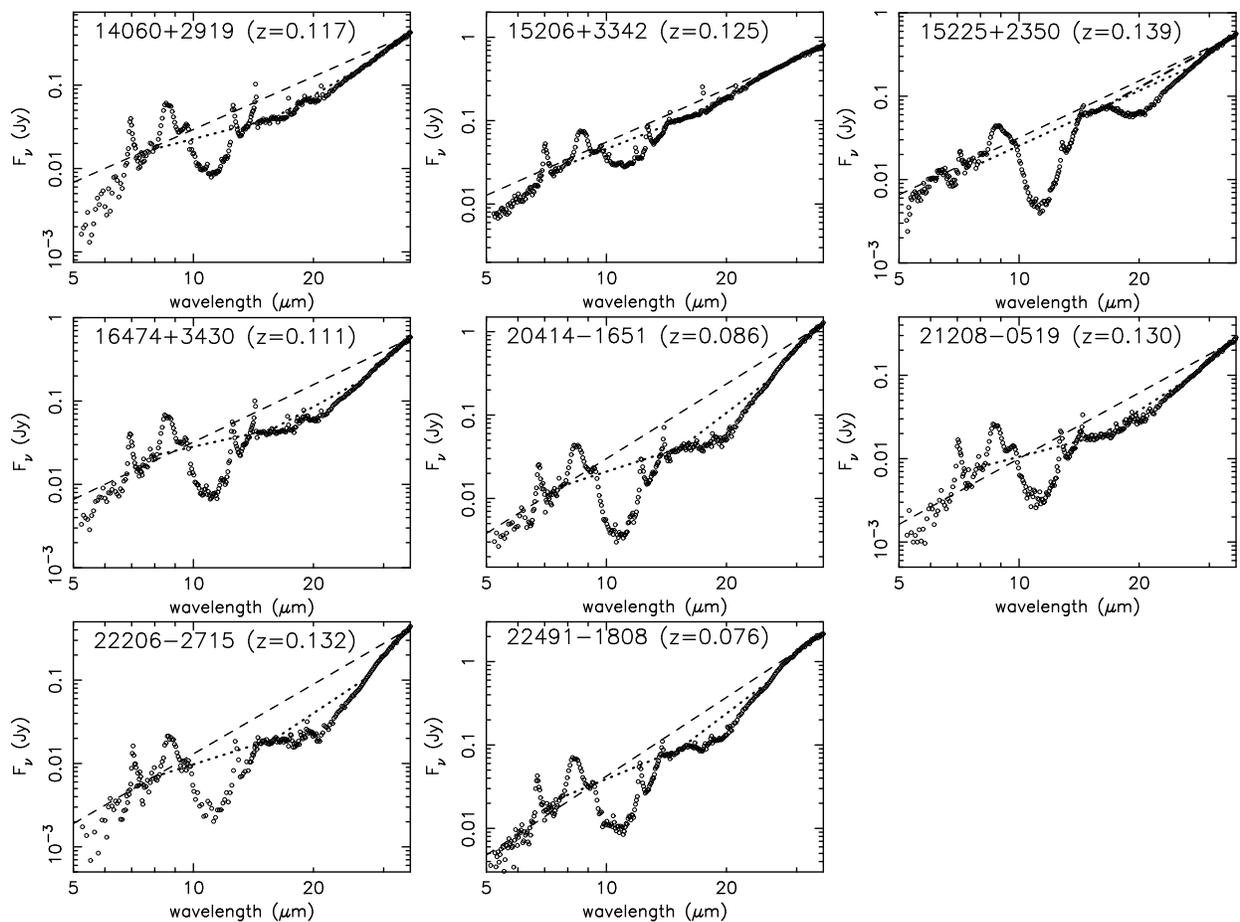

\includegraphics[angle=-90,scale=.24]{f4m.eps} 
\includegraphics[angle=-90,scale=.24]{f4n.eps}  
\includegraphics[angle=-90,scale=.24]{f4o.eps} \\
\includegraphics[angle=-90,scale=.24]{f4p.eps} 
\includegraphics[angle=-90,scale=.24]{f4q.eps} 
\includegraphics[angle=-90,scale=.24]{f4r.eps} \\ 
\includegraphics[angle=-90,scale=.24]{f4s.eps} 
\includegraphics[angle=-90,scale=.24]{f4t.eps} \\
\caption{Spectra of ULIRGs classified optically as HII-regions, shown
  in the same way as Figure 3. Symbols are also the same as Figure 3.
  For IRAS 10190+1322, SL spectra of the eastern and western nuclei
  are combined.  }
\end{figure}

\clearpage

\begin{figure}
\includegraphics[angle=-90,scale=.24]{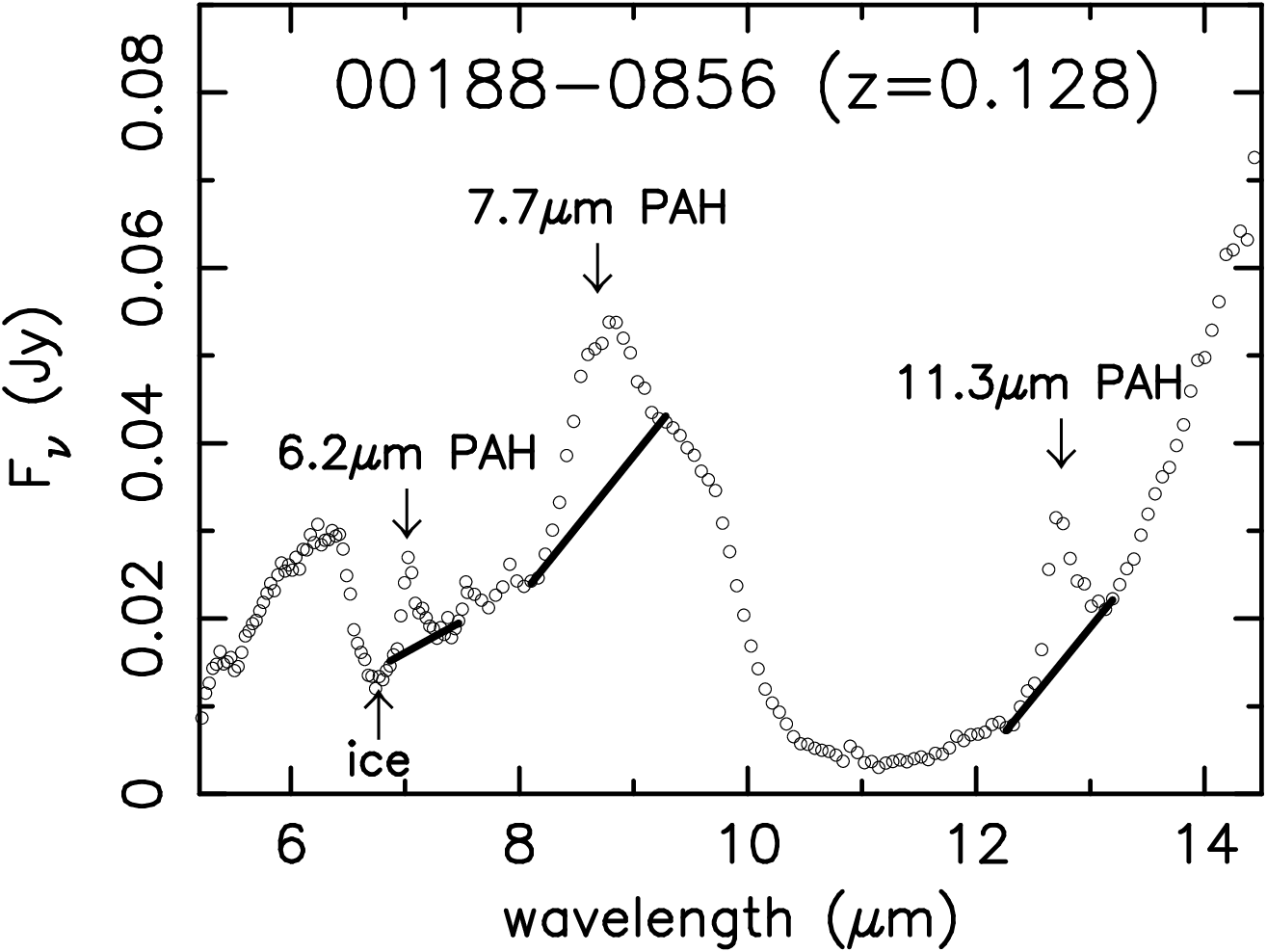} 
\includegraphics[angle=-90,scale=.24]{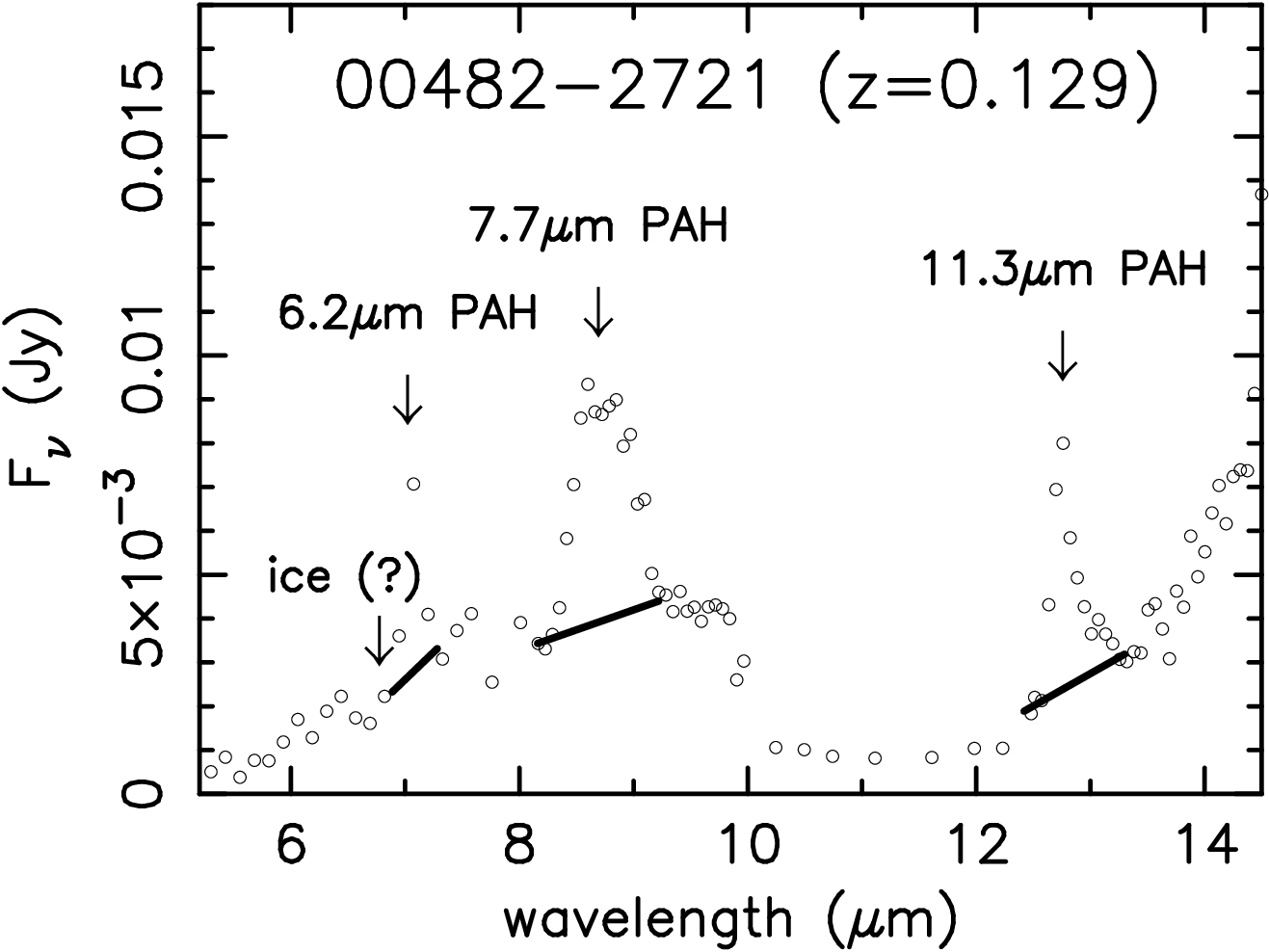} 
\includegraphics[angle=-90,scale=.24]{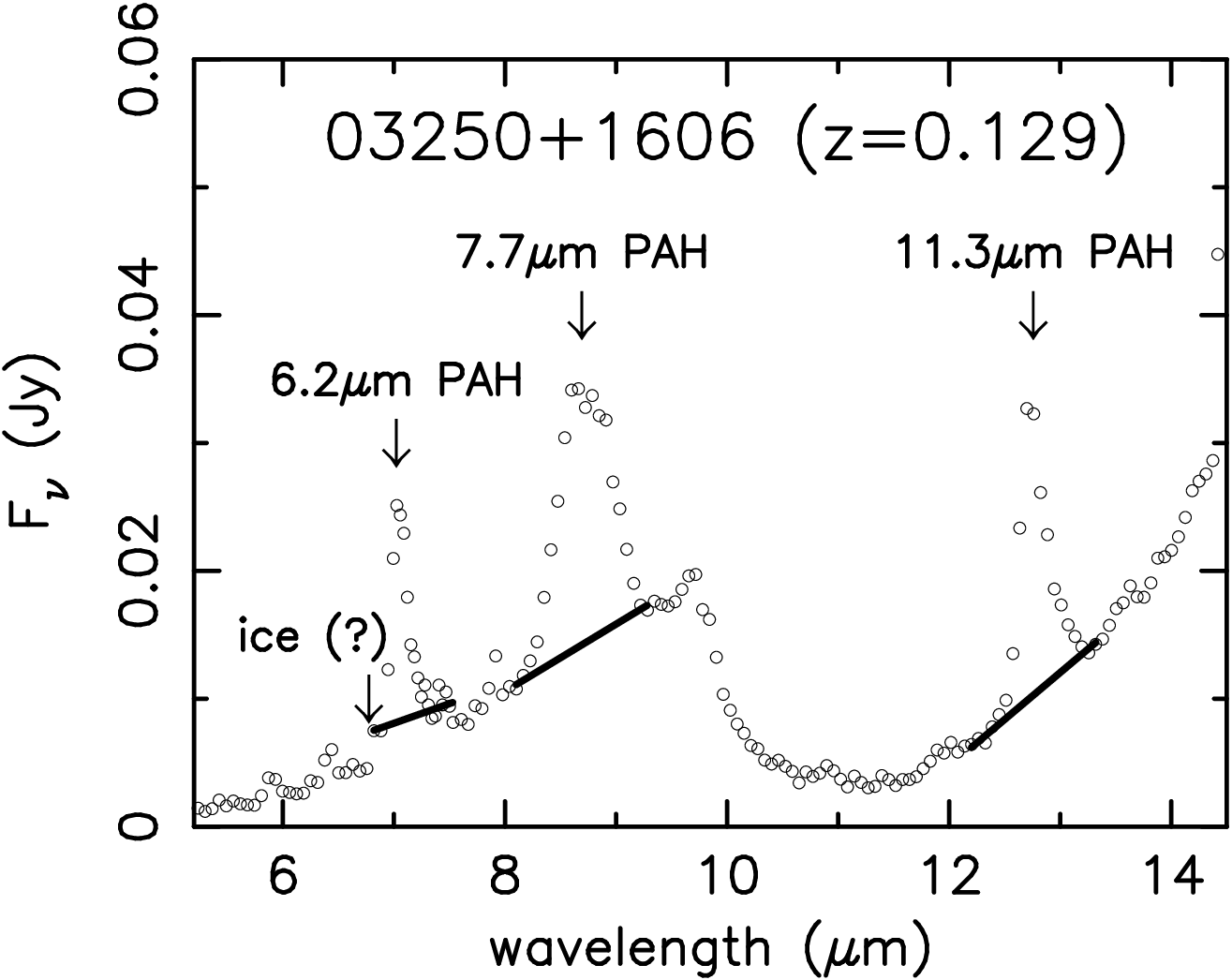} \\ 
\includegraphics[angle=-90,scale=.24]{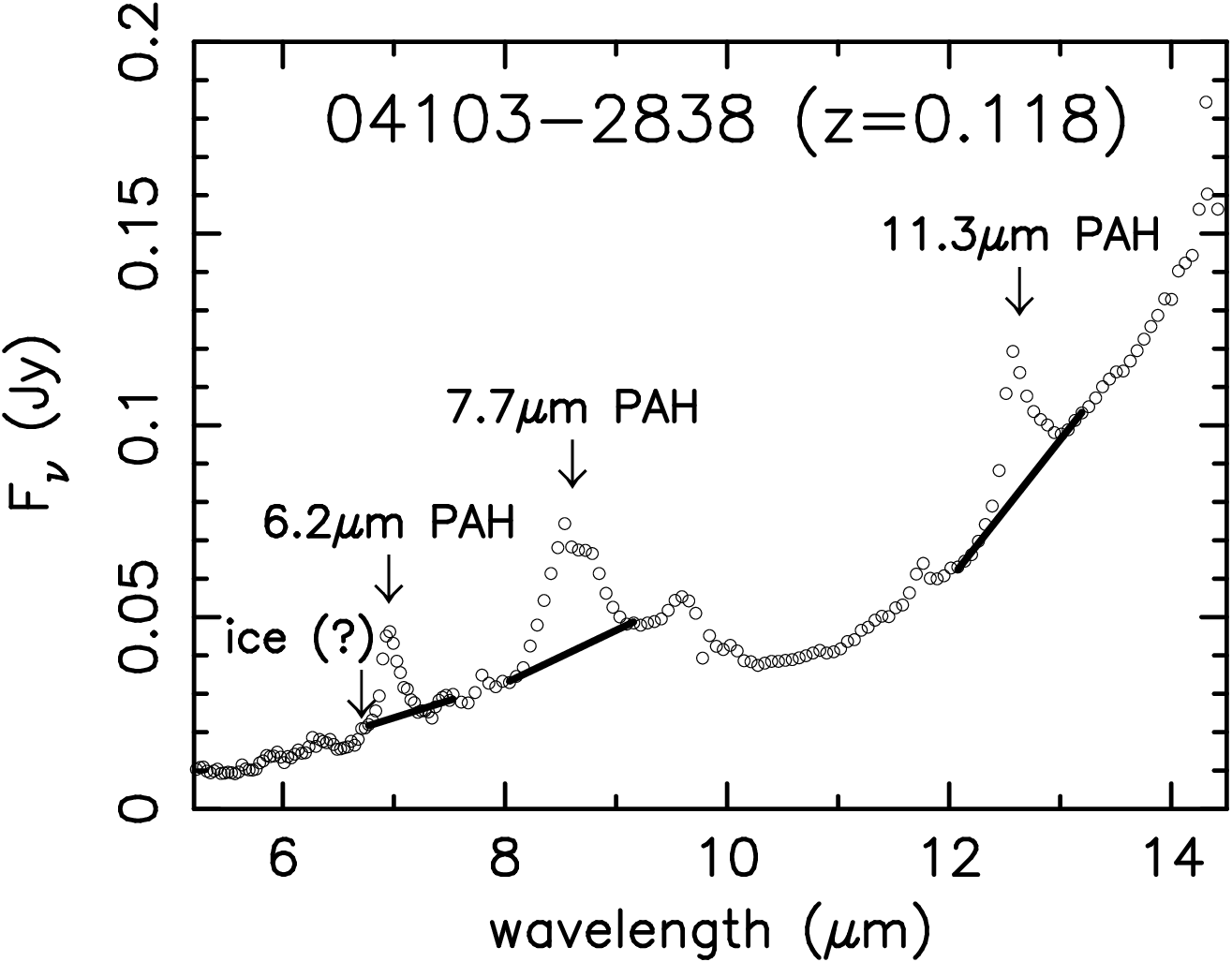} 
\includegraphics[angle=-90,scale=.24]{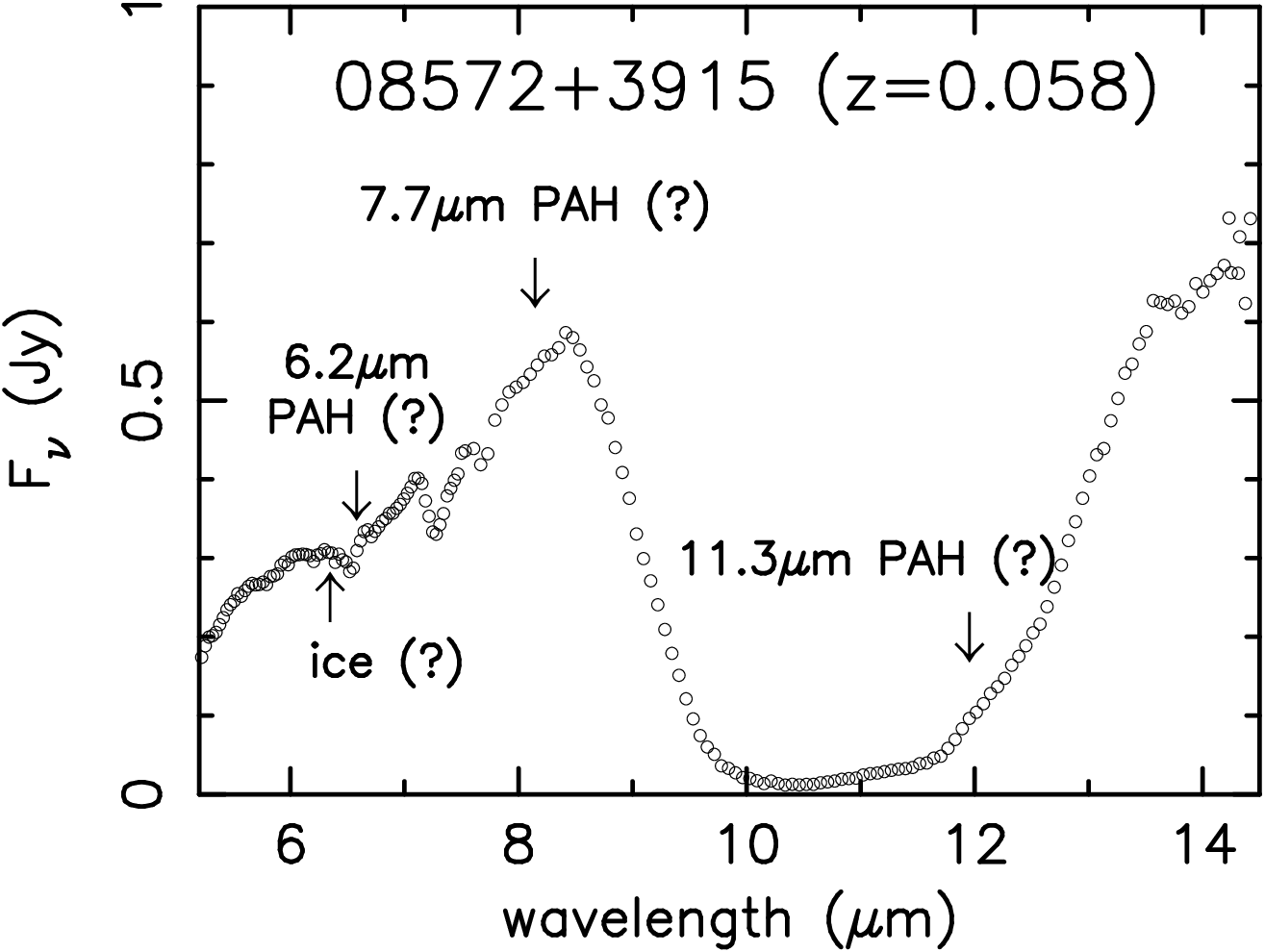} 
\includegraphics[angle=-90,scale=.24]{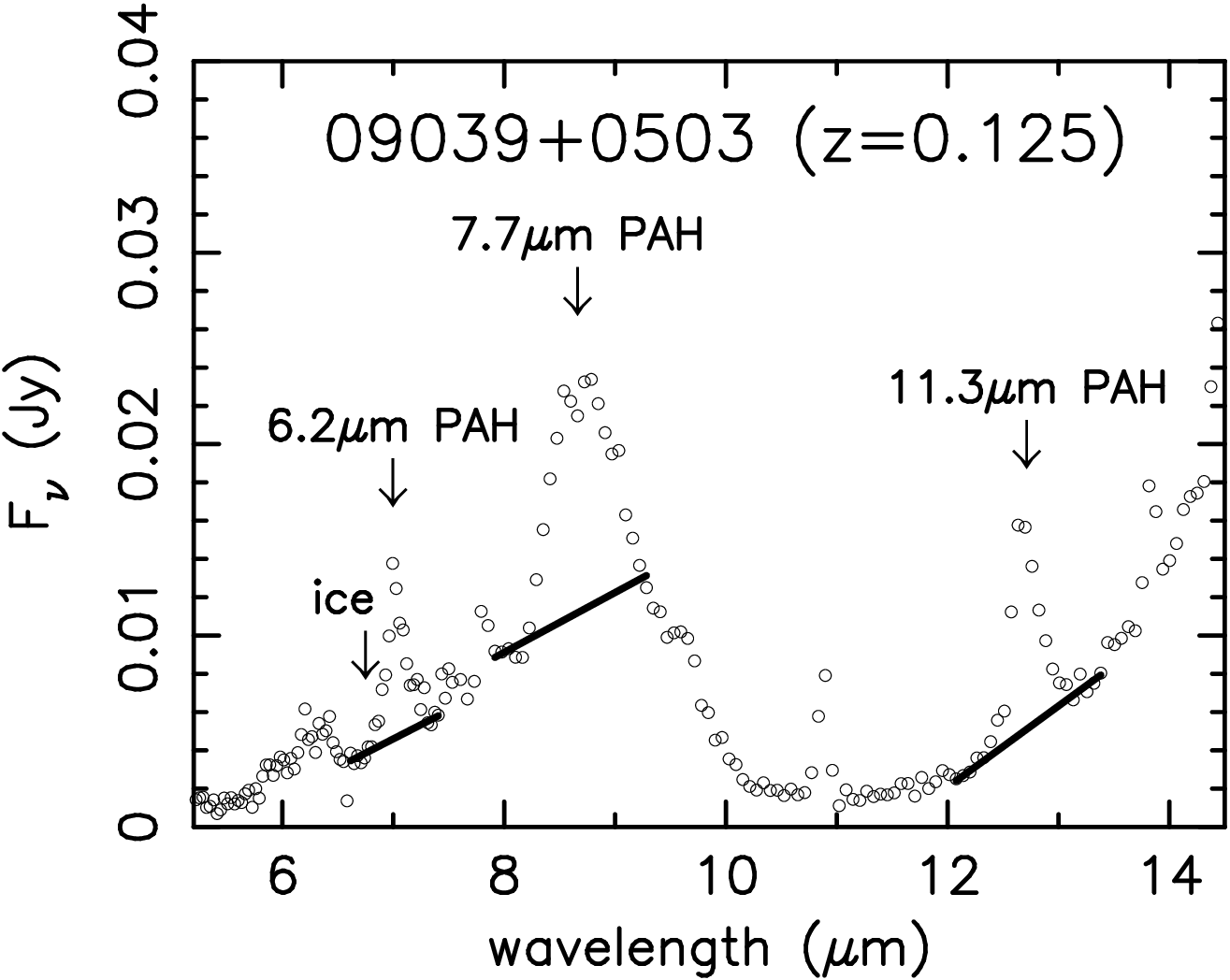} \\
\includegraphics[angle=-90,scale=.24]{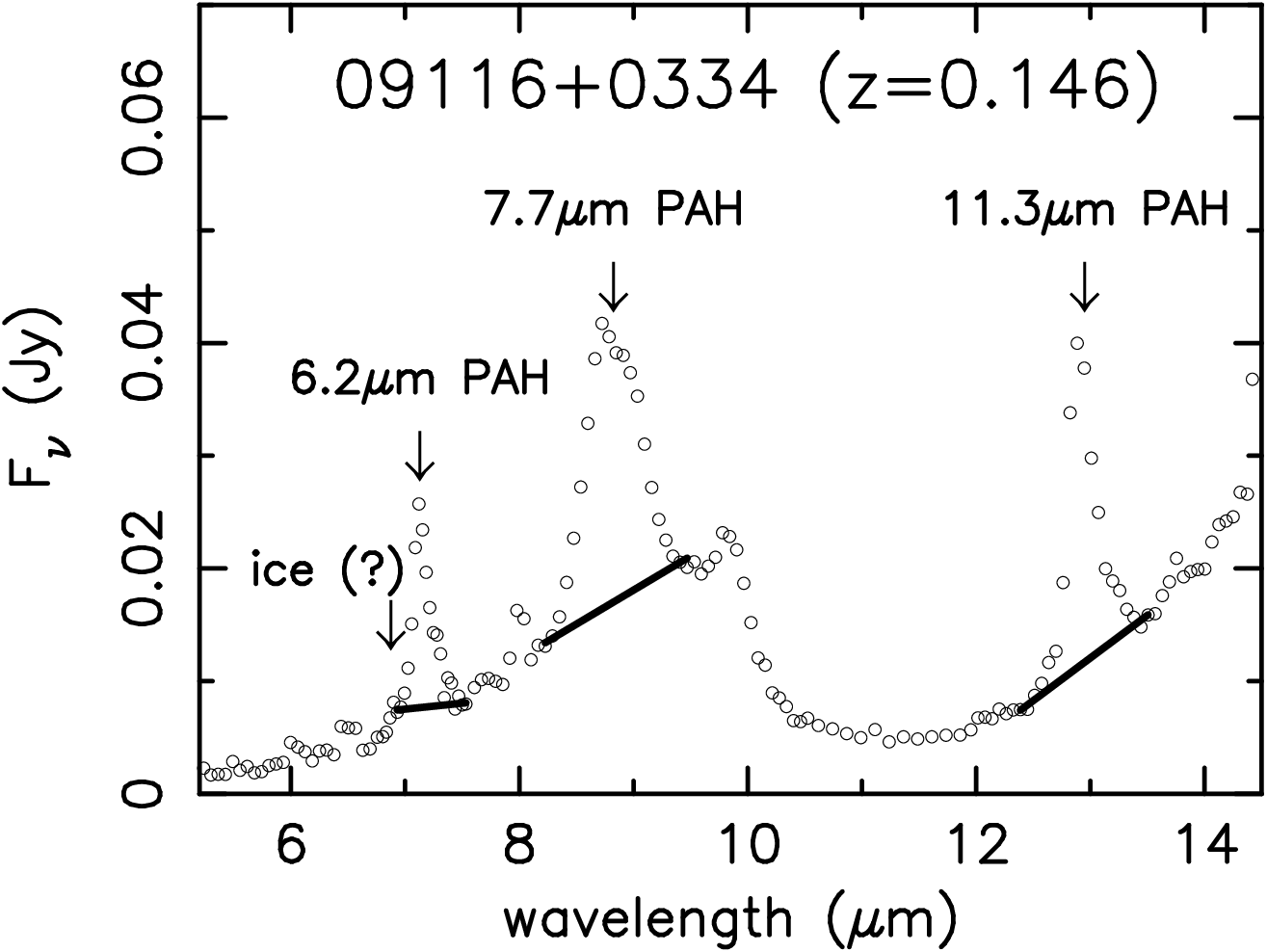} 
\includegraphics[angle=-90,scale=.24]{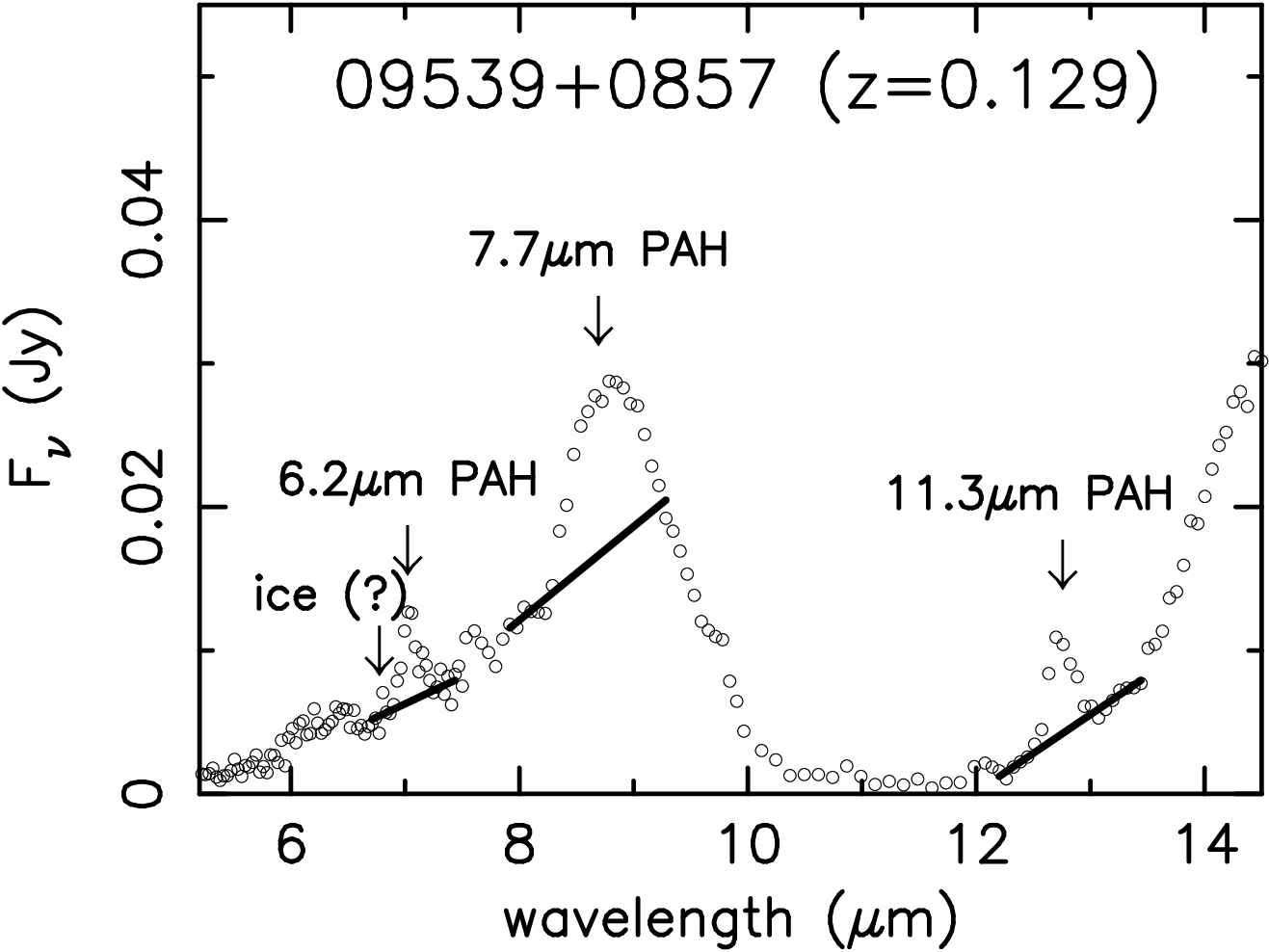} 
\includegraphics[angle=-90,scale=.24]{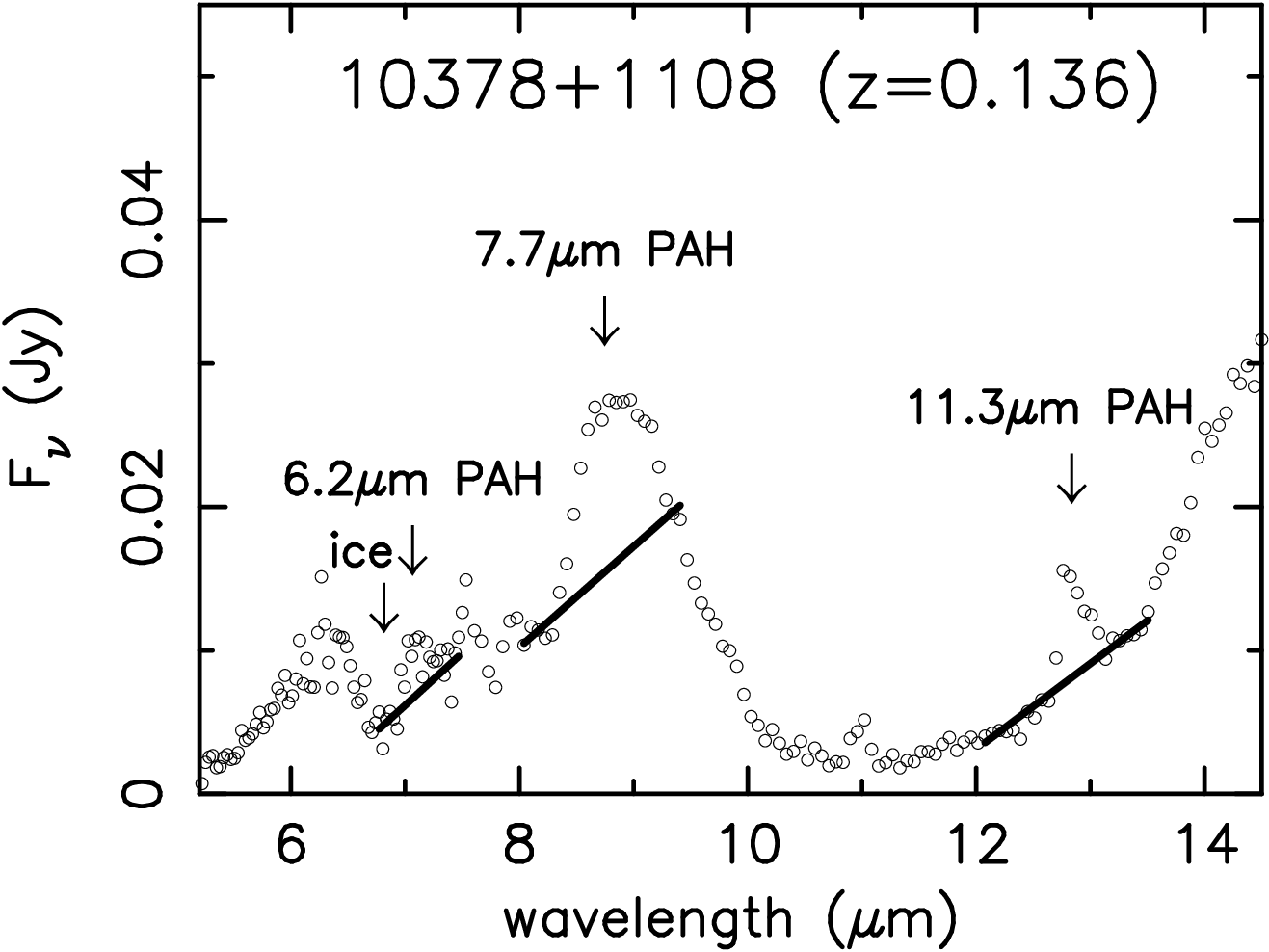} \\
\includegraphics[angle=-90,scale=.24]{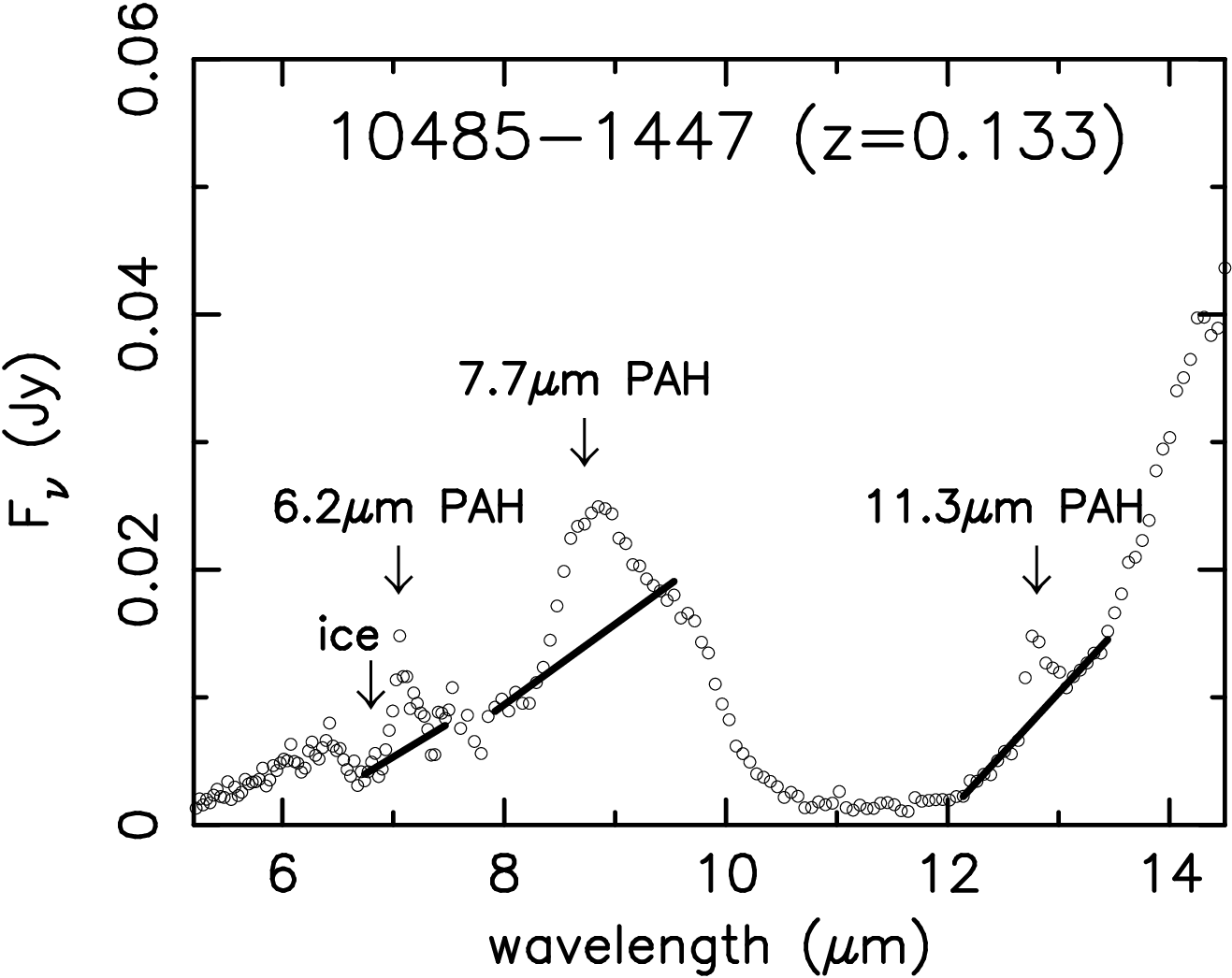} 
\includegraphics[angle=-90,scale=.24]{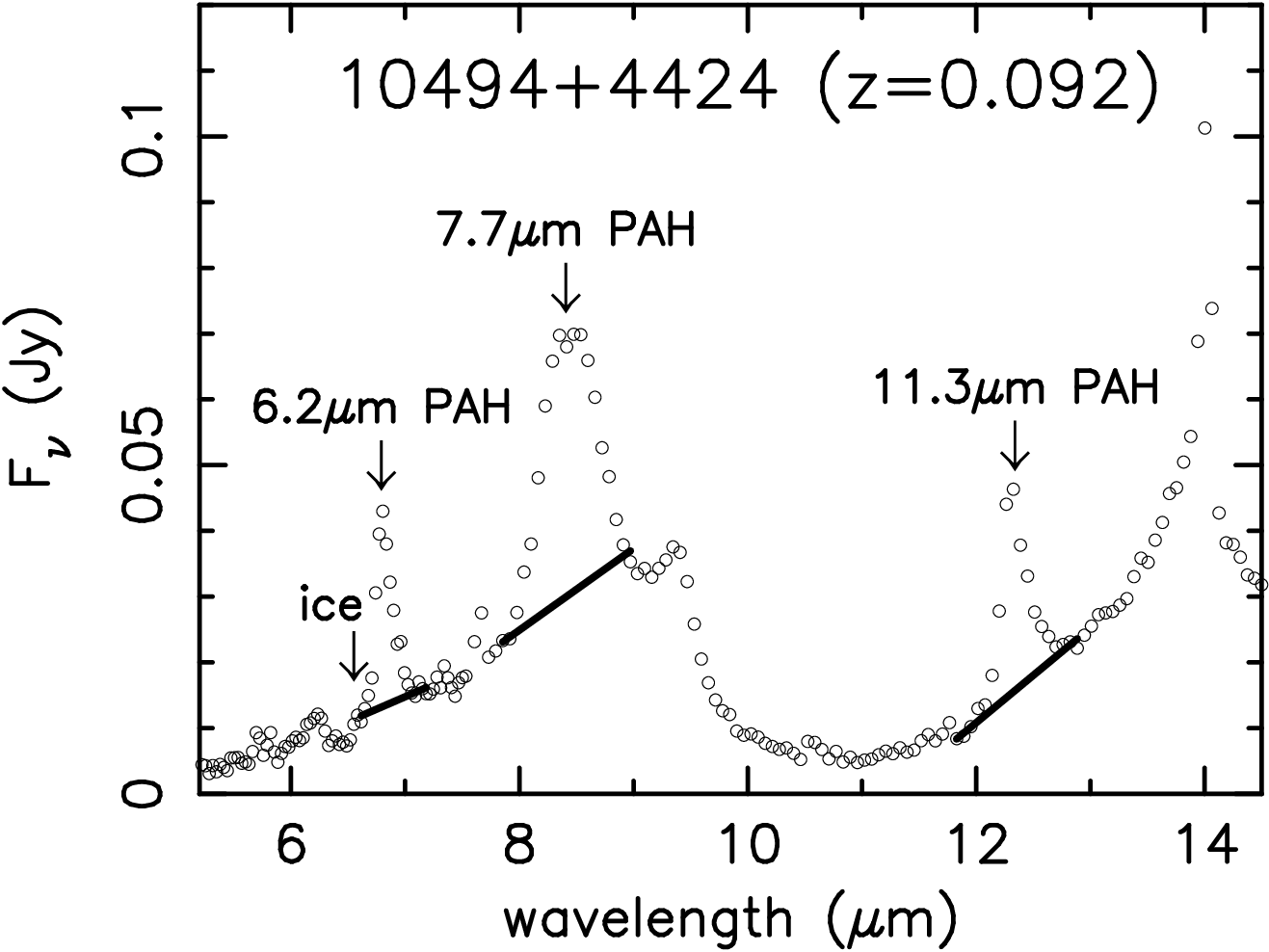} 
\includegraphics[angle=-90,scale=.24]{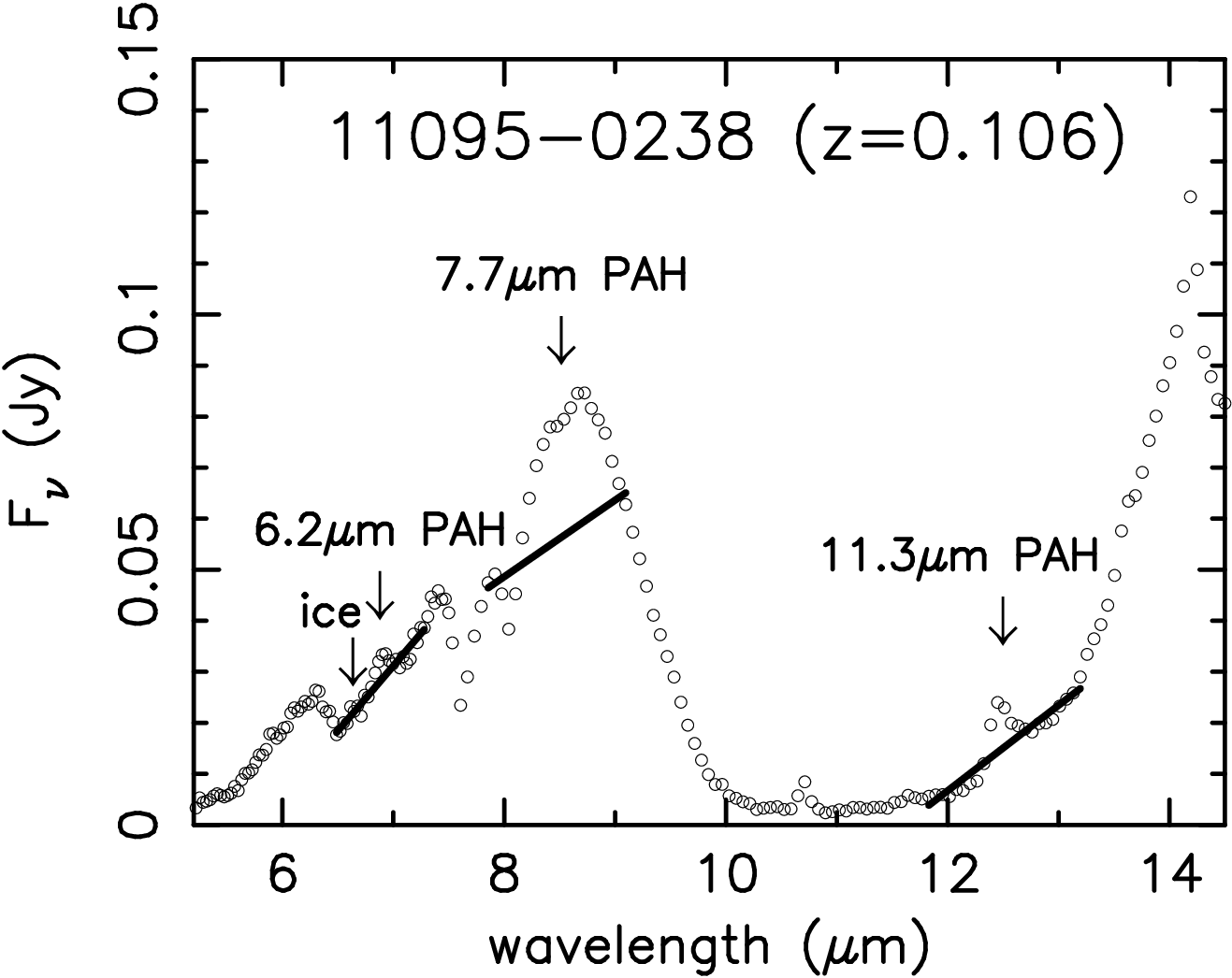} \\ 
\includegraphics[angle=-90,scale=.24]{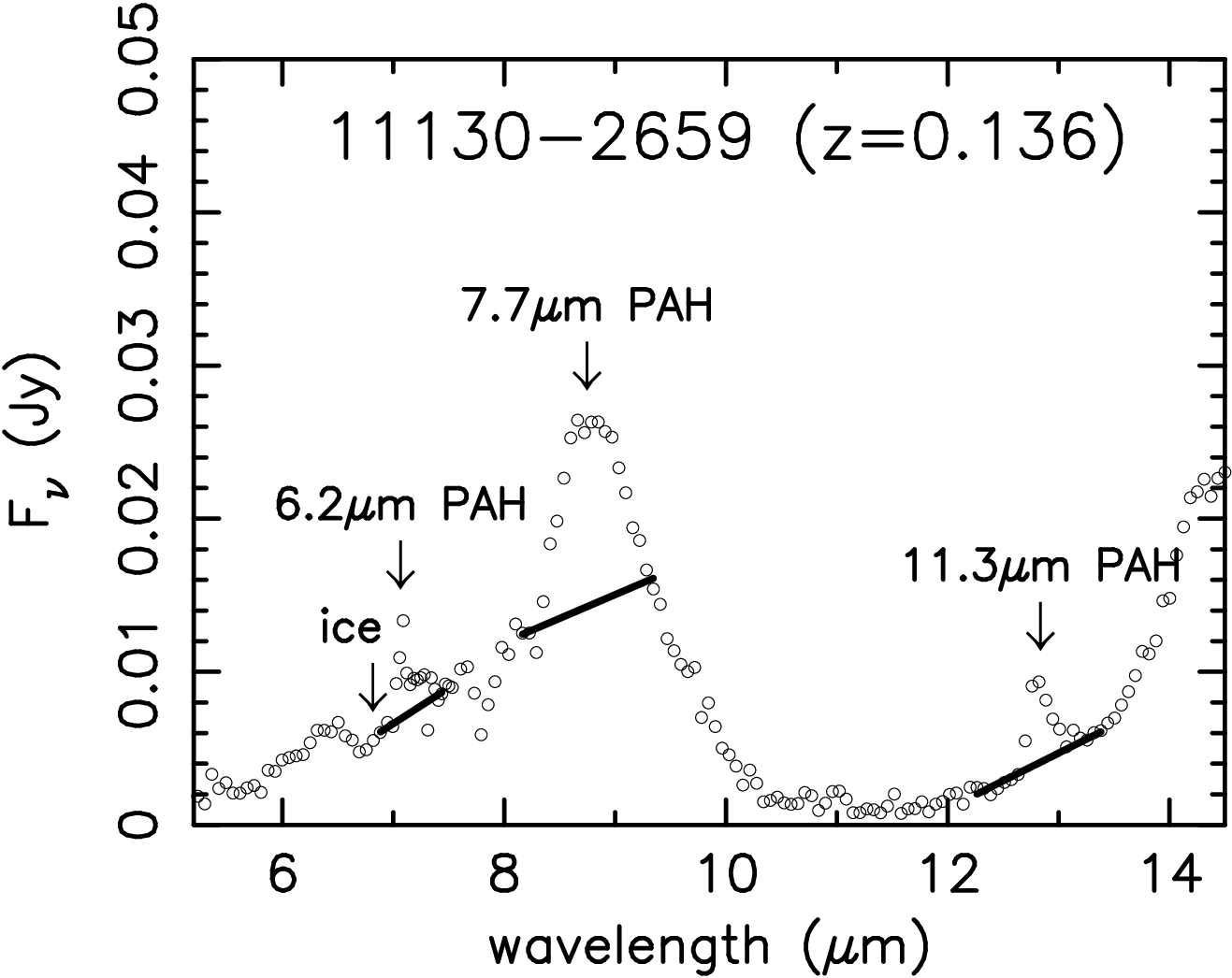} 
\includegraphics[angle=-90,scale=.24]{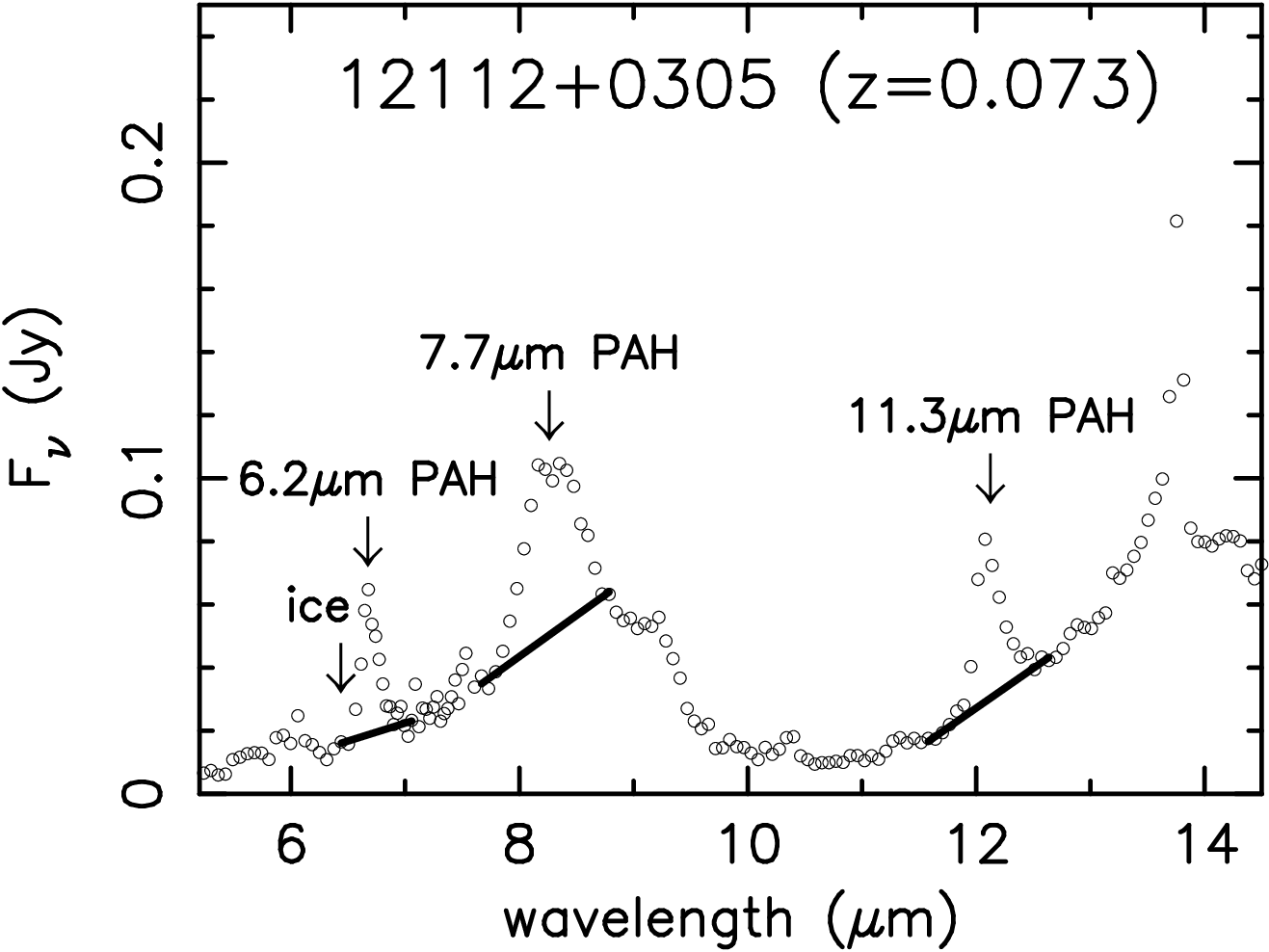} 
\includegraphics[angle=-90,scale=.24]{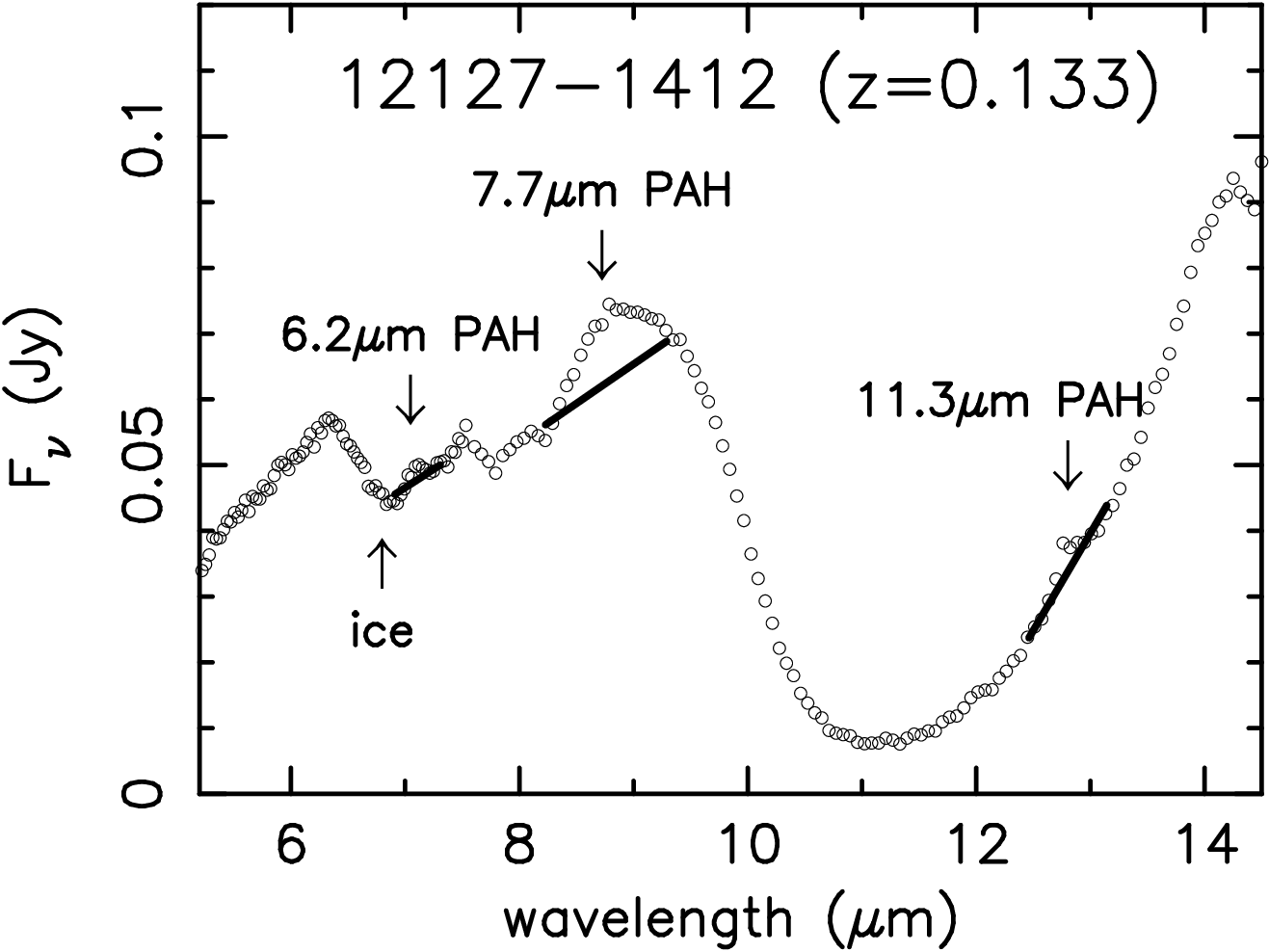} \\
\end{figure}

\clearpage

\begin{figure}
\includegraphics[angle=-90,scale=.24]{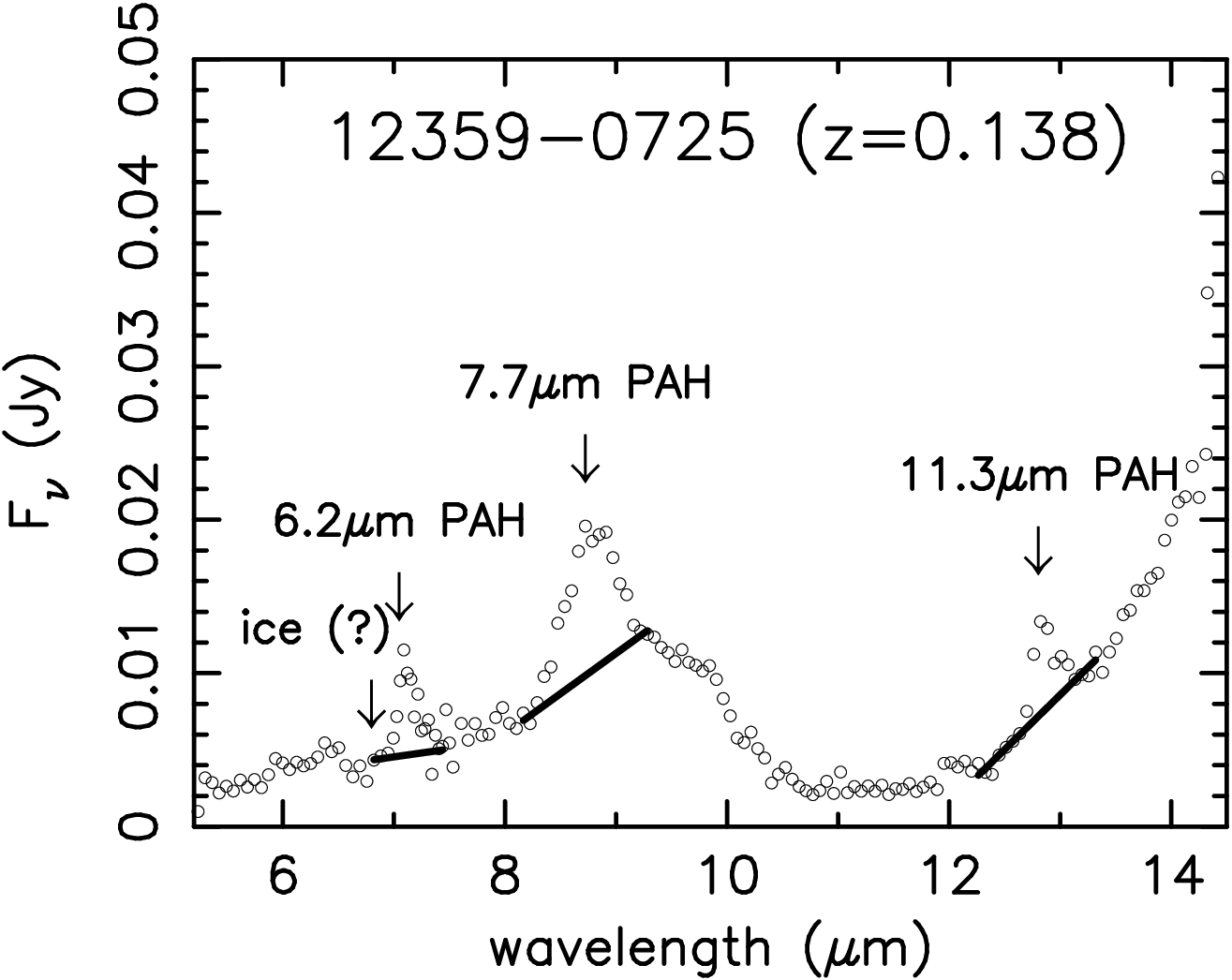} 
\includegraphics[angle=-90,scale=.24]{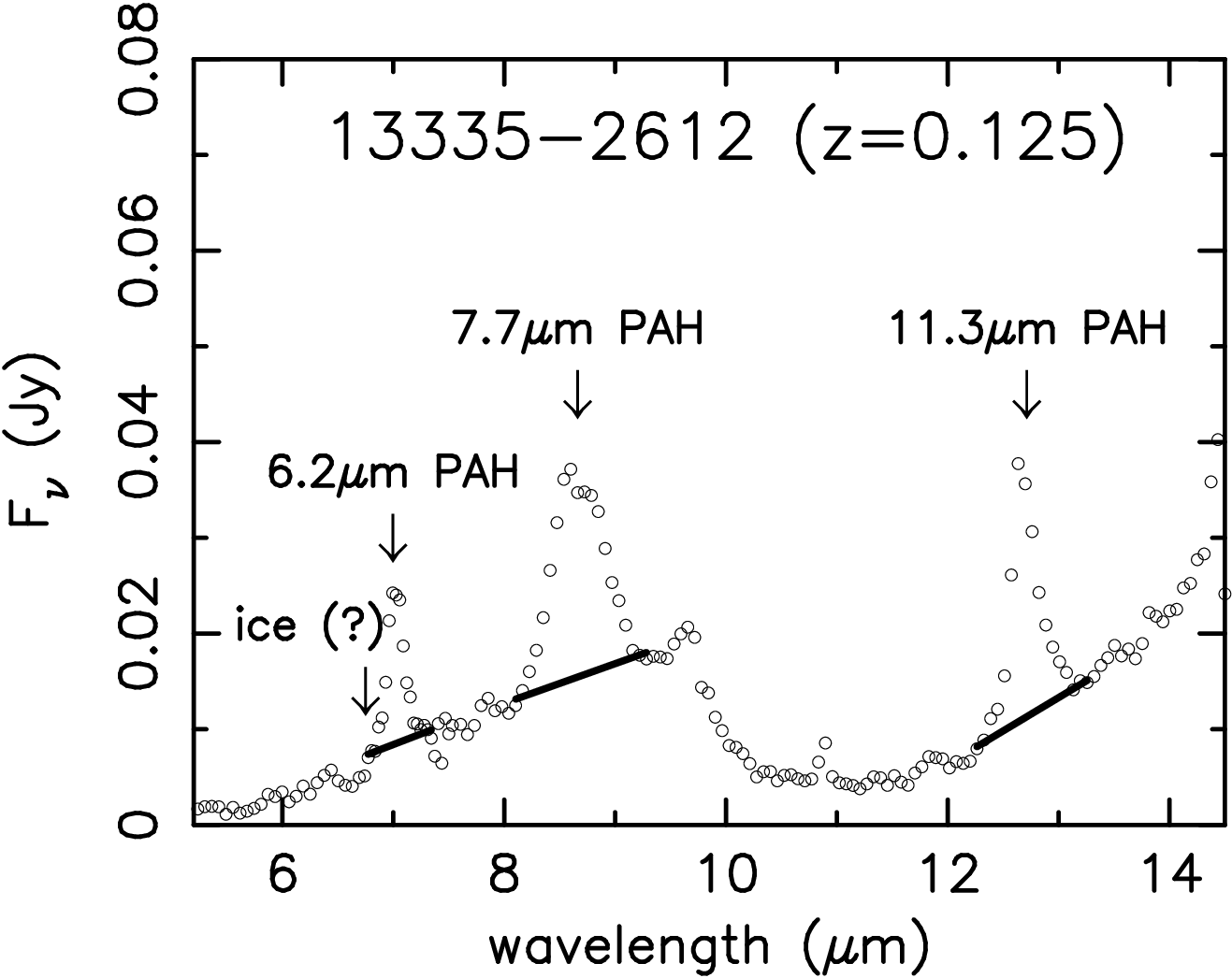} 
\includegraphics[angle=-90,scale=.24]{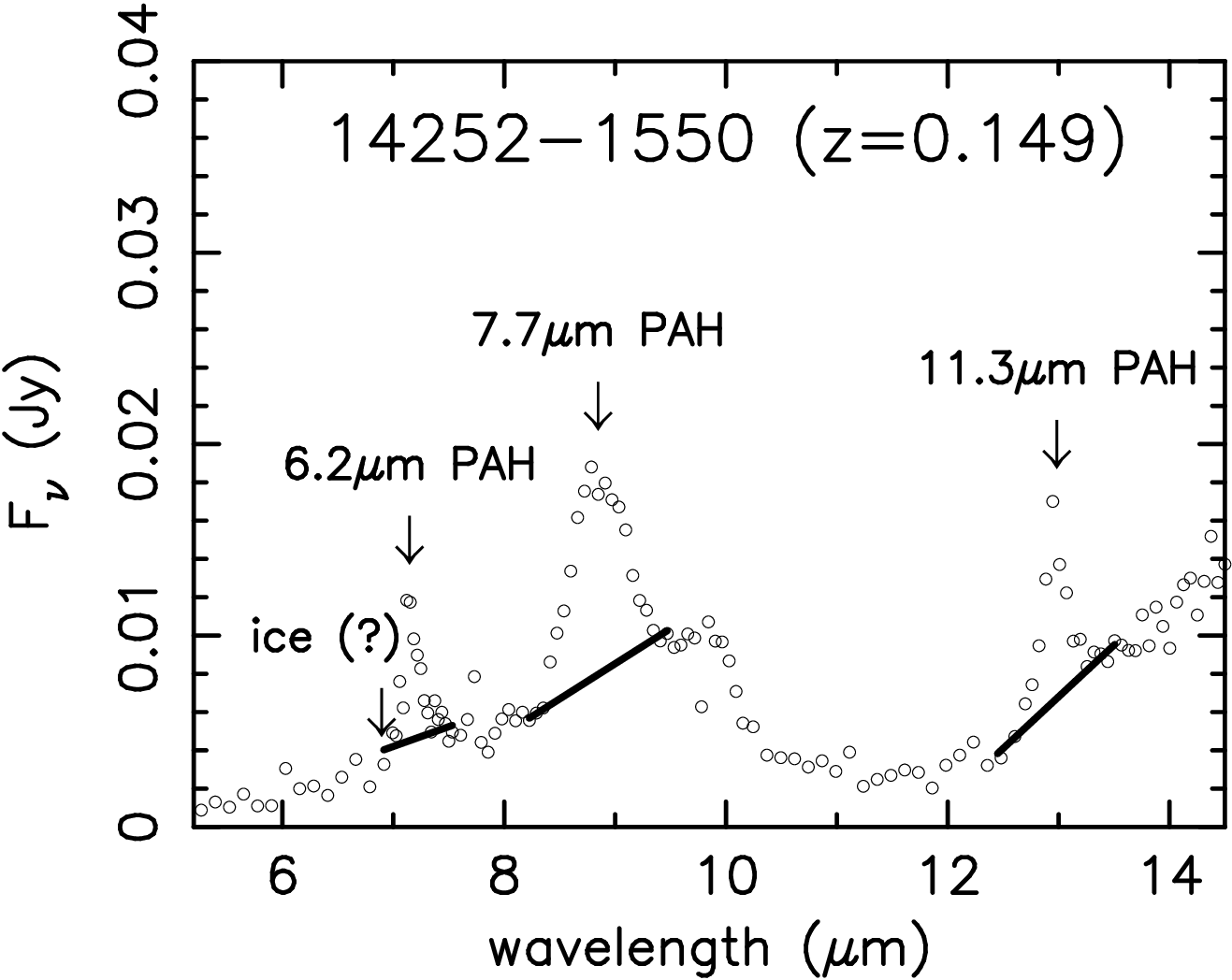} \\ 
\includegraphics[angle=-90,scale=.24]{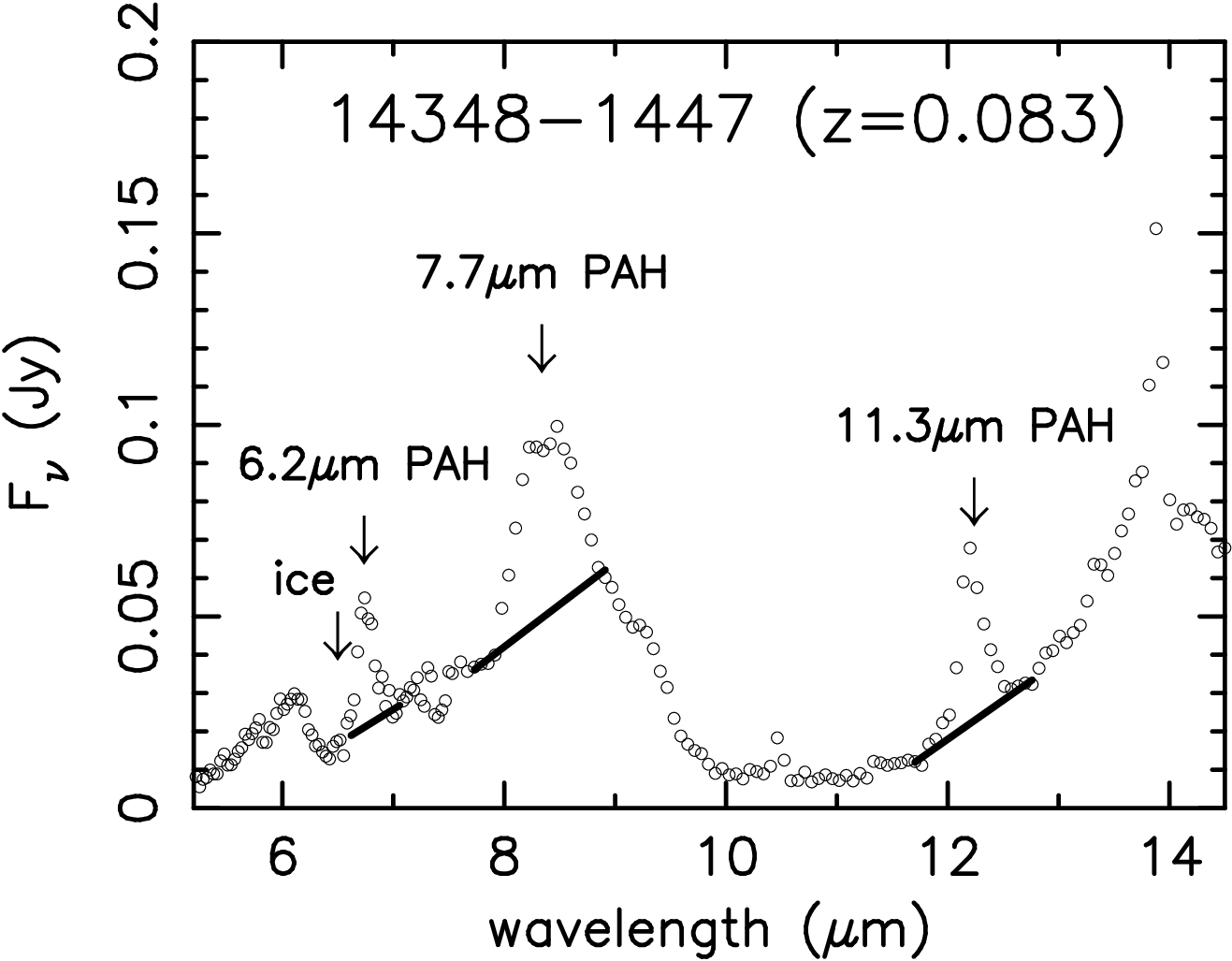} 
\includegraphics[angle=-90,scale=.24]{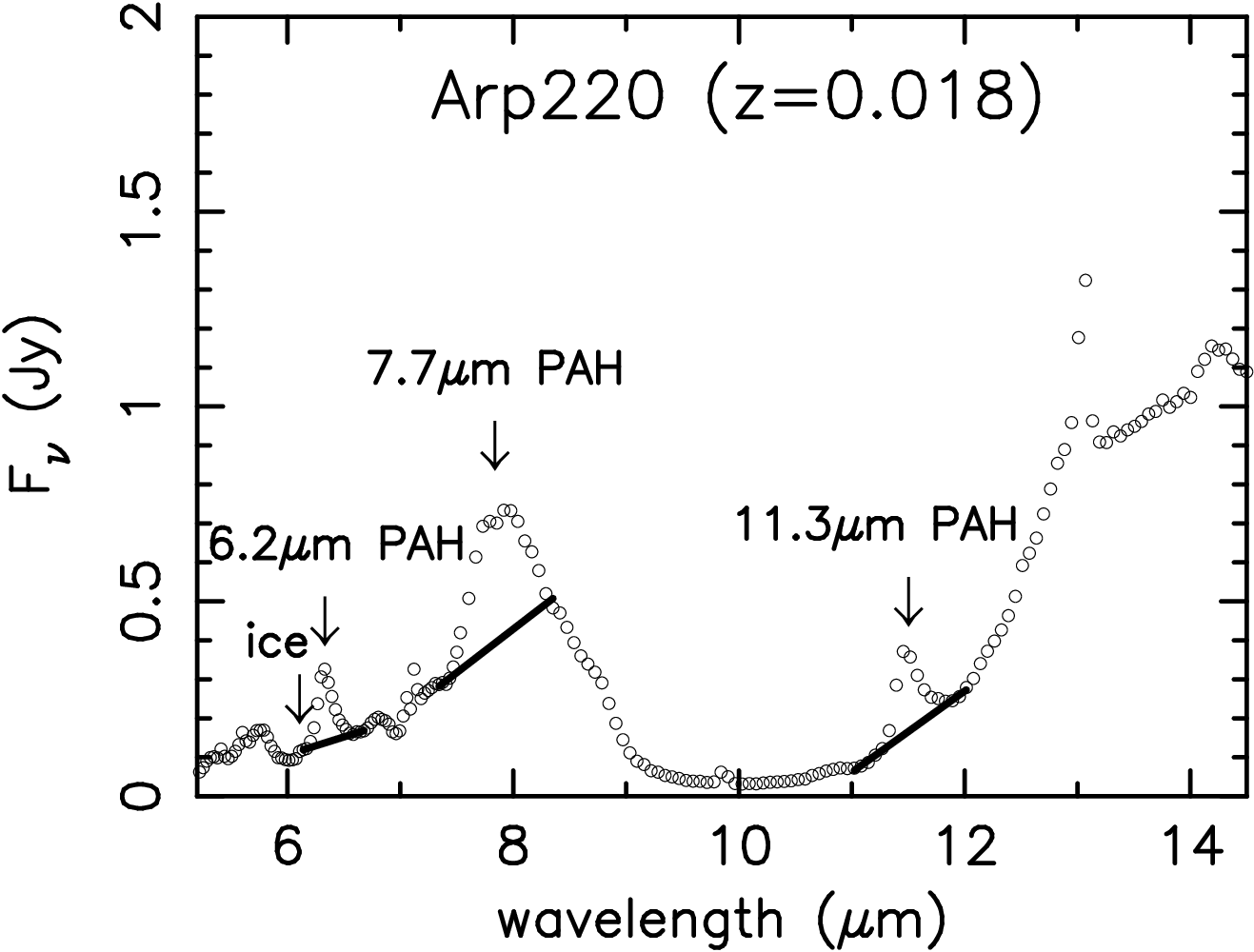}  
\includegraphics[angle=-90,scale=.24]{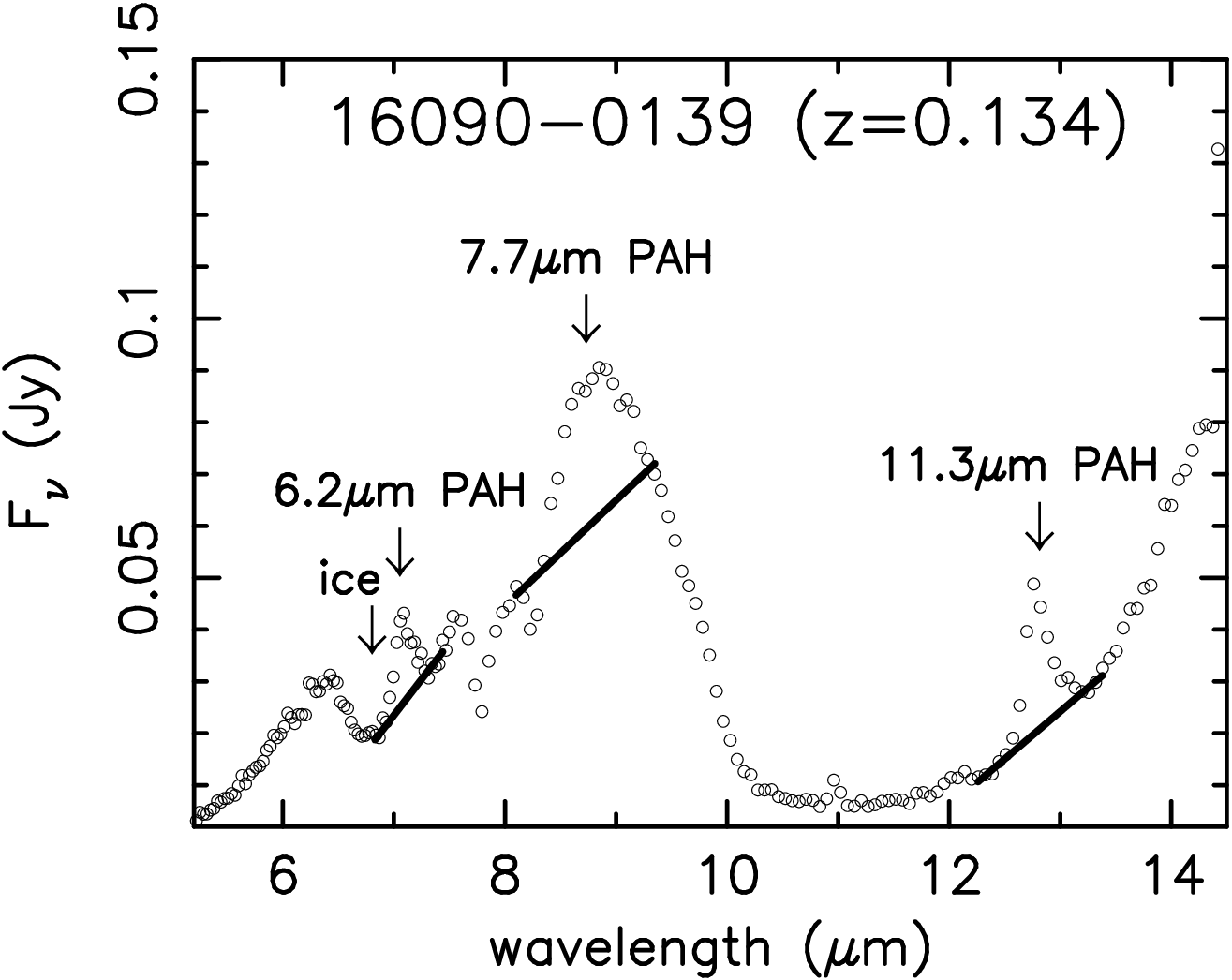}  \\
\includegraphics[angle=-90,scale=.24]{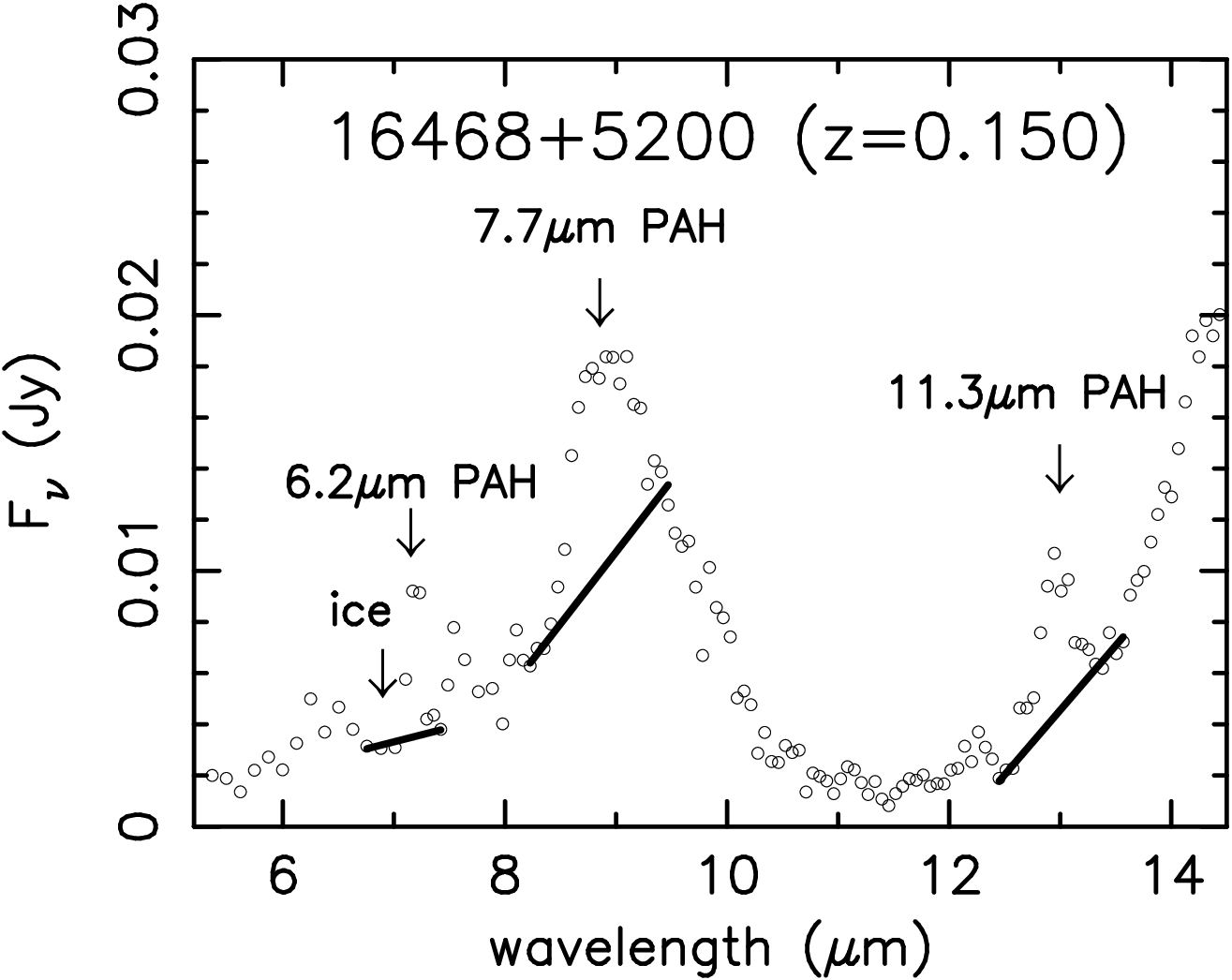} 
\includegraphics[angle=-90,scale=.24]{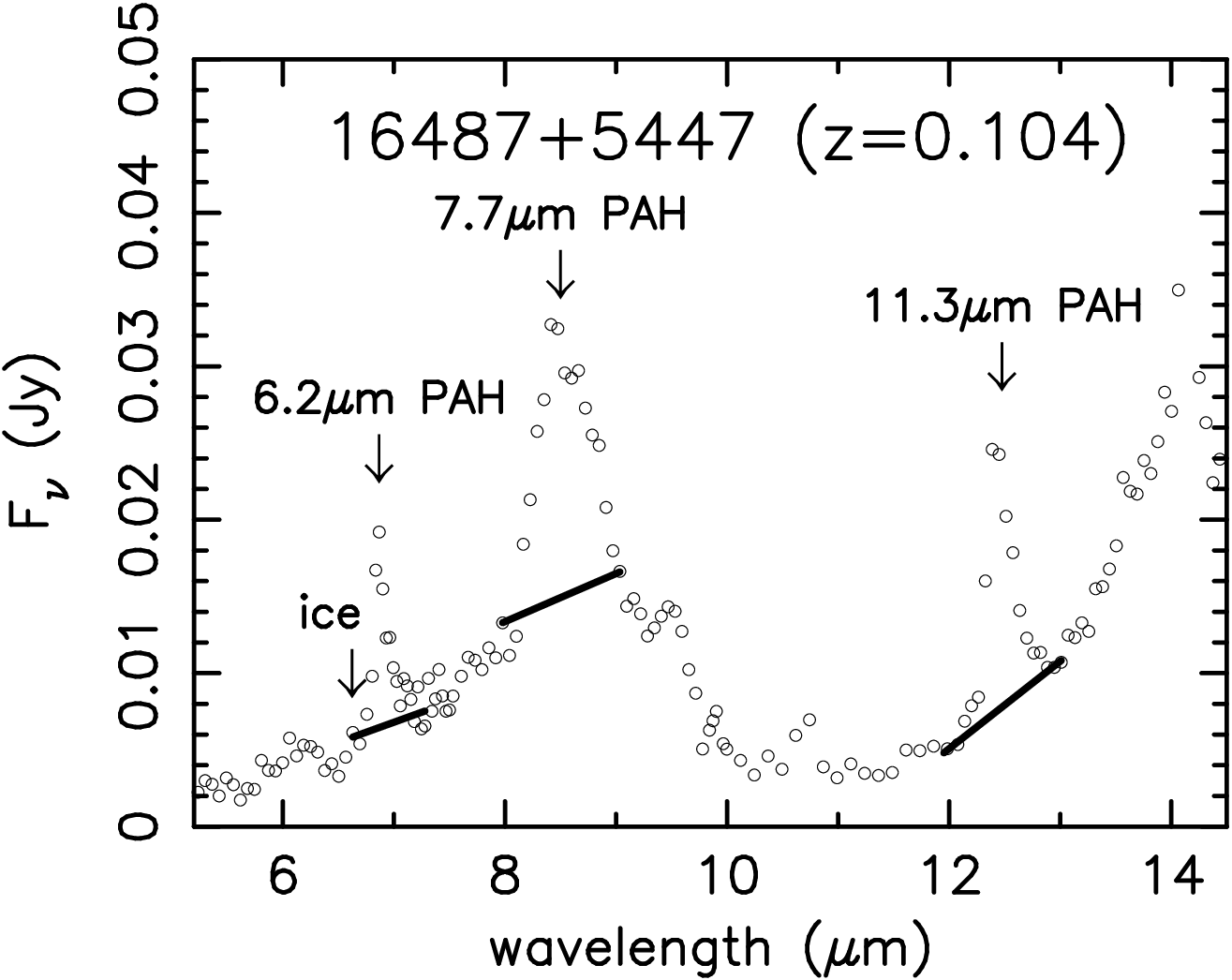} 
\includegraphics[angle=-90,scale=.24]{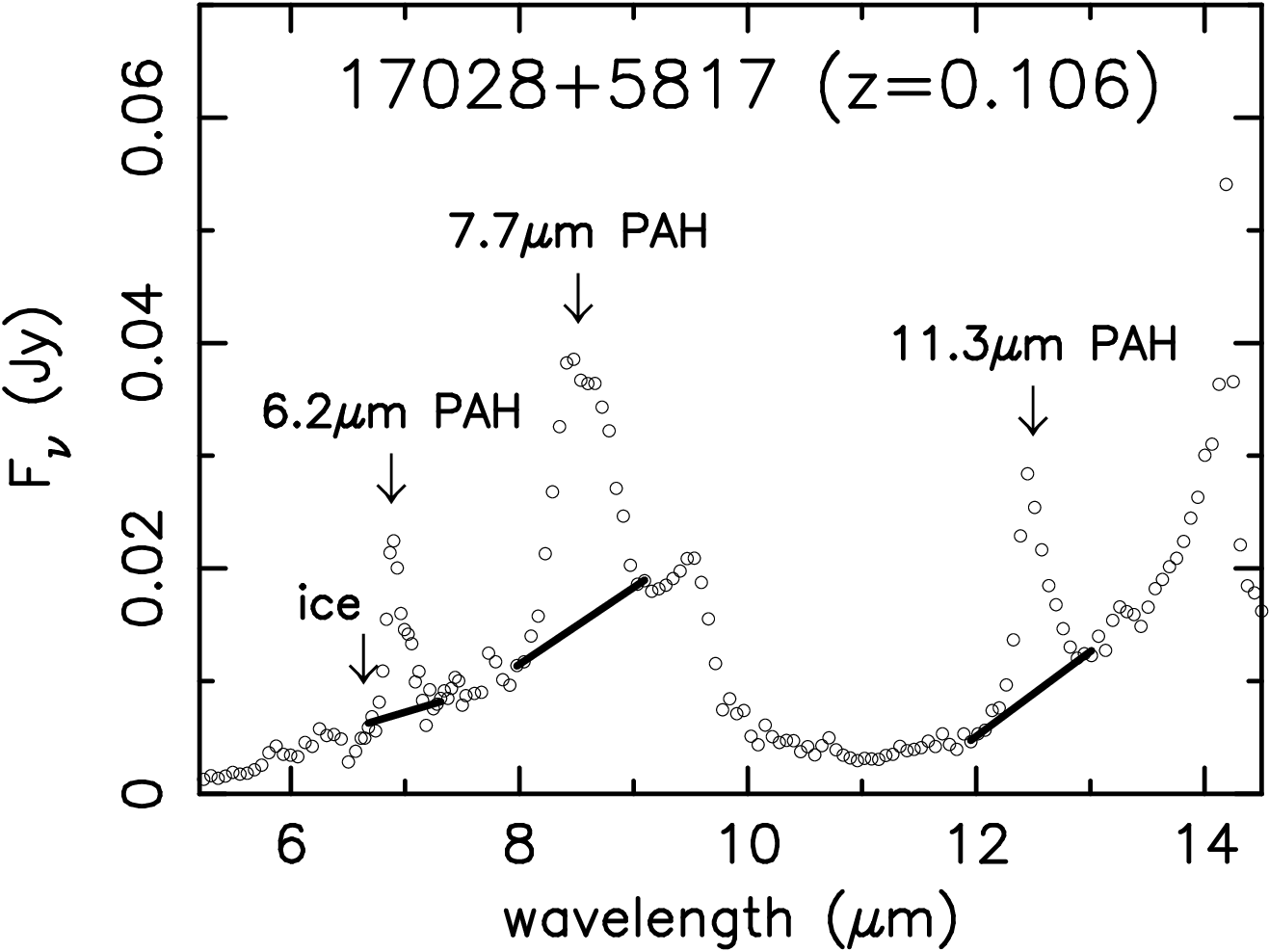} \\
\includegraphics[angle=-90,scale=.24]{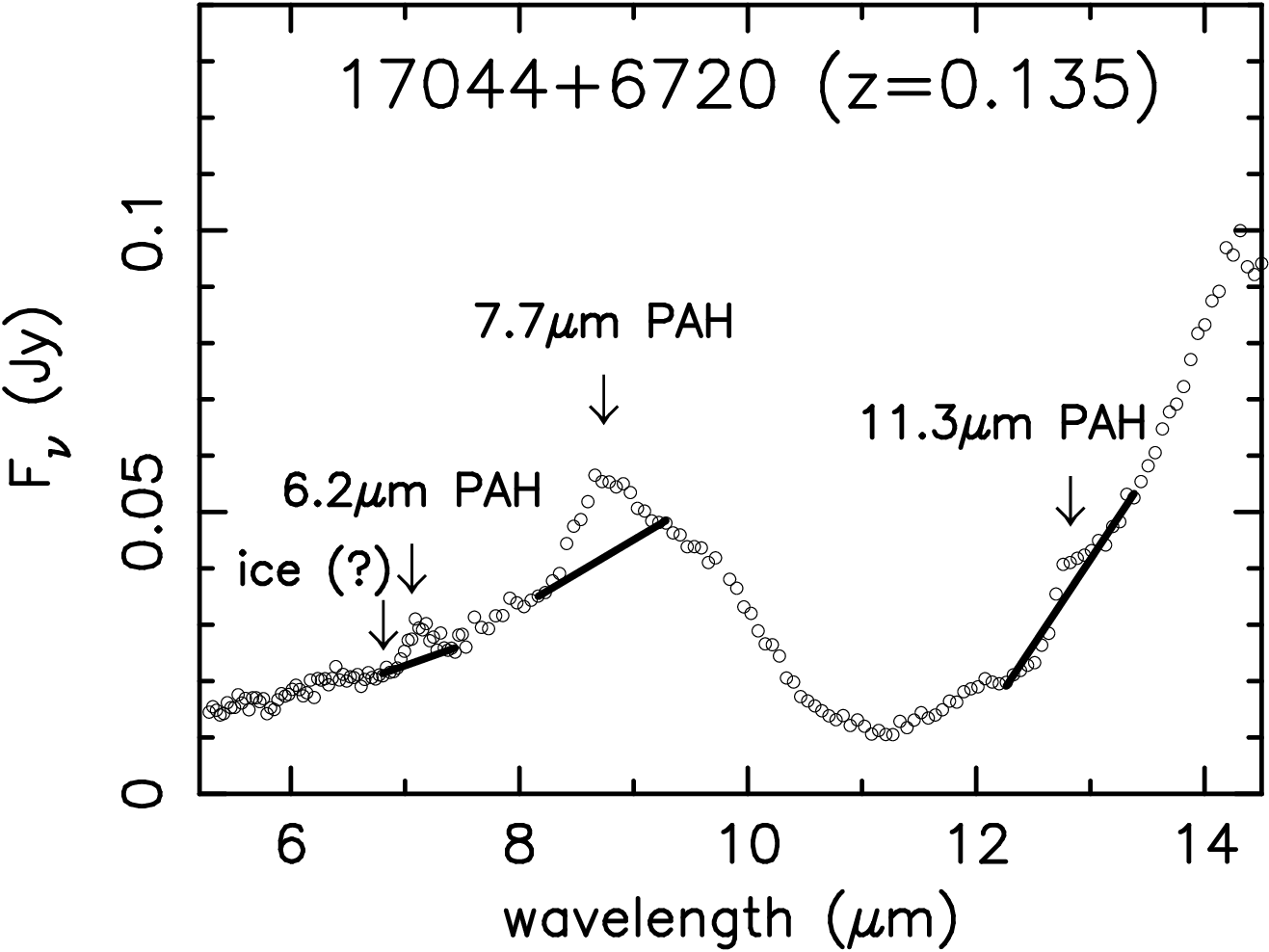} 
\includegraphics[angle=-90,scale=.24]{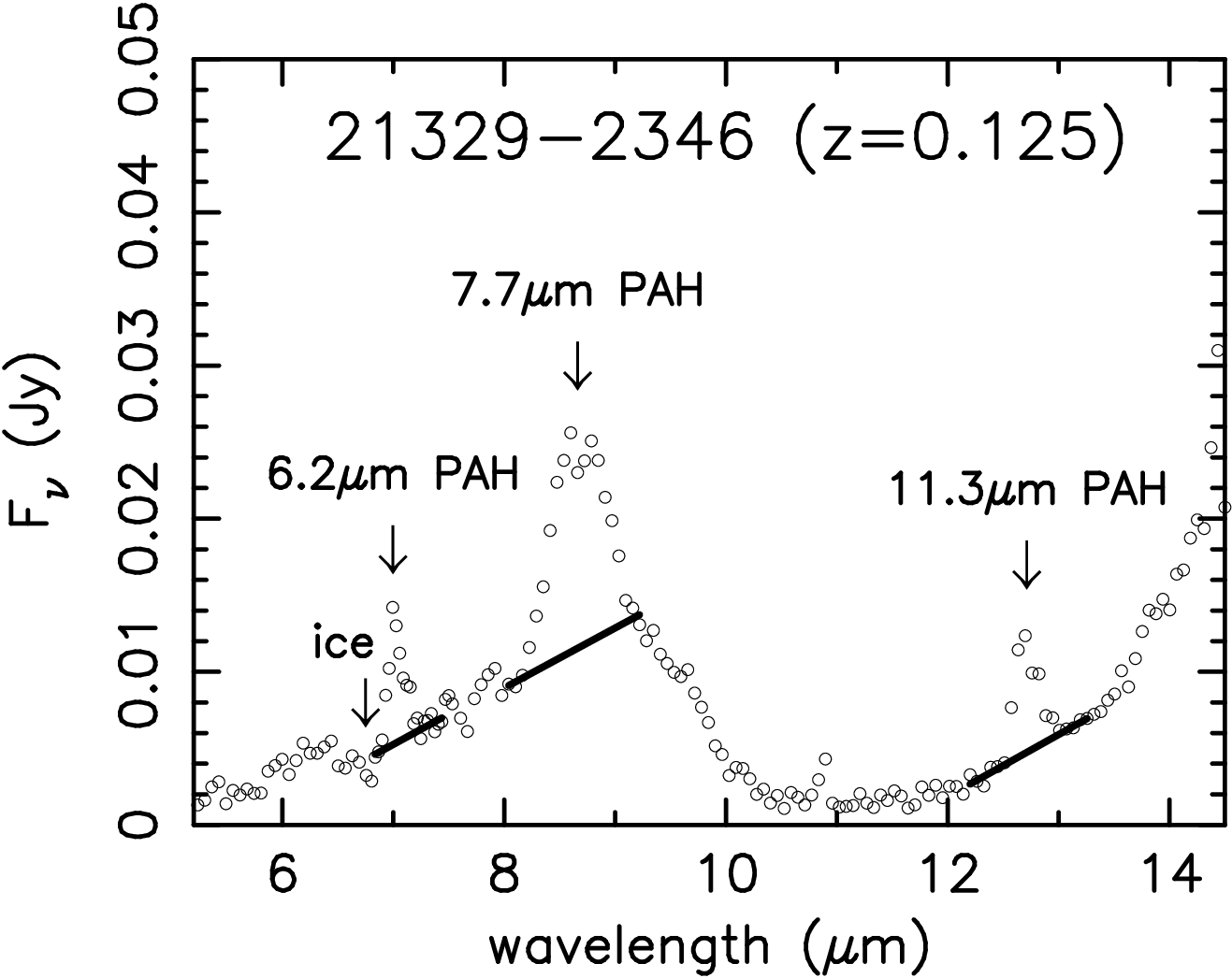} 
\includegraphics[angle=-90,scale=.24]{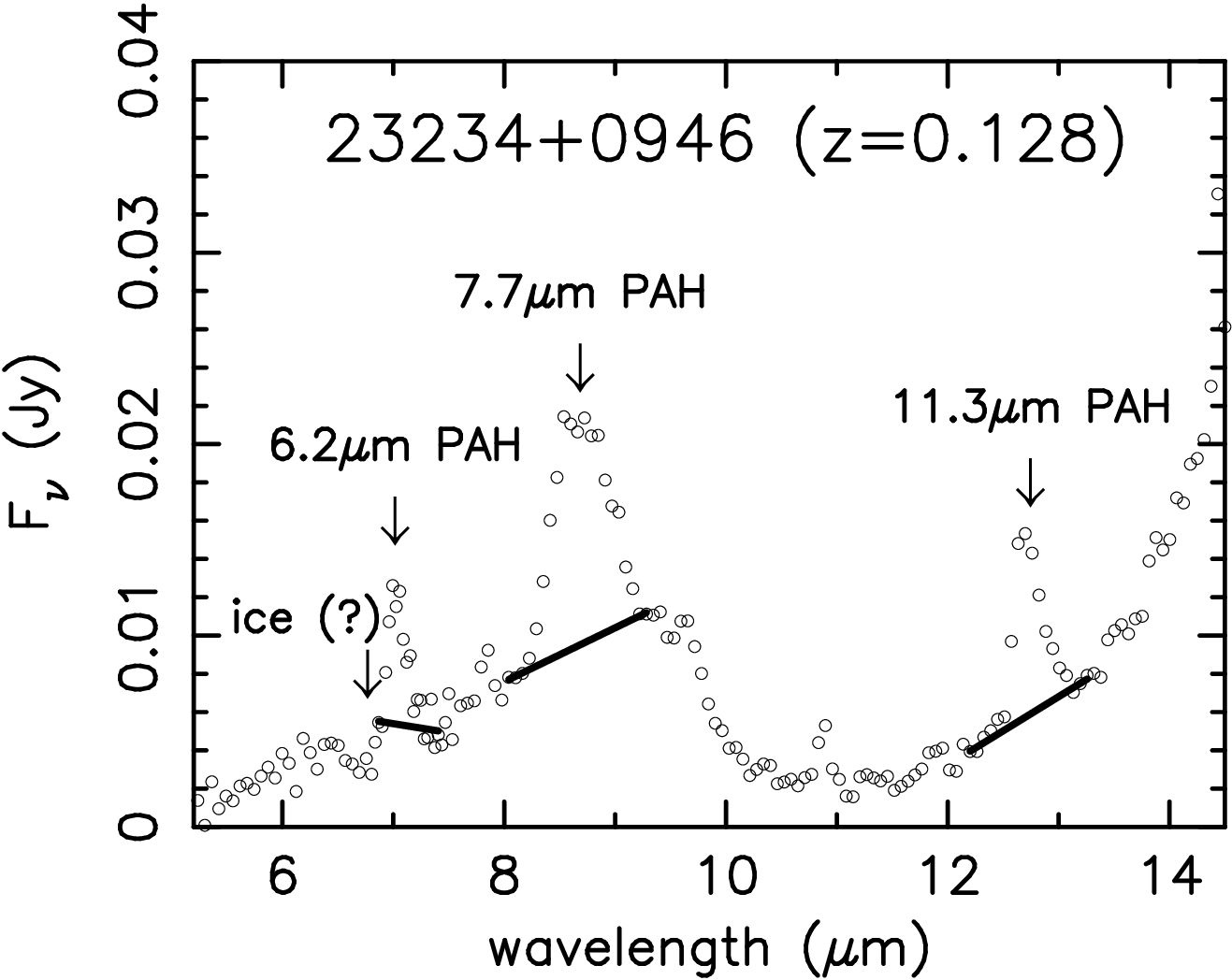} \\
\includegraphics[angle=-90,scale=.24]{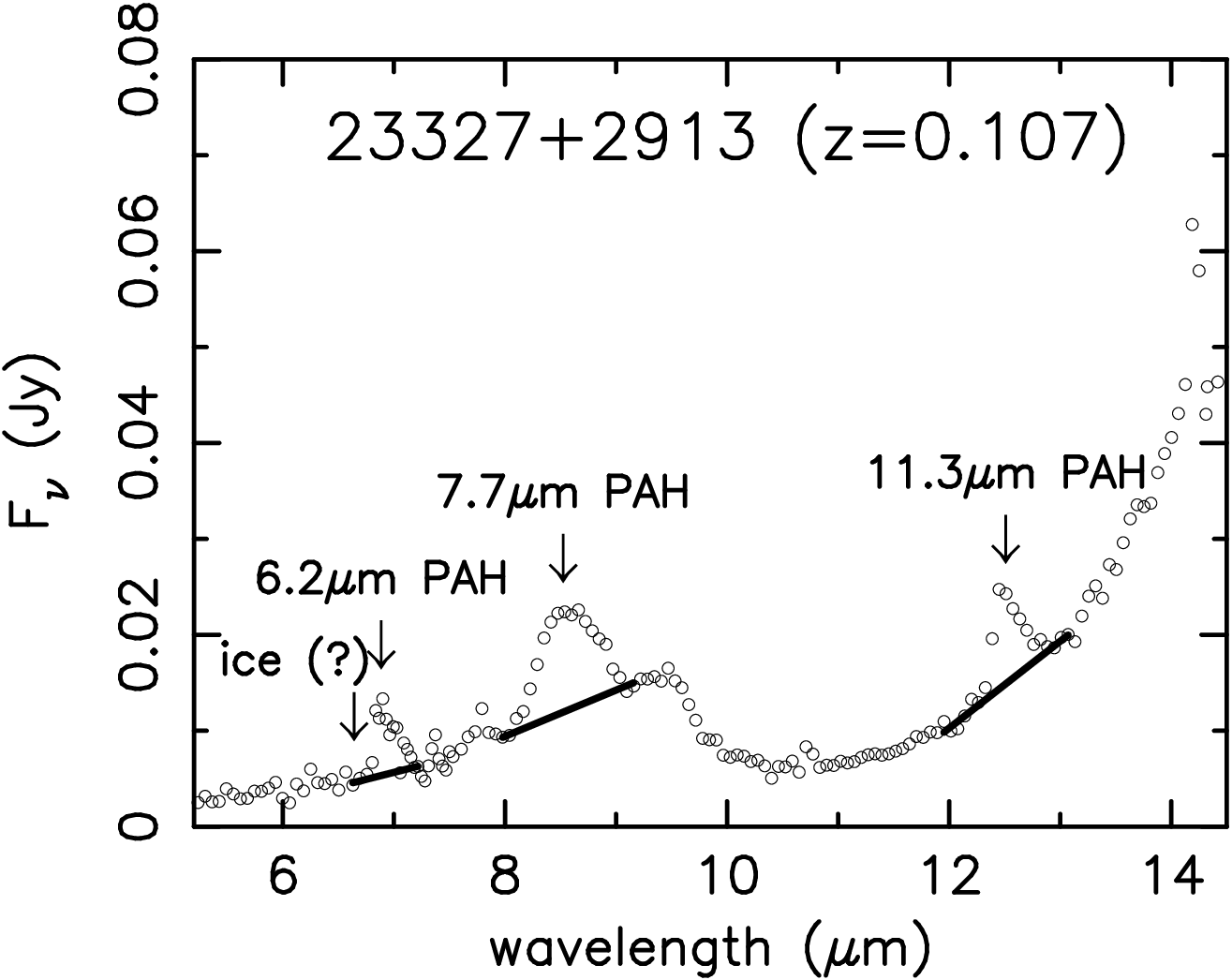} \\
\end{figure}

\clearpage

\begin{figure}
\includegraphics[angle=-90,scale=.24]{f5ac.eps} 
\includegraphics[angle=-90,scale=.24]{f5ad.eps} 
\includegraphics[angle=-90,scale=.24]{f5ae.eps} \\
\includegraphics[angle=-90,scale=.24]{f5af.eps} 
\includegraphics[angle=-90,scale=.24]{f5ag.eps} 
\includegraphics[angle=-90,scale=.24]{f5ah.eps} \\ 
\includegraphics[angle=-90,scale=.24]{f5ai.eps} 
\includegraphics[angle=-90,scale=.24]{f5aj.eps}
\includegraphics[angle=-90,scale=.24]{f5ak.eps} \\
\includegraphics[angle=-90,scale=.24]{f5al.eps} 
\includegraphics[angle=-90,scale=.24]{f5am.eps} 
\includegraphics[angle=-90,scale=.24]{f5an.eps} \\
\includegraphics[angle=-90,scale=.24]{f5ao.eps} 
\includegraphics[angle=-90,scale=.24]{f5ap.eps} 
\includegraphics[angle=-90,scale=.24]{f5aq.eps} \\
\end{figure}

\clearpage

\begin{figure}
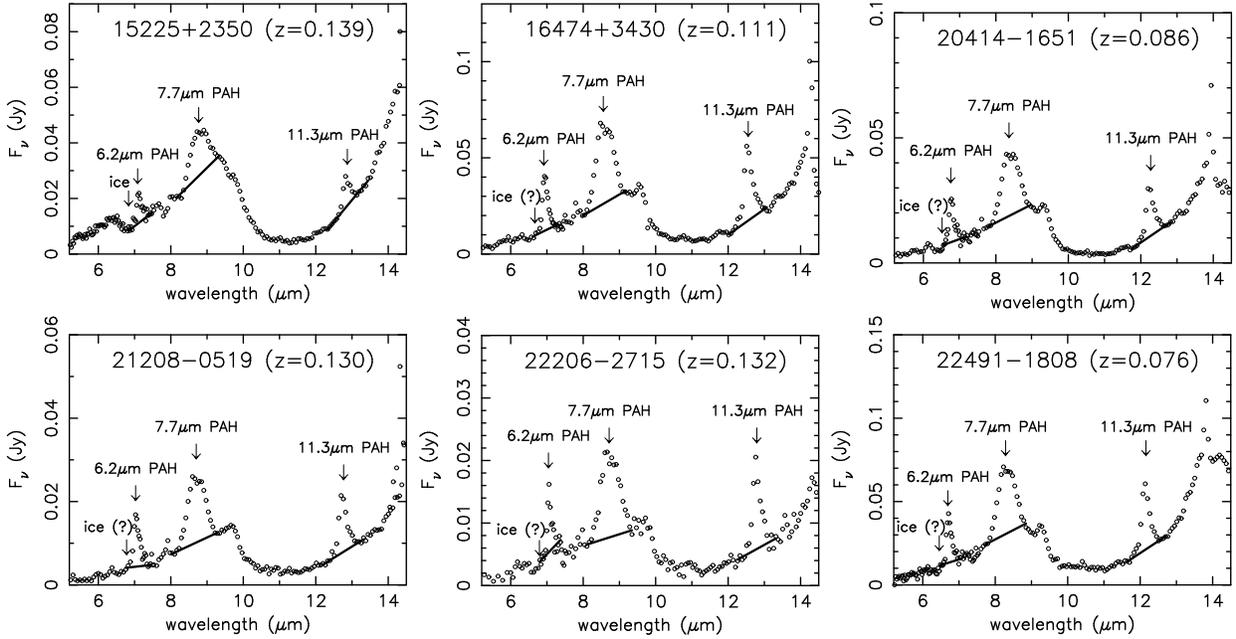

\includegraphics[angle=-90,scale=.24]{f5ar.eps} 
\includegraphics[angle=-90,scale=.24]{f5as.eps} 
\includegraphics[angle=-90,scale=.24]{f5at.eps} \\
\includegraphics[angle=-90,scale=.24]{f5au.eps} 
\includegraphics[angle=-90,scale=.24]{f5av.eps} 
\includegraphics[angle=-90,scale=.24]{f5aw.eps} \\
\caption{Spectra of all ULIRGs at $\lambda_{\rm obs}$ = 5.2--14.5
  $\mu$m, to zoom in on the PAH emission features and display them in
  detail.  The abscissa and ordinate are, respectively, observed
  wavelength in $\mu$m and flux in Jy, both shown at linear scale.
  For IRAS 10190+1322, the spectra of the eastern and western nuclei
  are shown separately.  The expected wavelengths of the 6.2 $\mu$m,
  7.7 $\mu$m, and 11.3 $\mu$m PAH emission features, as well as 6.0
  $\mu$m H$_{2}$O ice absorption feature, are indicated as down
  arrows.  The mark ``(?)'' is added, when detection is unclear.  The
  solid lines are adopted continuum levels to estimate the strengths
  of the detected PAH emission features.  }
\end{figure}

\clearpage

\begin{figure}
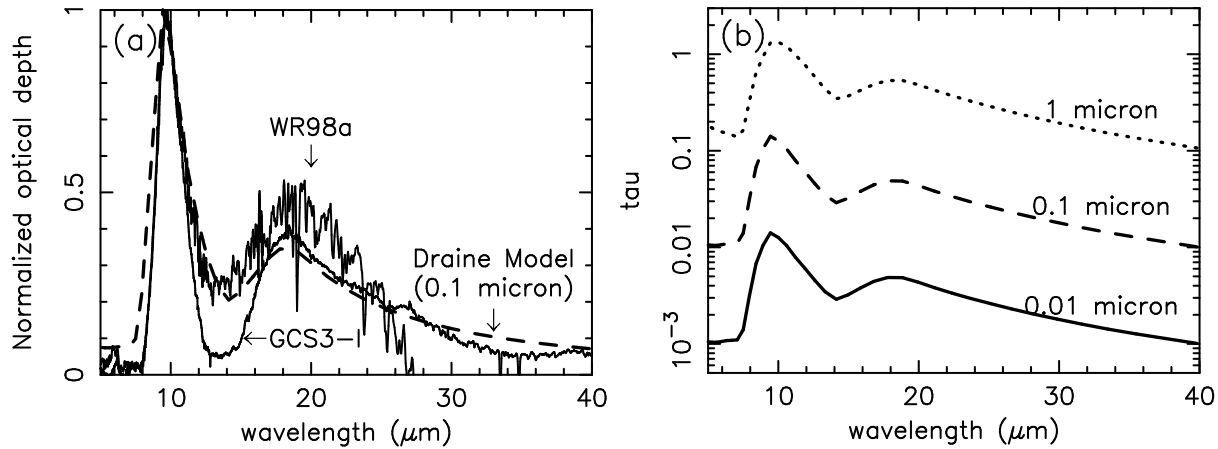

\includegraphics[angle=-90,scale=.35]{f6a.eps} 
\includegraphics[angle=-90,scale=.35]{f6b.eps} 
\caption{ Silicate dust absorption profile.  {\it (a)}: Solid lines
  are the observed silicate absorption profiles of the Galactic stars
  GCS 3-I and WR 98a \citep{chi06}.  The dashed line is the
  theoretical absorption profile of ``astronomical silicate'' grains
  with 0.1 $\mu$m in size \citep{dra84}.  All profiles are normalized at
  the 9.7 $\mu$m absorption peak.  The silicate absorption profile
  extends from 8 $\mu$m to $>$30 $\mu$m.  {\it (b)}: Theoretical
  absorption profile of ``astronomical silicate'' grains \citep{dra84}
  of different sizes, 1 $\mu$m (dotted line), 0.1 $\mu$m (dashed
  line), and 0.01 $\mu$m (solid line).  Adopted from
  http://www.astro.princeton.edu/$^{\sim}$draine/dust/dust.diel.html.
  The ordinate is in arbitrary units, but is proportional to the
  actual optical depth.  In this figure, only absorption is taken into
  account, and scattering is neglected, because scattering is
  negligible in this wavelength range (see the above webpage).  }
\end{figure}

\clearpage

\begin{figure}
\includegraphics[angle=-90,scale=.24]{f7a.eps} 
\includegraphics[angle=-90,scale=.24]{f7b.eps} 
\includegraphics[angle=-90,scale=.24]{f7c.eps} \\
\includegraphics[angle=-90,scale=.24]{f7d.eps} 
\includegraphics[angle=-90,scale=.24]{f7e.eps} 
\includegraphics[angle=-90,scale=.24]{f7f.eps} \\ 
\includegraphics[angle=-90,scale=.24]{f7g.eps} 
\includegraphics[angle=-90,scale=.24]{f7h.eps} 
\includegraphics[angle=-90,scale=.24]{f7i.eps} \\
\includegraphics[angle=-90,scale=.24]{f7j.eps}  
\includegraphics[angle=-90,scale=.24]{f7k.eps} 
\includegraphics[angle=-90,scale=.24]{f7l.eps} \\
\end{figure}

\clearpage

\begin{figure}
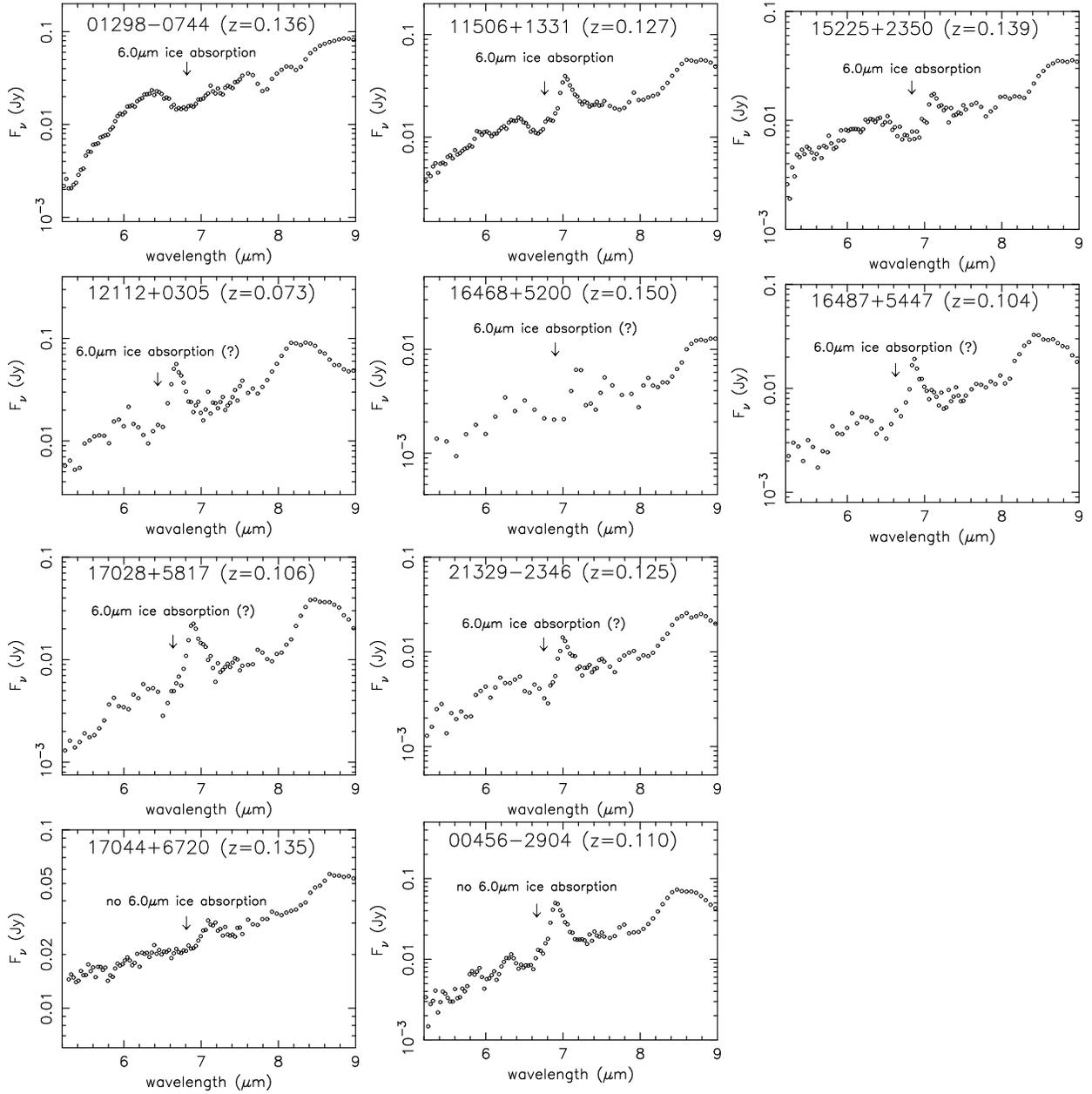


\includegraphics[angle=-90,scale=.24]{f7m.eps} 
\includegraphics[angle=-90,scale=.24]{f7n.eps} 
\includegraphics[angle=-90,scale=.24]{f7o.eps} \\ 
\includegraphics[angle=-90,scale=.24]{f7p.eps} 
\includegraphics[angle=-90,scale=.24]{f7q.eps} 
\includegraphics[angle=-90,scale=.24]{f7r.eps} \\
\includegraphics[angle=-90,scale=.24]{f7s.eps} 
\includegraphics[angle=-90,scale=.24]{f7t.eps} \\
\includegraphics[angle=-90,scale=.24]{f7u.eps} 
\includegraphics[angle=-90,scale=.24]{f7v.eps} 
\caption{IRS spectra at $\lambda_{\rm obs}$ = 5.2--9 $\mu$m for ULIRGs
  displaying clear 6.0 $\mu$m H$_{2}$O absorption feature (marked with
  ``6.0$\mu$m ice absorption'' in the first 15 plots).  Spectra of the
  next five ULIRGs [marked with ``6.0$\mu$m ice absorption (?)''] show
  possible signatures of the 6.0 $\mu$m ice absorption feature.
  Spectra of the last two ULIRGs, IRAS 17044+6720 and 00456$-$2904,
  marked with ``no 6.0$\mu$m ice absorption'' are shown as examples
  with undetected ice absorption features.  The abscissa is observed
  wavelength in $\mu$m in linear scale, and the ordinate is flux in Jy
  in decimal logarithmic scale.  }
\end{figure}

\clearpage

\begin{figure}
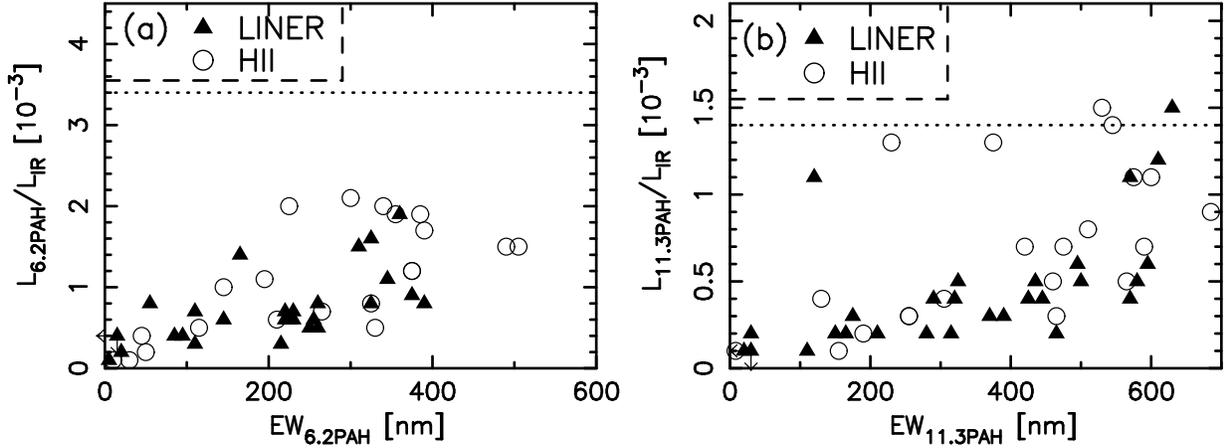

\includegraphics[angle=-90,scale=.35]{f8a.eps} 
\includegraphics[angle=-90,scale=.35]{f8b.eps} 
\caption{ {\it (a)}: Comparison of the rest-frame equivalent width of
  the 6.2 $\mu$m PAH emission (abscissa) and 6.2 $\mu$m PAH to
  infrared luminosity ratio (ordinate).  Filled triangle: ULIRGs
  classified optically as LINERs.  Open circle: ULIRGs classified
  optically as HII-regions.  The horizontal dotted line indicates the
  ratio for normal starburst galaxies with modest dust obscuration
  (L$_{\rm 6.2PAH}$/L$_{\rm IR}$ = 3.4 $\times$ 10$^{-3}$; see
  $\S$5.1).  In the upper-left small area separated by the dashed line,
  there are no data points.  {\it (b)}: The abscissa is the rest-frame
  equivalent width of the 11.3 $\mu$m PAH emission and the ordinate is
  the 11.3 $\mu$m PAH to infrared luminosity ratio.  The horizontal
  dotted line indicates the ratio for normal starburst galaxies with
  modest dust obscuration (L$_{\rm 11.3PAH}$/L$_{\rm IR}$ = 1.4
  $\times$ 10$^{-3}$; see $\S$5.1).  The larger scatter of L$_{\rm
    11.3PAH}$/L$_{\rm IR}$ in this Figure than for L$_{\rm
    6.2PAH}$/L$_{\rm IR}$ in Figure 8a may be due to intrinsically
  larger scatter of the 11.3 $\mu$m PAH strengths in normal starburst
  galaxies (see $\S$5.2.1).  }
\end{figure}

\clearpage

\begin{figure}
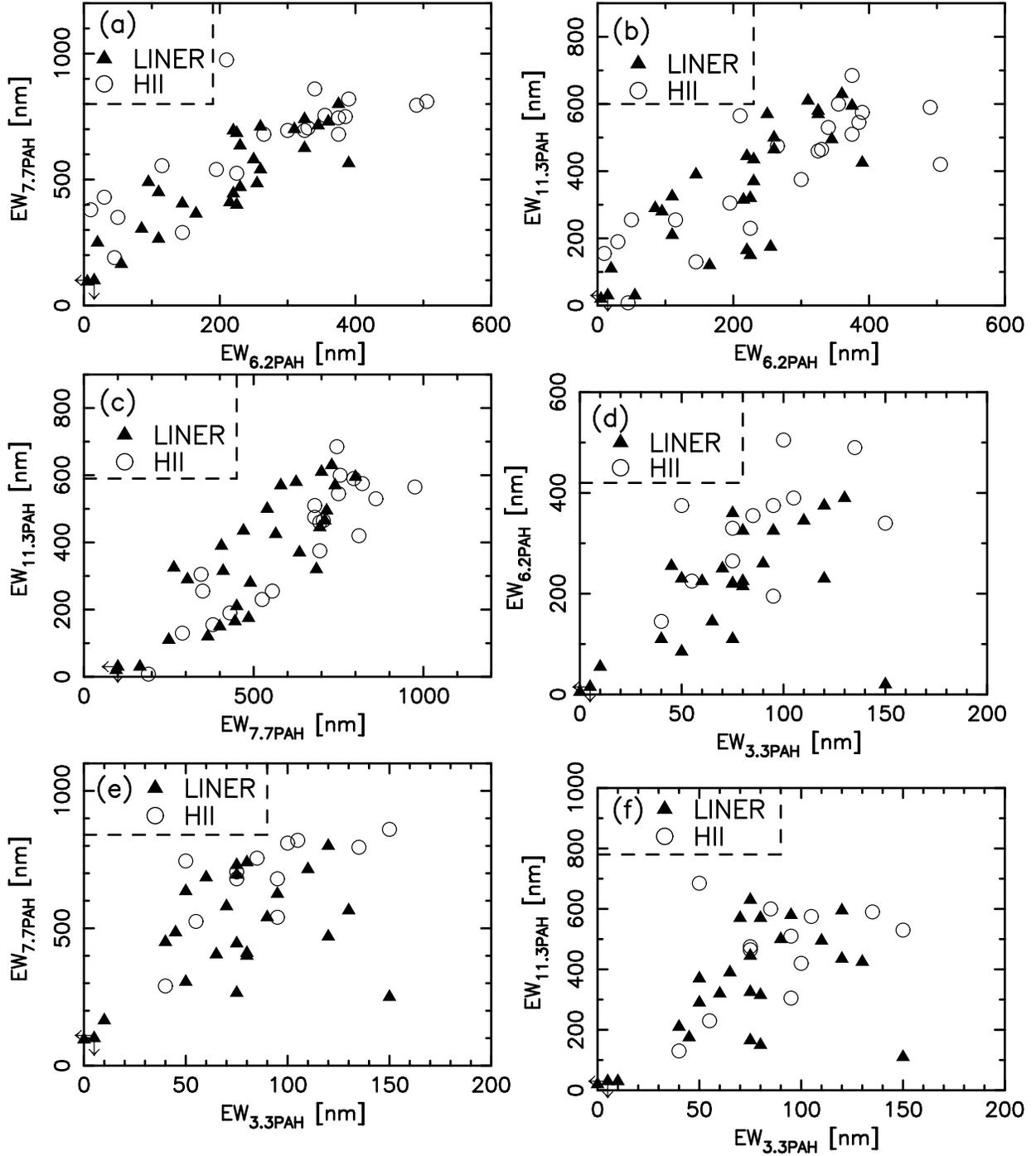

\includegraphics[angle=-90,scale=.35]{f9a.eps} 
\includegraphics[angle=-90,scale=.35]{f9b.eps} \\
\includegraphics[angle=-90,scale=.35]{f9c.eps} 
\includegraphics[angle=-90,scale=.35]{f9d.eps} \\
\includegraphics[angle=-90,scale=.35]{f9e.eps} 
\includegraphics[angle=-90,scale=.35]{f9f.eps} 
\caption{\small{
{\it (a)}: Comparison of the rest-frame equivalent widths of PAH
emission features in nm. 
The abscissa is the 6.2 $\mu$m PAH and the ordinate is the 7.7 $\mu$m
PAH emission. 
Filled triangle: ULIRGs classified optically as LINERs.
Open circle: ULIRGs classified optically as HII-regions.
In the upper-left small area separated with the dashed line, there are no
data points.
{\it (b)}: The abscissa is the 6.2 $\mu$m PAH and the ordinate is the
11.3 $\mu$m PAH emission. 
{\it (c)}: The abscissa is the 7.7 $\mu$m PAH and the ordinate is the
11.3 $\mu$m PAH emission.
{\it (d)}: The abscissa is the 3.3 $\mu$m PAH and the ordinate is the
6.2 $\mu$m PAH emission.
{\it (e)}: The abscissa is the 3.3 $\mu$m PAH and the ordinate is the
7.7 $\mu$m PAH emission.
{\it (f)}: The abscissa is the 3.3 $\mu$m PAH and the ordinate is the
11.3 $\mu$m PAH emission.
In the last three plots, the average EW$_{\rm 3.3PAH}$
value of both nuclei is plotted for IRAS 12112+0305 and 14348$-$1447
(Table 3).   
}}
\end{figure}

\clearpage

\begin{figure}
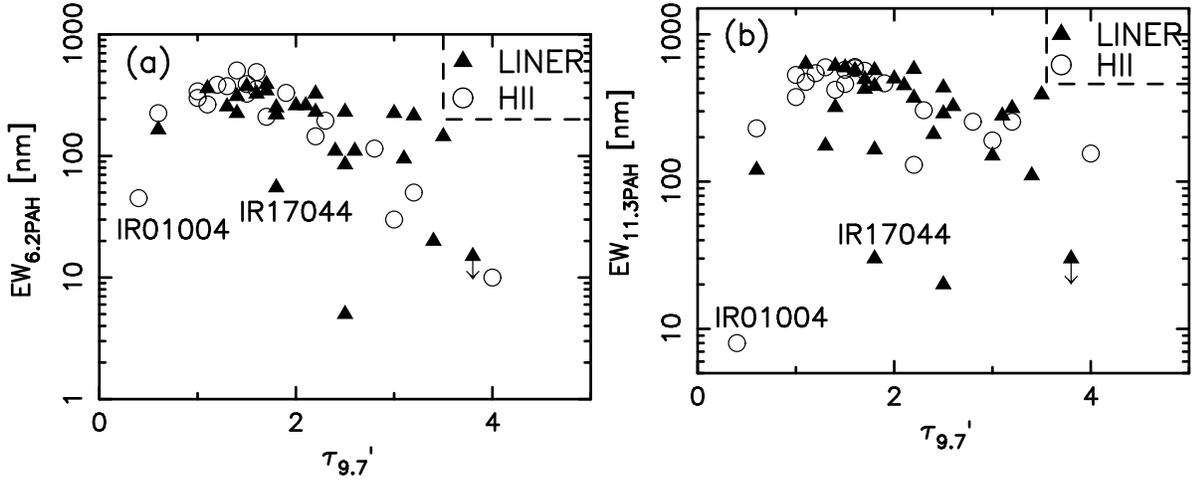

\includegraphics[angle=-90,scale=.35]{f10a.eps} 
\includegraphics[angle=-90,scale=.35]{f10b.eps} 
\caption{
{\it (a)}: Comparison of the strength of the 9.7 $\mu$m silicate dust
absorption ($\tau_{9.7}'$) (abscissa) and the rest-frame equivalent
widths of the 6.2 $\mu$m PAH emission feature (ordinate). 
Filled triangle: ULIRGs classified optically as LINERs.
Open circle: ULIRGs classified optically as HII-regions.
In the upper-right small area separated with the dashed line, there are no
data points.
{\it (b)}: The ordinate is the rest-frame equivalent width of the 11.3
$\mu$m PAH emission feature.
In both plots, the two outliers IRAS 01004$-$2237 and 17044+6720 
(weakly obscured AGNs; $\S$5.2.4) are indicated as ``IR01004'' and
``IR17044'', respectively.  
For IRAS 10190+1322, the average value of both nuclei (Table 3) are
plotted for PAH equivalent widths in the ordinate. 
}
\end{figure}

\begin{figure}
\begin{center}
\includegraphics[angle=-90,scale=.35]{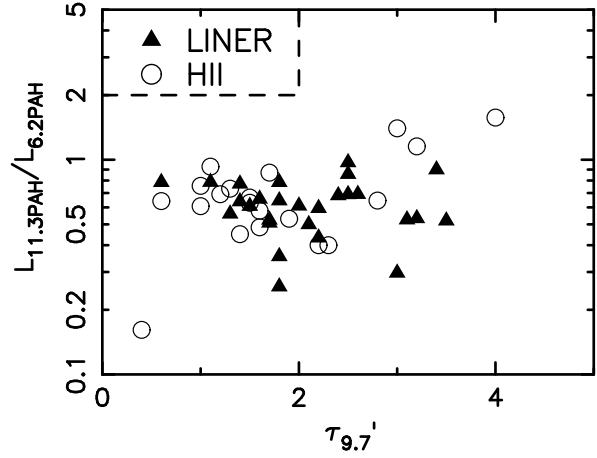} 
\end{center}
\caption{
Comparison of the strength of the 9.7 $\mu$m silicate dust absorption
($\tau_{9.7}'$) (abscissa) and 11.3 $\mu$m to 6.2 $\mu$m PAH luminosity
ratio (ordinate).
Filled triangle: ULIRGs classified optically as LINERs.
Open circle: ULIRGs classified optically as HII-regions.
In the upper-left small area separated with the dashed line, there are no
data points.
For IRAS 10190+1322, the 11.3 $\mu$m to 6.2 $\mu$m PAH luminosity
ratio in the ordinate is derived by combining spectra of both nuclei.
IRAS 08572+3915 is not plotted, because both the 6.2 $\mu$m and 11.3
$\mu$m PAH emission luminosities are upper limits (Table 3). 
}
\end{figure}

\begin{figure}
\begin{center}
\includegraphics[angle=-90,scale=.35]{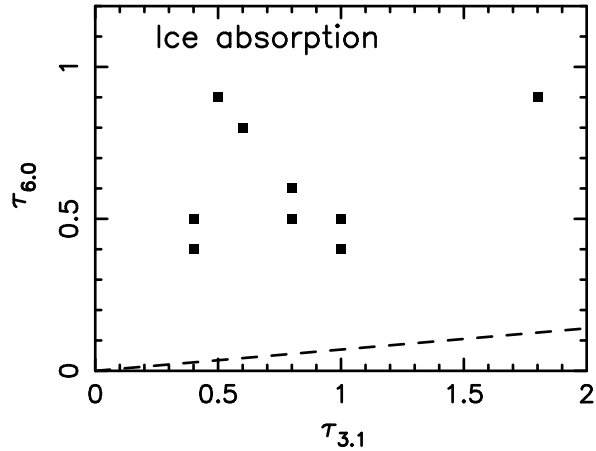} 
\end{center}
\caption{Comparison of the optical depths of the H$_{2}$O ice
  absorption features at 3.1 $\mu$m (abscissa) and 6.0 $\mu$m
  (ordinate).  The dashed line is the expected ratio from a laboratory
  H$_{2}$O ice sample at 10 K \citep{kea01}.  }
\end{figure}

\begin{figure}
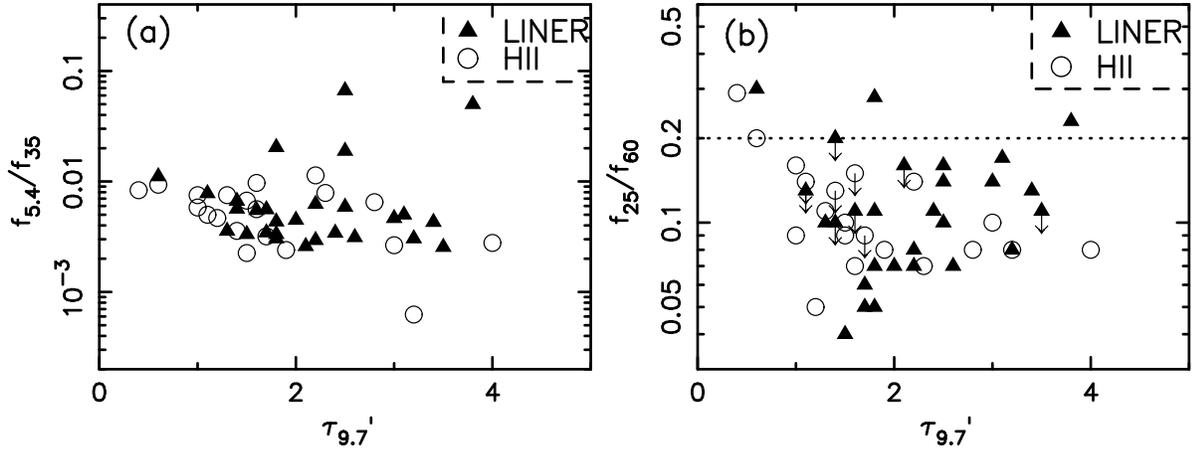

\includegraphics[angle=-90,scale=.35]{f13a.eps} 
\includegraphics[angle=-90,scale=.35]{f13b.eps} 
\caption{
Comparison of the strength of the 9.7 $\mu$m silicate dust absorption
and infrared continuum colors.
{\it (a)}:  The abscissa is the $\tau_{9.7}'$ value and the ordinate 
is the 5.4 $\mu$m to 35 $\mu$m flux ratio measured from the {\it Spitzer} IRS
spectra in Figures 3 and 4. 
Filled triangle: ULIRGs classified optically as LINERs.
Open circle: ULIRGs classified optically as HII-regions.
In the upper-right small area separated with the dashed line, there are no
data points.
{\it (b)}: The ordinate is the {\it IRAS} 25 $\mu$m to 60 $\mu$m flux ratio,
taken from Table 1.
The horizontal dotted line indicates the separation between warm and cool
ULIRGs \citep{san88b}. 
}
\end{figure}

\end{document}